\documentclass[useAMS,usenatbib]{mn2e}

\usepackage{graphicx}
\usepackage{times}%
\usepackage{natbib}
\usepackage{graphics}
\usepackage{epsfig}
\usepackage{array}

\newcommand{\gsim}{\,\lower2truept\hbox{${>\atop\hbox{\raise4truept\hbox{$\sim$}}}$}\,}
\newcommand{\pp}{~~~.}

\newcommand{\be}{\begin{equation}}
\newcommand{\ee}{\end{equation}}
\newcommand{\bea}{\begin{eqnarray}}
\newcommand{\eea}{\end{eqnarray}}

\renewcommand{\vec}[1]{ {\bmath #1} } 

\def\ltsima{$\; \buildrel < \over \sim \;$}
\def\simlt{\lower.5ex\hbox{\ltsima}}
\def\gtsima{$\; \buildrel > \over \sim \;$}
\def\simgt{\lower.5ex\hbox{\gtsima}}

\title[Hydrodynamical N-body simulations of coupled dark energy cosmologies]{Hydrodynamical N-body simulations of coupled dark energy cosmologies}

\author[M. Baldi, V. Pettorino, G. Robbers, V. Springel]{Marco Baldi$^{1}$, Valeria Pettorino$^{2}$, Georg Robbers$^{2}$, Volker Springel$^{1}$
\\ $^{1}$ Max-Planck-Institut f\"{u}r Astrophysik, Karl-Schwarzschild Strasse 1, D-85748 Garching, Germany.
\\ $^{2}$ Institut f\"{u}r Theoretische Physik, Universit\"{a}t Heidelberg, Philosophenweg 16, D-69120 Heidelberg, Germany.
}

\begin{document}


\pagerange{\pageref{firstpage}--\pageref{lastpage}} \pubyear{2008}

\maketitle

\label{firstpage}

\begin{abstract}

  If the accelerated expansion of the Universe at the present epoch is driven
  by a dark energy scalar field, there may well be a non-trivial coupling
  between the dark energy and the cold dark matter (CDM) fluid.  Such
  interactions give rise to new features in cosmological structure growth,
  like an additional long-range attractive force between CDM particles, or
  variations of the dark matter particle mass with time.  We have implemented
  these effects in the N-body code {\small GADGET-2} and present results of a
  series of high-resolution N-body simulations where the dark energy component
  is directly interacting with the cold dark matter.  As a consequence of the
  new physics, CDM and baryon distributions evolve differently both
  in the linear and in the nonlinear regime of structure formation.  Already
  on large scales a linear bias develops between these two components, which
  is further enhanced by the nonlinear evolution.  We also find, in contrast
  with previous work, that the density profiles of CDM halos are less
  concentrated in coupled dark energy cosmologies compared with $\Lambda $CDM,
  and that this feature does not depend on the initial conditions setup, but
  is a specific consequence of the extra physics induced by the
  coupling. Also, the baryon fraction in halos in the coupled models is
  significantly reduced below the universal baryon fraction. These features
  alleviate tensions between observations and the $\Lambda $CDM model on small
  scales. Our methodology is ideally suited to explore the predictions of
  coupled dark energy models in the fully non-linear regime, which can provide
  powerful constraints for the viable parameter space of such scenarios.

\end{abstract}



\section{Introduction}
\label{i}

The last decade has seen an astonishing amount of new cosmological data from
many different experiments, ranging from large-scale structure surveys
\citep[e.g.]{Percival_etal_2001} to Cosmic Microwave Background (CMB)
\citep{wmap5} and Type Ia Supernovae \citep{Riess_etal_1998,
  Perlmutter_etal_1999, SNLS} observations. These experiments all consistently
show that the Universe is almost spatially flat with a current expansion rate
of about $70\,{\rm km\, s^{-1} Mpc^{-1}}$, and contains a total amount of
matter that accounts for only $\sim 24\%$ of the total energy density.  From a
theoretical point of view, understanding the remaining $76\%$ -- which must be
in the form of some dark energy (DE) component able to drive an accelerated
expansion -- is a serious challenge.

A simple cosmological constant would be in agreement with a large number of
observational datasets, but would also raise two fundamental questions
concerning its fine-tuned value and the coincidence of its domination over
cold dark matter (CDM) only at a relatively recent cosmological epoch.  An
alternative consists of identifying the dark energy component with a dynamic
scalar field \citep{Wetterich_1988, Ratra_Peebles_1988}, thereby seeking
an explanation of the fundamental problems challenging the cosmological
constant in the properties of the dynamic evolution of such scalar field.

An interesting suggestion recently developed concerns the investigation of
possible interactions between the dark energy scalar field and other matter
species in the Universe \citep{Wetterich_1995, Amendola_2000, Farrar2004,
  Gubser2004, Farrar2007}. The existence of such a coupling could provide a
unique handle for a deeper understanding of the DE problem
\citep{Quartin:2008px, Brookfield:2007au, Gromov:2002ek, Mangano:2002gg,
  Anderson:1997un}. It is then crucial to understand in detail what effects
such a coupling imprints on observable features like, for example, the CMB and
structure formation \citep{LaVacca_etal_2009, Bean:2008ac, Bertolami:2007zm, Matarrese_etal_2003, Wang:2006qw,
  Guo:2007zk, Mainini:2007ft, Lee:2006za}.

In this paper, we perform the first fully self-consistent high-resolution
hydrodynamic N-body simulations of cosmic structure formation for a selected
family of coupled DE cosmologies.  The interaction between DE and CDM is
expected to imprint characteristic features in linear and nonlinear
structures, and could possibly open up new ways to overcome a series of
observational challenges for the $\Lambda $CDM concordance cosmology, ranging
from the satellite abundance in CDM halos \citep{Navarro_Frenk_White_1995}, to
the observed low baryon fraction in large galaxy clusters
\citep{Allen_etal_2004, Vikhlinin_etal_2006, LaRoque_etal_2006}, and to the
so-called  ``cusp-core'' problem for the density
profiles of CDM halos, that was first raised in relation to observations of dwarf and low-surface-brightness galaxies \citep{Moore_1994, Flores_Primack_1994, Simon_etal_2003}, but that more recent studies seem to extend to a wider range of objects including Milky-Way type spiral galaxies \citep{Navarro_Steinmetz_2000, Binney_Evans_2001} up to galaxy groups and clusters \citep{Sand_etal_2002, Sand_etal_2004, Newman_etal_2009}.

In the following we will consider DE as the energy associated with a scalar
field $\phi$ \citep{Wetterich_1988, Ratra_Peebles_1988} whose energy density
and pressure are defined respectively as the $(0,0)$ and $(0,i)$ components of
its stress energy tensor, so that\footnote{In this work we always denote with
  a prime the derivative with respect to the conformal time $\tau $, and with
  an overdot the derivative with respect to the cosmic time $t$. The two time
  variables are related via the equation: ${\rm d}\tau = {\rm d}t / a$.}:
\be \label{rho_phi} \rho_\phi \equiv \frac{1}{2}\frac{{\phi '}^2}{a^2} +
U(\phi) \,, \ee \be \label{p_phi} p_\phi \equiv \frac{1}{2}\frac{{\phi
    '}^2}{a^2} - U(\phi) \,, \ee with $U(\phi)$ being the self interaction
potential.  
The models we investigate in this work by means of detailed high-resolution N-body simulations are physically identical to the ones already studied by \citet{Maccio_etal_2004}.
With respect to this previous work, however, our analysis 
significantly improves on the statistics
of the number of cosmic structures analyzed in each of our different
cosmological models and on the dynamic range of the simulations.  We also
extend the analysis to a wider range of observable effects arising from the
new physics of the DE-CDM interaction, and we include hydrodynamics.
Despite the physical models investigated be identical, our outcomes are starkly different from the ones found in \citet{Maccio_etal_2004}.

This work is organized as follows. In Section~{\ref{cde}} we detail the
coupled DE cosmologies under investigation, both regarding the background
evolution and the perturbations, and we introduce a numerical package to
integrate the full set of background and perturbation equations up to linear
order.  In Section~\ref{sim}, we describe the numerical methods we use, and
the particular set of simulations we have run.  We then present and discuss
our results for N-body simulations for a series of coupled DE models in
Section~\ref{results}.  Finally, in Section~\ref{concl} we draw our
conclusions.

\section{Coupled dark energy cosmologies} \label{cde}

Coupled cosmologies can be described following the consideration
\citep{Kodama_Sasaki_1984} that in any multicomponent system, though the total
stress energy tensor ${T{^\mu }}_{\nu}$ is conserved
\begin{equation} \label{tensor_conserv_total}
 \sum_{\alpha} \nabla_\nu T^{ \nu}_{(\alpha) \mu} = 0 \,,
\end{equation}
the ${T{^\mu }}_{\nu (\alpha)}$ for each species $\alpha$ is, in general, not
conserved and its divergence has a source term $Q_{(\alpha)\mu}$ representing
the possibility that species are coupled:
\begin{equation} \label{tensor_conserv_alpha}
 \nabla_\nu T^{ \nu}_{(\alpha) \mu} = Q_{(\alpha) \mu} \,,
\end{equation}
with the constraint
\begin{equation} \label{Q_conserv_total}
 \sum_{\alpha} Q_{(\alpha) \mu} = 0 \pp 
\end{equation}
Furthermore, we will assume a flat Friedmann-Robertson-Walker (FRW) cosmology,
in which the line element can be written as $ {\rm d}s^2 = a^2(\tau)(-{\rm
  d}\tau^2 + \delta_{ij}{\rm d}x^i {\rm d}x^j) $ where $a(\tau)$ is the scale
factor.

\subsection{Background} \label{bkg}

The Lagrangian of the system is of the form: \be \label{L_phi} {\cal L} =
-\frac{1}{2}\partial^\mu \phi \partial_\mu \phi - U(\phi) -
m(\phi)\bar{\psi}\psi + {\cal L}_{\rm kin}[\psi] \,, \ee in which the mass of
matter fields $\psi$ coupled to the DE is a function of the scalar field $\phi
$. In the following we will consider the case in which the DE is only coupled
to cold dark matter (CDM, hereafter denoted with a subscript $c$). The choice
$m(\phi)$ specifies the coupling and as a consequence the source term
$Q_{(\phi) \mu}$ via the expression: \be Q_{(\phi) \mu} = \frac{\partial
  \ln{m(\phi)}}{\partial \phi} \rho_c \, \partial_\mu \phi . \ee Due to the
constraint (\ref{Q_conserv_total}), if no other species is involved in the
coupling, $Q_{(c) \mu} = - Q_{(\phi) \mu}$.

The zero-component of equation (\ref{tensor_conserv_alpha}) provides the
conservation equations for the energy densities of each species: 
\bea
\label{rho_conserv_eq_phi} \rho_{\phi}' &=& -3 {\cal{H}} \rho_{\phi} (1 + w_\phi) - Q_{(\phi)0} \,\,\,\, , \\
\label{rho_conserv_eq_c} \rho_{c}' &=& -3 {\cal{H}} \rho_{c} + Q_{(\phi)0} \pp
\nonumber \eea Here we have treated each component as a fluid with
${T^\nu}_{(\alpha)\mu} = (\rho_\alpha + p_\alpha) u_\mu u^\nu + p_\alpha
\delta^\nu_\mu$, where $u_\mu = (-a, 0, 0, 0)$ is the fluid 4-velocity and
$w_\alpha \equiv p_\alpha/\rho_\alpha$ is the equation of state.  The class of
models considered here corresponds to the choice: \be \label{coupling_const}
m(\phi) = m_0 e^{-\beta(\phi) \frac{\phi}{M}} \,, \ee with the coupling term
equal to \be Q_{(\phi)0} = - \frac{\beta(\phi)}{M} \rho_c \phi' \,. \ee This
set of cosmologies has been widely investigated, for $\beta(\phi )$ given by a
constant, both in its background and linear perturbation features
\citep{Wetterich_1995, Amendola_2000} as well as with regard to the effects on
structure formation \citep{Amendola_2004, Pettorino_Baccigalupi_2008} , and
via a first N-body simulation \citep{Maccio_etal_2004}.

\subsection{Linear perturbations} \label{prt}

We now perturb the quantities involved in our cosmological framework up to the
first order in the perturbations \citep{Kodama_Sasaki_1984,
  Ma_Bertschinger_1995}. The perturbed metric tensor can then be written as
\be \tilde{g}_{\mu \nu} (\tau, \bf{x}) = g_{\mu \nu}(\tau) + \delta g_{\mu
  \nu}(\tau, \bf{x}) \,, \ee where $\delta g_{\mu \nu} \ll 1$ is the linear
metric perturbation, whose expression in Fourier space is given by:
\bea \delta g_{00} = -2 a^2 A Y  \ \ , \ \  \delta g_{0i} = -a^2 B Y_i \ &\,,& \nonumber \\
\ \ \delta g_{ij} = 2 a^2 [H_L Y \delta_{ij} + H_T Y_{ij}] &\,,&\eea where $A,
B, H_L, H_T$ are functions of time and of the wave vector
$\bf{k}$, and $Y_i, Y_{ij}$ are the vector and tensor harmonic
functions obtained by deriving Y, defined as the solution of the Laplace
equation $ \delta^{ij}\nabla_i \nabla_j Y = -|k|^2 Y$.  Analogously, the
perturbed stress energy tensor for each fluid $(\alpha)$ can be written as
$\tilde{T}^{\mu}_{(\alpha) \nu} = T^{\mu}_{(\alpha) \nu} + \delta
T^{\mu}_{(\alpha) \nu}$ where the perturbations read as: \bea
\label{T_fluid_pert} 
&\delta T^{0}_{(\alpha) 0}& = - \rho_{(\alpha)} \delta_{(\alpha)} Y  \,,\\ 
&\delta T^{0}_{(\alpha) i}& = h_{(\alpha)}( v_{(\alpha)} - B) Y_i   \,, \nonumber \\ 
&\delta T^{i}_{(\alpha) 0}& = - h_{(\alpha)} v_{(\alpha)} Y^i  \,, \nonumber \\ 
&\delta T^{i}_{(\alpha) j}& = p_{(\alpha)} \left[ \pi_{L (\alpha)} Y \delta^i_j + \pi_{T (\alpha)} Y^i_j \right]  \,. \nonumber 
\eea
The perturbed conservation equations then become:
\be 
\label{ev_delta} 
(\rho_\alpha \delta_\alpha)' + 3 {\cal H} \rho_\alpha \delta_\alpha + h_\alpha (k v_\alpha + 3 H'_L) + 3 {\cal H} p_\alpha \pi_{L \alpha} = - \delta Q_{(\alpha)0} 
\ee
for the energy density perturbation $\delta_\alpha = \delta \rho_\alpha / \rho_\alpha$, and:
\bea 
\label{ev_v} 
& &\left[ h_{\alpha}(v_\alpha - B) \right]' + 4 {\cal H} h_\alpha (v_\alpha - B) \nonumber \\
& &- k p_\alpha \pi_{L \alpha} - h_\alpha k A + \frac{2}{3}k p \pi_{T \alpha} = \delta Q_{(\alpha)i} 
\eea
for the velocity perturbation $v_\alpha$.

The scalar field $\phi$ can also be perturbed, yielding  in  Fourier space \be \label{phi_pert} \tilde{\phi} = \phi + \delta \phi = \phi + \chi(\tau) Y \,.\ee
Furthermore, we can express the perturbations of the source as:
\be 
\delta Q_{(\phi)0} = -\frac{\beta(\phi)}{M} \rho_c \delta \phi' - \frac{\beta(\phi)}{M} \phi' \delta \rho_c - \frac{\beta_{,\phi}}{M}\phi' \rho_c \delta \phi \,,
\ee
\be 
\delta Q_{(\phi)i} = k\frac{\beta(\phi)}{M} \rho_c \delta \phi \,.
\ee
In the Newtonian gauge $(B=0, H_T=0, H_L = {\bf{\Phi}}, A = \bf{\Psi})$ the set of equations for the density and velocity perturbations for DE and CDM read:

\bea 
\label{eq_newtonian_gauge} 
\delta \rho_\phi' + 3 {\cal H} (\delta \rho_\phi + \delta p_\phi) + k h_\phi v_\phi + 3 h_\phi {\bf{\Phi}}' = & & \\
 \frac{\beta(\phi)}{M}\rho_c \delta \phi' + \frac{\beta(\phi)}{M} \phi' \delta \rho_c + \frac{\beta_{,\phi}}{M} \phi' \delta \phi \rho_c \,,& & \nonumber \\ 
\delta \rho_c' + 3 {\cal H} \delta \rho_c + k \rho_c v_c + 3 \rho_c {\bf{\Phi}}' = & & \nonumber \\
-\frac{\beta(\phi)}{M} \rho_c \delta \phi' - \frac{\beta(\phi)}{M} \phi' \delta \rho_c - \frac{\beta_{,\phi}}{M} \phi' \delta \phi \rho_c \,, & & \nonumber \\ 
h_\phi v_\phi' + \left(h_\phi' + 4 {\cal H} h_\phi \right) v_\phi - k \delta p_\phi - k h_\phi {\bf{\Psi}} = 
k \frac{\beta(\phi)}{M} \rho_c \delta \phi \,, & & \nonumber \\ 
v_c' + \left( {\cal H} - \frac{\beta(\phi)}{M} \phi' \right) v_c - k {\bf{\Psi}} = 
- k \frac{\beta(\phi)}{M} \delta \phi \,. \nonumber & & 
\eea
The perturbed Klein Gordon equation in Newtonian gauge reads:
\bea 
\delta \phi'' + 2{\cal H} \delta \phi' + \left(k^2 + a^2 U_{,\phi\phi} \right) \delta \phi - \phi' \left({\bf{\Psi}}' - 3{\bf{\Phi}'} \right) &+& \nonumber \\
 2a^2 U_{,\phi} {\bf{\Psi}} = 3 {\cal H}^2 \Omega_c \left[ \beta(\phi) \delta_c + 2 \beta(\phi) {\bf{\Psi}} + \beta_{,\phi}(\phi) \delta \phi \right] & & \,.
\eea

For the N-body implementation we are interested in, the Newtonian limit holds,
for which $\lambda \equiv {\cal H}/k \ll 1$. In this case we have \be
\label{deltaphi_newt_limit} 
\delta \phi \sim 3 \lambda^2 \Omega_c \beta(\phi) \delta_c \,.
\ee
In this limit, the gravitational potential is approximately given by
\be 
\label{grav_pot_newt_limit} {\bf{\Phi}} \sim \frac{3}{2} \frac{\lambda^2}{M^2}
\sum_{\alpha \neq \phi} \Omega_\alpha \delta_\alpha \,.  \ee We can then
define an effective gravitational potential \be {\bf{\Phi_c}} \equiv
{\bf{\Phi}} + \frac{\beta(\phi)}{M} \delta \phi , \ee which also reads, in
real space and after substituting the expressions for ${\bf{\Phi}}$
(Eqn.~\ref{grav_pot_newt_limit}) and for $\delta \phi$
(Eqn.~\ref{deltaphi_newt_limit}): \be \nabla^2 {\bf{\Phi_c}} = -\frac{a^2}{2}
\rho_c \delta_c \left(1+2 \beta^2(\phi)\right) -\frac{a^2}{2} \sum_{\alpha
  \neq \phi, c} \rho_\alpha \delta_\alpha \,, \ee where the last term takes
into account the case in which other components not coupled to the DE are present
in the total energy budget of the Universe.  Cold dark matter then feels an
effective gravitational constant \be
\label{G_eff} 
\tilde{G}_{c} = G_{N}[1+2\beta^2(\phi)], \ee where $G_{N}$ is the usual
Newtonian value.  Therefore, the strength of the gravitational interaction is
not a constant anymore if $\beta$ is a function of the scalar
field $\phi$.  The last equation in (\ref{eq_newtonian_gauge}), written in
real space and in terms of the effective gravitational potential, gives a
modified Euler equation of the form: \bea
& &\vec{\nabla }\vec{v}_c' + \left({\cal H} - \frac{\beta(\phi)}{M} \phi' \right) \vec{\nabla }\vec{v}_c + \nonumber \\
& &\frac{3}{2} {\cal H}^2 \left[\Omega_c \delta_c + 2 \Omega_c \delta_c
  \beta^2(\phi) + \sum_{\alpha \neq \phi,c} \Omega_\alpha \delta_\alpha
\right] = 0 \,.  \eea

As in \citet{Amendola_2004}, if we assume that the CDM is concentrated in one
particle of mass $m_c$ at a distance $r$ from a particle of mass $M_c$ at the
origin, we can rewrite the CDM density contribution as \be
\label{mass} 
\Omega_c \delta_c = \frac{8 \pi G M_c e^{-\int{\beta(\phi){\rm d}\phi}}
  {\delta}(0)} {3{\cal H}^2 a}\ , \ee where we have used the fact that a
non-relativistic particle at position $\vec{r}$ has a density given by $m_c n
{\delta(\vec{r})}$ (where ${\delta(\vec{r})}$ stands for the Dirac
distribution) with mass given by $m_{c} \propto e^{-\int{\beta(\phi){\rm
      d}\phi}}$, formally obtained from equation (\ref{rho_conserv_eq_c}). We
have further assumed that the density of the $M_c$ mass particle is much
larger than $\rho_c$.  The Euler equation in cosmic time (${\rm d}t = a\, {\rm
  d}\tau$) can then be rewritten in the form of an acceleration equation for
the particle at position $\vec{r}$:
\begin{equation}
\label{CQ_euler}
\dot{\vec{v}}_{c} = -\tilde{H}\vec{v}_{c} - \vec{\nabla }\frac{\tilde{G}_{c}\tilde{M}_{c}}{r} \,,
\end{equation}
where we explicitely see that the usual equation is modified in three ways.

First, the velocity-dependent term now contains an additional contribution given by the
second term of the expression defining $\tilde{H}$: \be
\label{friction_term} 
\tilde{H} \equiv H \left(1 - \frac{\beta(\phi)}{M}
  \frac{\dot{\phi}}{H}\right)\ .  \ee Second, the CDM particles feel an effective
gravitational constant $\tilde{G}$ given by (\ref{G_eff}).  Third, the CDM
particles have an effective mass, varying with time, given by: \be
 \label{effective_mass} 
 \tilde{M}_c \equiv M_c e^{-\int{\beta(\phi)\frac{{\rm d}\phi}{{\rm d}a}{\rm d}a}} \pp 
\ee 
Eqn.~\ref{CQ_euler} is very important for our discussion since it represents the starting point for the implementation of coupled DE models in an N-body code. We will discuss in detail how this implementation is realized in Sec.~\ref{methods}, but it is important to stress here that Eqn.~\ref{CQ_euler} is written in a form that explicitely highlights its vectorial nature, which has not been presented in previous literature. The vectorial nature of Eqn.~\ref{CQ_euler} is a key point in its numerical implementation and therefore needs to be properly taken into account.

In the N-body analysis carried out in the present paper, we consider $\beta$
to be a constant, so that the effective mass formally reads $\tilde{M}_c
\equiv M_c e^{-\beta\left(\phi-\phi_0\right)} $. We defer an investigation of
variable $\beta(\phi)$ to future work.  We have numerically solved the full
background and linear perturbation equations with a suitably modified version
of {\small CMBEASY} \citep{CMBEASY}, briefly described in the following
section.

\subsection{Background and linear perturbation integration of coupled DE
  models: a modified {\small CMBEASY} package}\label{CMBEASY_description}

We have implemented the full background and linear perturbation equations
derived in the previous section in the Boltzmann code {\small CMBEASY}
\citep{CMBEASY} for the general case of a DE component coupled to dark matter
via a coupling term given by Eqn.~(\ref{tensor_conserv_alpha}).  The form of
this coupling, as well as the evolution of the DE (either modelled as a scalar
field or purely as a DE fluid), can be freely specified in our implementation.

Compared to the standard case of uncoupled DE, the modifications include a
modified behaviour of the background evolution of CDM and DE given by
Eqs.~(\ref{rho_conserv_eq_phi})~and~(\ref{rho_conserv_eq_c}), as well as the
implementation of the linear perturbations described by
Eqn.~(\ref{eq_newtonian_gauge}), and their corresponding adiabatic initial
conditions.  The presence of a CDM-DE coupling furthermore complicates the
choice of suitable initial conditions even for the background quantities,
since dark matter no longer scales as $a^{-3}$, and so cannot simply be
rescaled from its desired value today. For each of the models considered here,
we choose to set the initial value of the scalar field close to its tracker
value in the uncoupled case, and then adjust the value of the potential
constant $\Lambda$ (see Eqn.~(\ref{U_def}) below) and the initial CDM energy
density such that we obtain the desired present-day values.

\section{The simulations} \label{sim}

The aim of this work is to investigate the effects that a coupling between
DE and other cosmological components, as introduced in Sec.~\ref{cde}, can
have on cosmic structure formation, with a particular focus on the nonlinear
regime which is not readily accessible by the linear analytic approach
described in Sec.~\ref{prt}.  To this end we study a set of cosmological
N-body simulations performed with the code {\small GADGET-2} \citep{gadget-2},
that we suitably modified for this purpose.

The required modifications of the standard algorithms of the N-body code for
simulating coupled DE cosmologies are extensively described in
Sec.~\ref{methods}. Interestingly, they require to reconsider and in part drop
several assumptions and approximations that are usually adopted in N-body
simulations. We note that previous attempts to use cosmological N-body
simulations for different flavours of modified Newtonian gravity have been
discussed, for example, in
\citet{Maccio_etal_2004,Nusser_Gubser_Peebles_2005,Stabenau_Jain_2006,Kesden_Kamionkowski_2006,Springel2007,Laszlo_Bean_2008,
  Sutter_Ricker_2008, Oyaizu_2008,Keselman_Nusser_Peebles_2009} but to our knowledge \citet{Maccio_etal_2004} is the
only work to date focusing on the properties of nonlinear structures in models
of coupled quintessence.

Therefore, with our modified version of {\small GADGET-2} we first ran a set
of low-resolution cosmological simulations ($L_{\rm box} = 320 h^{-1}{\rm
  Mpc}$, $N_{\rm part}= 2 \times 128^{3}$) for the same coupled DE models
investigated in \citet{Maccio_etal_2004}, but with cosmological parameters
updated to the latest results from WMAP \citep{wmap5} for a $\Lambda $CDM
cosmological model.  In the coupled models we consider, the role of DE is
played by a quintessence scalar field with a Ratra-Peebles
\citep{Ratra_Peebles_1988} self-interaction potential of the form:
\be \label{U_def} U(\phi) = \frac{\Lambda^{4+\alpha}}{\phi^\alpha} \,, \ee
where $\Lambda$ and $\alpha$ fix the DE scale in agreement with observations,
and with a constant coupling to CDM particles only, as described in
Sec.~\ref{bkg}; we label them as RP1-RP6 in analogy with
\citet{Maccio_etal_2004}.

For four of these models ($\Lambda$CDM, RP1, RP2, RP5) we then ran
high-resolution simulations in a smaller cosmological box ($L_{\rm box} = 80
h^{-1} $ Mpc, $N_{\rm part}= 2 \times 512^{3}$), and we investigated the
properties of collapsed objects for this set of simulations.  In addition to
these four high-resolution simulations we ran other three simulations with the
same resolution ($\Lambda$CDM-NO-SPH, RP5-NO-SPH, RP5-NO-GF), whose features
will be described below.  The cosmological parameters for our models are
listed in Table~\ref{cosmological_parameters}, and the physical parameters of
each model together with the details of the corresponding N-body simulations
are listed in Table~\ref{Simulations_Table}.
\begin{center}
\begin{table}
\begin{center}
\begin{tabular}{cc}
\hline
Parameter & Value\\
\hline
$\Omega _{\rm CDM} $ & 0.213 \\
$H_{0}$ & 71.9 km s$^{-1}$ Mpc$^{-1}$\\
$\Omega _{\rm DE} $ & 0.743 \\
$\sigma_{8}$ & 0.769\\
$ \Omega _{b} $ & 0.044 \\
$n$ & 0.963\\
\hline
\end{tabular}
\end{center}
\caption{Cosmological parameters for our set of N-body simulations, 
  consistent with the WMAP 5 year results for a $\Lambda $CDM cosmology \citep{wmap5}. }
\label{cosmological_parameters}
\end{table}
\begin{table*}
\begin{tabular}{ccccccllc}
\hline
Model & $\alpha $ & $\beta _{b}$ &  $\beta _{c}$ & Box Size ($h^{-1}$ Mpc) &
Number of particles & $M_{b}$ ($h^{-1} M_{\odot }$) & $M_{\rm CDM}$ ($h^{-1} M_{\odot }$) & $\epsilon _{s}$ ($h^{-1}$ kpc) \\
\hline
$\Lambda$CDM (low)& 0 & 0 & 0 & 320 & 2 $\times $ 128$^{3}$ & 1.9 $\times $ 10$^{11} $& 9.2 $\times $ 10$^{11} $& 50.0\\
$\Lambda$CDM (high)& 0 & 0 & 0 & 80 & 2 $\times $ 512$^{3}$ & 4.7 $\times $ 10$^{7}$ & 2.3 $\times $ 10$^{8}$& 3.5\\
$\Lambda$CDM (high - no SPH)& 0 & 0 & 0 & 80 & 2 $\times $ 512$^{3}$ & 4.7 $\times $ 10$^{7}$ & 2.3 $\times $ 10$^{8}$& 3.5\\
RP1 (low)& 0.143 & 0 & 0.04 & 320 & 2 $\times $ 128$^{3}$ & 1.9 $\times $ 10$^{11} $& 9.2 $\times $ 10$^{11} $& 50.0\\
RP1 (high)& 0.143 & 0 & 0.04 & 80 & 2 $\times $ 512$^{3}$ & 4.7 $\times $ 10$^{7}$ & 2.3 $\times $ 10$^{8}$& 3.5\\
RP2 (low)& 0.143 & 0 & 0.08 & 320 & 2 $\times $ 128$^{3}$ & 1.9 $\times $ 10$^{11} $& 9.2 $\times $ 10$^{11} $& 50.0\\
RP2 (high)& 0.143 & 0 & 0.08 & 80 & 2 $\times $ 512$^{3}$ & 4.7 $\times $ 10$^{7}$ & 2.3 $\times $ 10$^{8}$& 3.5\\
RP3 (low)& 0.143 & 0 & 0.12 & 320 & 2 $\times $ 128$^{3}$ & 1.9 $\times $ 10$^{11} $& 9.2 $\times $ 10$^{11} $& 50.0\\
RP4 (low)& 0.143 & 0 & 0.16 & 320 & 2 $\times $ 128$^{3}$ & 1.9 $\times $ 10$^{11} $& 9.2 $\times $ 10$^{11} $& 50.0\\
RP5 (low)& 0.143 & 0 & 0.2 & 320 & 2 $\times $ 128$^{3}$ & 1.9 $\times $ 10$^{11} $& 9.2 $\times $ 10$^{11} $& 50.0\\
RP5 (high)& 0.143 & 0 & 0.2 & 80 & 2 $\times $ 512$^{3}$ & 4.7 $\times $ 10$^{7}$ & 2.3 $\times $ 10$^{8}$& 3.5\\
RP5 (high - no SPH)& 0.143 & 0 & 0.2 & 80 & 2 $\times $ 512$^{3}$ & 4.7 $\times $ 10$^{7}$ & 2.3 $\times $ 10$^{8}$& 3.5\\
RP5 (high - no GF)& 0.143 & 0 & 0.2 & 80 & 2 $\times $ 512$^{3}$ & 4.7 $\times $ 10$^{7}$ & 2.3 $\times $ 10$^{8}$& 3.5\\
RP6 (low)& 2.0 & 0 & 0.12 & 320 & 2 $\times $ 128$^{3}$ & 1.9 $\times $ 10$^{11} $& 9.2 $\times $ 10$^{11} $& 50.0\\
\hline 
\end{tabular}
\caption{List of the different simulations performed with our modified version
  of {\small GADGET-2}. The simulations have different force and mass
  resolution, and are accordingly labelled as {\em low} or {\em high}
  resolution. Notice that the cited values of the coupling listed here are different
  from the ones adopted in \citet{Maccio_etal_2004} due to the different
  definition of the coupling function (\ref{coupling_const}). 
  However, the models in effect have {\em identical}
  coupling strength to those investigated in \citet{Maccio_etal_2004}. }
\label{Simulations_Table}
\end{table*}
\end{center}

\subsection{Methods} \label{methods}

The presence of a direct coupling between the DE scalar field $\phi $ and
other cosmic fluids -- in the fashion described by
Eqs.~(\ref{tensor_conserv_alpha}, \ref{rho_conserv_eq_phi},
\ref{rho_conserv_eq_c}) -- introduces new features in the cosmic evolution as
well as additional physical processes that must be taken into account in
N-body models.  In the following, we describe these features and their
implementation in {\small GADGET-2} one by one, recalling and further
emphasising the results described in \citet{Maccio_etal_2004} and in
\citet{Pettorino_Baccigalupi_2008}.

\subsubsection{Modified expansion rate}\label{modified_expansion}

\begin{figure*}
\includegraphics[scale=0.4]{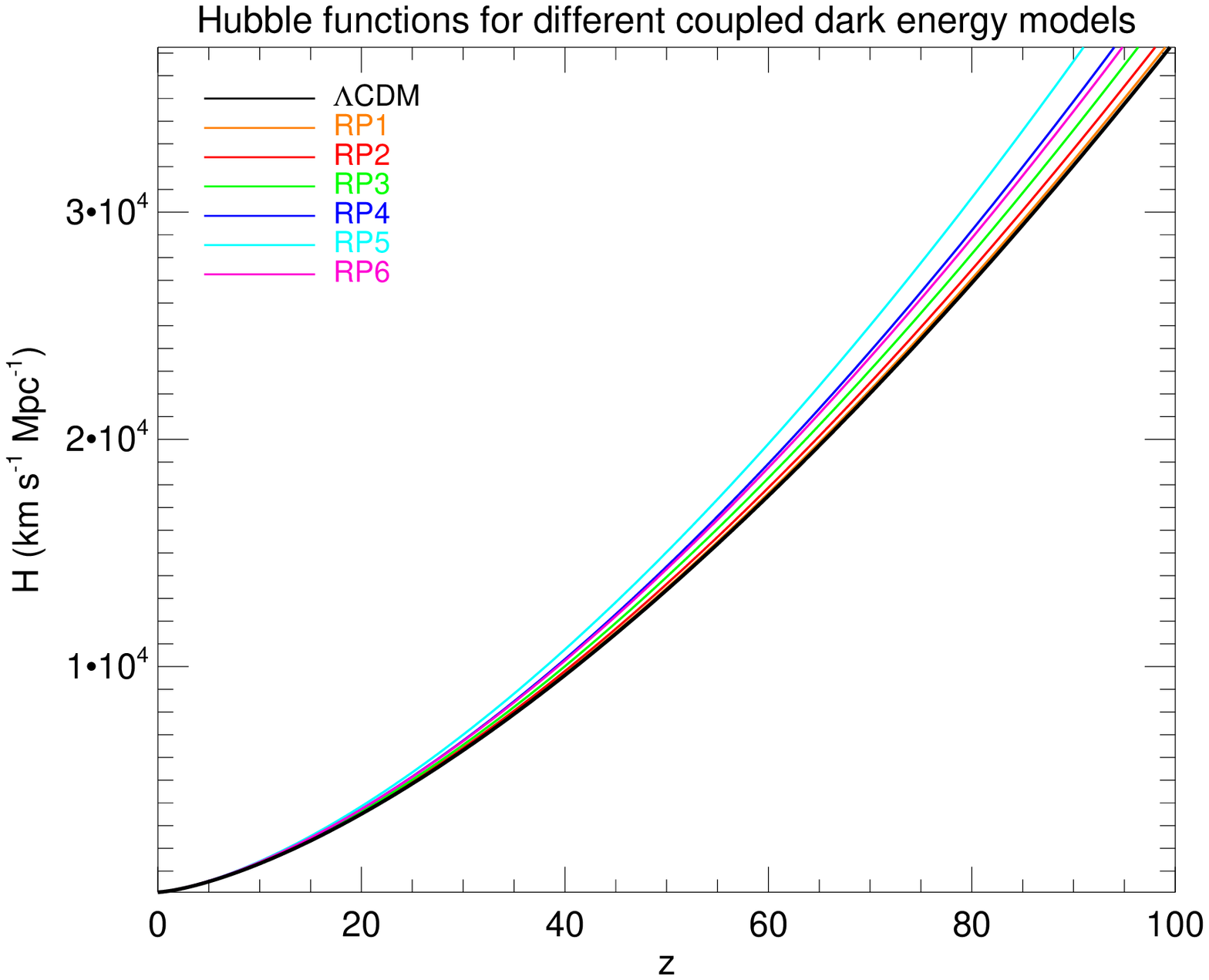}
\includegraphics[scale=0.4]{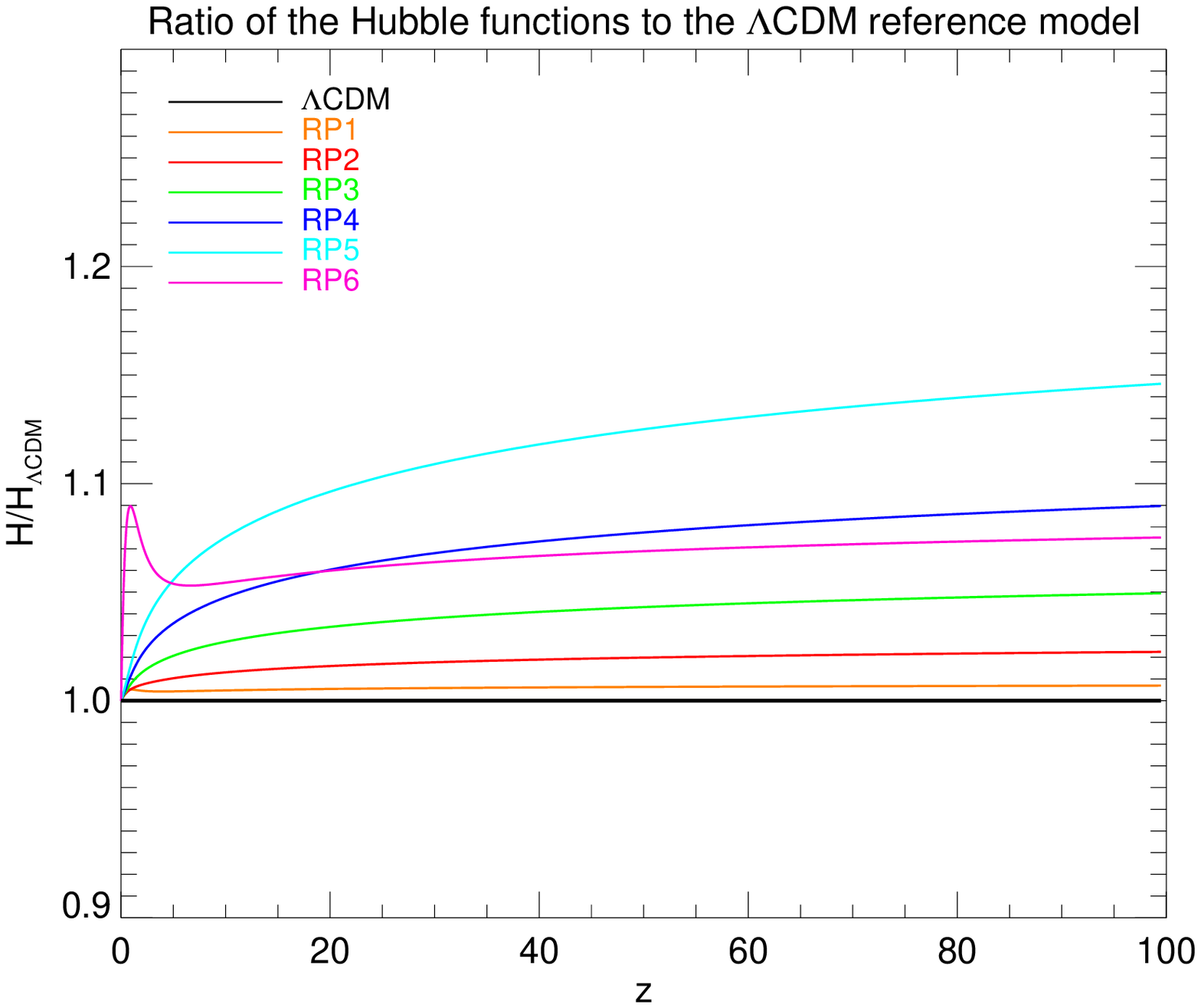}
  \caption{{\em Left panel}: Hubble functions as a function of redshift for the
    different coupled DE models with constant coupling investigated in this
    work and described in Table~\ref{Simulations_Table} as compared to
    $\Lambda $CDM (black curve).  {\em Right panel}: Ratio of the Hubble
    functions of each coupled DE model to the Hubble function for the
    reference $\Lambda $CDM cosmology as a function of redshift. }
\label{hubble_plot}
\end{figure*}

As pointed out in Sec.~\ref{bkg}, the coupling modifies the background
evolution through the existence of a phase -- corresponding to the so called
$\phi $MDE era in \citet{Amendola_2000} -- in which the coupled matter fluid
(CDM in our case) and the DE scalar field evolve with a constant energy
density ratio (here we always assume the Universe to be flat such that $\Omega
_{\rm tot} = 1$).  This leads to the presence of a non negligible Early DE
component \citep[EDE,][]{EDE1, EDE2} during the entire epoch of structure
formation. The effect of such an EDE is to change the expansion history of the
Universe, which has to be properly taken into account for the N-body time
integration.  In order to do so, we replaced the computation of the Hubble
function $H(a)$ in {\small GADGET-2} by a linear interpolation from a table of
values of $H(a)$ precomputed for each model under investigation with the
modified version of {\small CMBEASY} described above.  The effect of the
coupling on the expansion is shown in Fig.~\ref{hubble_plot}.  We note that
the same approach has also been adopted for the other relevant quantities
described in Table~\ref{input_functions}, which were first computed
numerically using {\small CMBEASY}, and then used as an input for our modified
version of {\small GADGET-2}.

\normalsize

\begin{center}
\begin{table}
\begin{center}
\begin{tabular}{cl}
\hline
Function & Meaning\\
\hline
$H(a)$ & Hubble function \\
 $\Delta G(a)$ & Possible global variation of the gravitational constant \\
$\beta _{b}(\phi )$ & Coupling function for the baryons \\
$\beta _{c}(\phi )$ & Coupling function for CDM \\
$\Delta m_{b}$ & Correction term for baryon particle masses\\
$\Delta m_{c}$ & Correction term for CDM particle masses\\
$\Omega _{kin}(\phi )$ & Dimensionless kinetic energy  density of the scalar field \\
\hline
\end{tabular}
\end{center}
\caption{List of input functions for the coupled DE implementation in {\small GADGET-2}.}
\label{input_functions}
\end{table}
\end{center}

\subsubsection{Mass variation}\label{mass_variation}

As described in Sec.~\ref{bkg}, the coupled species feature an effective
particle mass that changes in time.  Consequently, the corresponding
cosmological densities $\rho _{c}$ or $\rho _{b}$ will not be scaling anymore as
$a^{-3}$, but will have a correction term arising from the variation of
particle masses on top of the pure volume dilution.  This correction 
depends on the scalar field dynamics, and takes the form of
Eqn.~(\ref{effective_mass}).  Then, if the particle masses in the simulation
box are computed according to the cosmic densities $\Omega _{c,0}$ and $\Omega
_{b,0}$ at the present time, they will need to be corrected at each timestep
by a factor
\begin{eqnarray}
\label{mass_correctionc}
\Delta m_{c} (a)&=& e^{-\int_{a}^{1}{\beta _{c}(\phi)\frac{{\rm d}\phi}{{\rm
        d}a}{\rm d}a}} \,, \\
\label{mass_correctionb}
\Delta m_{b} (a)&=& e^{-\int_{a}^{1}{\beta _{b}(\phi)\frac{{\rm d}\phi}{{\rm
        d}a}{\rm d}a}}\,.
\end{eqnarray}
\begin{figure}
\includegraphics[scale=0.4]{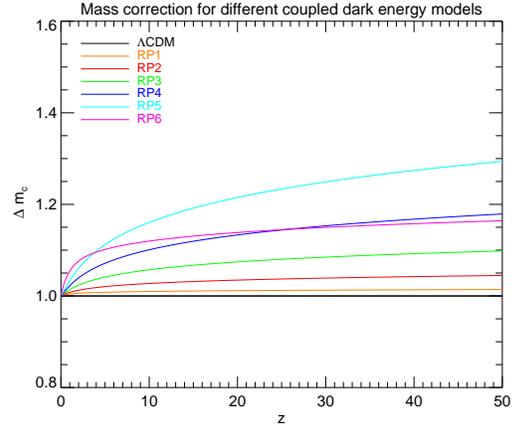}
  \caption{Mass correction as a function of redshift for the coupled DE models
    with constant coupling.}
\label{mass_corr_plot}
\end{figure}
In Fig.~\ref{mass_corr_plot}, we show the evolution with redshift of the
correction term (\ref{mass_correctionc}) for our set of models.

Although the coupled DE implementation for {\small GADGET-2} that we present
here in principle allows for a coupling also to baryons, such a coupling is
tightly constrained by experimental tests of gravity on solar system scales
\citep{Damour_Gibbons_Gundlach_1990}. Therefore, in the coupled DE models we
consider in this work, no coupling to baryons is considered, and baryon masses
are always constant.  However, even a very tiny coupling to baryons, in the
range allowed by present bounds, could possibly play a significant role in
cosmological models with multiple dark matter families
\citep{Huey_Wandelt_2006}, like for example the {\em Growing Neutrino}
scenario introduced by \citet{Amendola_Baldi_Wetterich_2008}. We plan to study
detailed N-body simulations of this kind of models in forthcoming work.

\subsubsection{Cosmological extra velocity-dependent acceleration}\label{extra_friction}

A further modification to the {\small GADGET-2} code concerns the extra
cosmological velocity-dependent term induced by the coupling in the Euler equation, as
shown in Equations (\ref{CQ_euler}, \ref{friction_term}).  In standard
cosmological simulations with {\small GADGET-2} the usual cosmological
velocity-dependent term (first term of Eqn.~\ref{friction_term}) is not directly computed
because the choice of variables removes it from the acceleration equation.
If we denote with $\vec{x}$ comoving coordinates and with
$\vec{r}=a(t)\vec{x}$ physical coordinates, we have that:
\begin{equation}
\dot{\vec{r}}= H\vec{r} + \vec{v}_{p} \,, \qquad \vec{v}_{p}\equiv a(t)\dot{\vec{x}}\,.
\end{equation}
Instead of using the peculiar velocity $\vec{v}_{p}$, {\small GADGET-2} makes
use of the variable $\vec{p} \equiv a^{2}(t)\dot{\vec{x}}$ (see
\citet{gadget-2} for more details), for which the following relation holds:
\begin{equation}
\label{relation_p_v}
\dot{\vec{v}}_{p} = \frac{1}{a}\dot{\vec{p}} - \frac{H}{a}\vec{p} \,.
\end{equation}
It is then straightforward, by using Eqn.~(\ref{relation_p_v}), to find a
generalization of Eqn.~(\ref{CQ_euler}) to a system of $N$ particles, in terms of
the new velocity variable $\vec{p}$:
\begin{equation}
\label{gadget-acceleration-equation}
\dot{\vec{p}}_{i} = \frac{1}{a}\left[ \beta _{\gamma}(\phi )\frac{\dot{\phi }}{M} a \vec{p}_{i} + \sum_{j \ne i}\frac{\tilde{G}_{ij}m_{j}\vec{x}_{ij}}{|\vec{x}_{ij}|^{3}} \right] \,,
\end{equation}
where $i$ and $j$ are indices that span over all the particles of the
simulation, $\gamma = c,b$ for CDM or baryons respectively, and
$\tilde{G}_{ij}$ is the effective gravitational constant between the $i$-th
and the $j$-th particles, as determined in Eqn.~(\ref{G_eff}) and whose
implementation will be discussed below.

It is evident from Eqn.~(\ref{gadget-acceleration-equation}) that for zero
coupling no cosmological velocity-dependent term is present in the acceleration
equation, which is then just Newton's law in comoving coordinates.  In
general, however, whenever a coupling is present, the additional term
\begin{equation}
\label{drag-term}
a(\dot{\vec{p}} + \Delta \dot{\vec{p}}) \equiv a \dot{\vec{p}} + \beta _{\gamma }(\phi )\frac{\dot{\phi }}{M} a \vec{p}_{i}
\end{equation}
has to be explicitely added to the Newtonian acceleration of every particle.
This term does not depend on the matter distribution in the rest of the
Universe. It is therefore a purely cosmological drag that would be present
also in absence of any gravitational attraction.  It is interesting to notice
that in the case of constant positive coupling and a monotonic scalar field
potential as investigated here, the extra cosmological velocity-dependent term induces
an acceleration in the same direction as the velocity of the particle.  It
therefore is effectively a ``dragging'' term, speeding up the motion of
the particles.

Scalar field models where the dynamics of the field or the evolution of the
coupling induce a change in the sign of $\beta _{i}(\phi )\dot{\phi }$,
thereby changing the direction of this extra velocity-dependent force, will be studied
in future work \citep{Baldi_Maccio_inprep}.

\subsubsection{Fifth force implementation}

One of the most important modifications introduced by the coupling between DE
and CDM is the presence of a modified gravitational constant, formally written
as in Eqn.~(\ref{G_eff}), for the gravitational interaction of CDM particles.  In
fact, if in general the substitution \citep{Amendola_2004}
\begin{equation}
\label{modified-G}
G_{N}\rightarrow\tilde{G}_{lm}=G_{N}\cdot (1 + 2\beta _{l}\beta _{m}) \,,
\end{equation}
holds for each pair $(l,m)$ of particles, with $l$ and $m$ denoting the
species of the particle, in our case only CDM-CDM interaction is affected,
while baryon-CDM or baryon-baryon interactions remain unchanged since $\beta_b
= 0$.  The dependence of this modified gravitational interaction on the
particle type requires an N-body code to distinguish among different particle
types in the gravitational force calculation.  In {\small GADGET-2}, the
gravitational interaction is computed by means of a TreePM hybrid method
\citep[see][for details about the TreePM algorithm]{gadget-2}, so that both
the tree and the particle-mesh algorithms have to be modified in order to
account for this new phenomenology.

{\bf Tree algorithm modifications} -- In a standard tree algorithm, each node
of the tree carries informations about the position of its centre of mass, its
velocity, and its total mass. The decision whether to compute the force
exerted on a target particle by the whole node or to further divide it into
eight smaller nodes is made based on a specific opening criterion, which sets
the accuracy threshold for approximating the gravitational potential of a
distribution of particles with its low-order multipole expansion. Since in
uncoupled cosmological models all particles interact with the same
gravitational strength, as soon as the opening criterion is fulfilled the
force is computed assigning all the mass contained in the node to its centre
of mass.  For coupled quintessence cosmologies, this is no longer accurate
enough given that the different particle species will contribute differently
to the gravitational force acting on a target particle.  This means that
besides the total mass and the total centre of mass position and velocity,
each node has to carry information about the mass and centre-of-mass position
and velocity of each particle species with different coupling.

{\bf Particle-Mesh algorithm modifications} -- In the Tree-PM algorithm, the
long-range part of the gravitational force is computed by means of Fourier
techniques.  For coupled DE models, where different particle species interact
with an effectively different gravitational force, the PM procedure has to be
repeated as many times as there are differently interacting particle types,
each time assigning to the cartesian grid only the mass in particles of a
single type, and then computing the gravitational potential and the
acceleration deriving from the spatial distribution of that particle species
alone. In this way, the total force is built up as a sum of several partial
force fields from each particle type.

\subsubsection{Initial conditions}

The initial conditions of a cosmological N-body simulation need to specify the
positions and velocities of all the particles in the cosmological box at the
starting redshift $z_{i}$ of the simulation.  These quantities are usually
computed by setting up a random-phase realization of the power spectrum of the
studied cosmological model according to the Zel'dovich approximation
\citep{Zeldovich_1970}.  The normalization amplitude of the power spectrum is
adjusted such that the linearly evolved rms-fluctuations $\sigma_{8}$ on a
top-hat scale of $8\,h^{-1}{\rm Mpc}$ at a given redshift $z_{\rm norm}$ (usually
chosen to be $z_{\rm norm} = 0$) have a prescribed amplitude.

\begin{figure}
\includegraphics[scale=0.4]{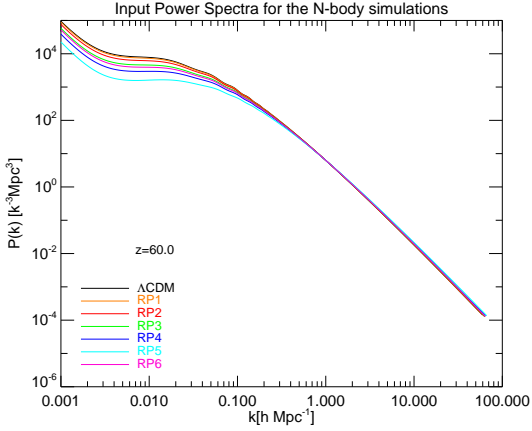}
  \caption{Matter power spectra at $z=60$ for 
    interacting DE models with constant coupling as computed by {\small CMBEASY}.}
\label{powerspectra}
\end{figure}

The coupling between DE and CDM can have a strong impact on the transfer
function of matter density fluctuations, as first pointed out by
\citet{Mainini_Bonometto_2007}. For this reason we compute the required
initial power spectrum directly with the modified Boltzmann code {\small
  CMBEASY}, because the phenomenological parameterizations of the matter power
spectrum available for the $\Lambda$CDM cosmology \citep[e.g.][]{BBKS,
  Eisenstein_Hu_1997} would not be accurate enough. The resulting effect on
the power spectrum is shown in Fig.~\ref{powerspectra} for the different
models considered in our set of simulations.

Once the desired density field has been realized with this procedure, the
displacements of the particles from the grid points need to be rescaled with
the linear growth factor $D_{+}$ for the cosmological model under
investigation between the redshifts $z_{\rm norm}$ and $z_{i}$ in order to set
the correct amplitude of the power spectrum at the starting redshift of the
simulation. Also, the velocities of the particles are related to the local
overdensities according to linear perturbation theory, via the following
relation, here written in Fourier space:
\begin{equation}
\vec{v}(\vec{k},a) = i f(a) a H\delta(\vec{k},a)\frac{\vec{k}}{k^{2}} \,,   
\end{equation}
where the growth rate $f(a)$ is defined as
\begin{equation} \label{f_def}
f(a) \equiv \frac{d \ln D_{+}}{d \ln a}\,. 
\end{equation}
This requires an accurate calculation of the linear growth function $D_{+}(z)$
for the coupled model, which we again compute numerically with {\small
  CMBEASY}.

\begin{figure*}
\includegraphics[scale=0.32]{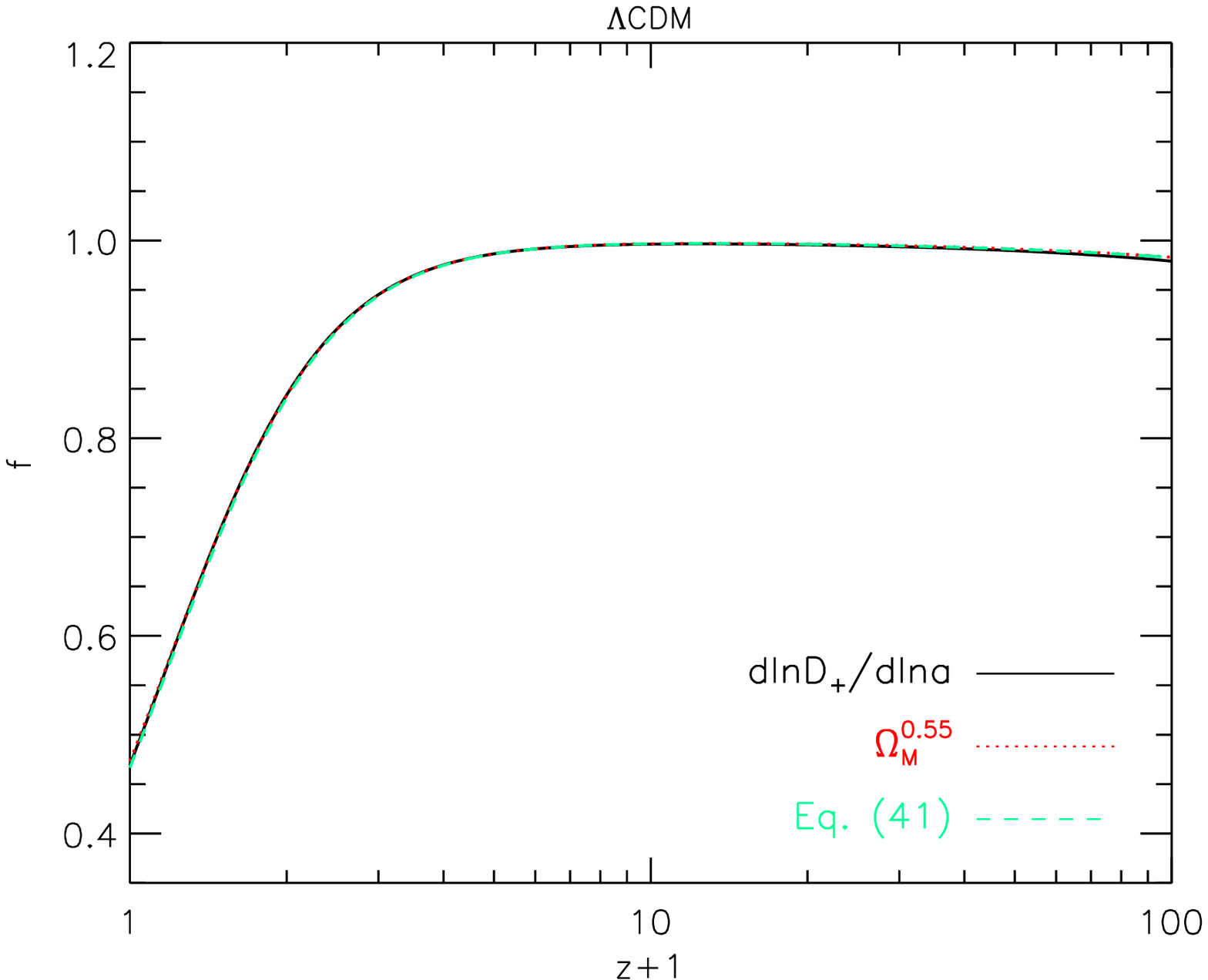}
\includegraphics[scale=0.32]{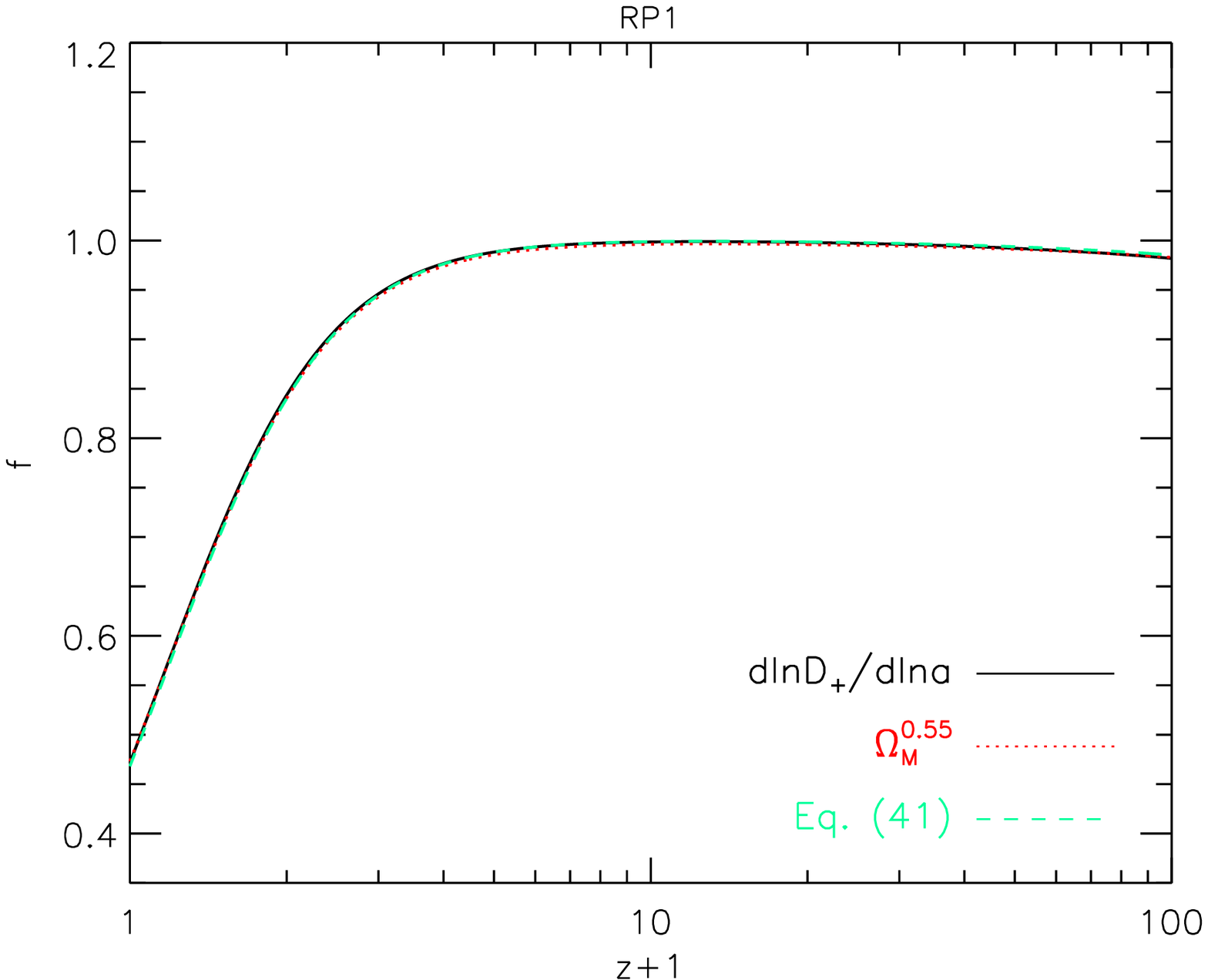}
\includegraphics[scale=0.32]{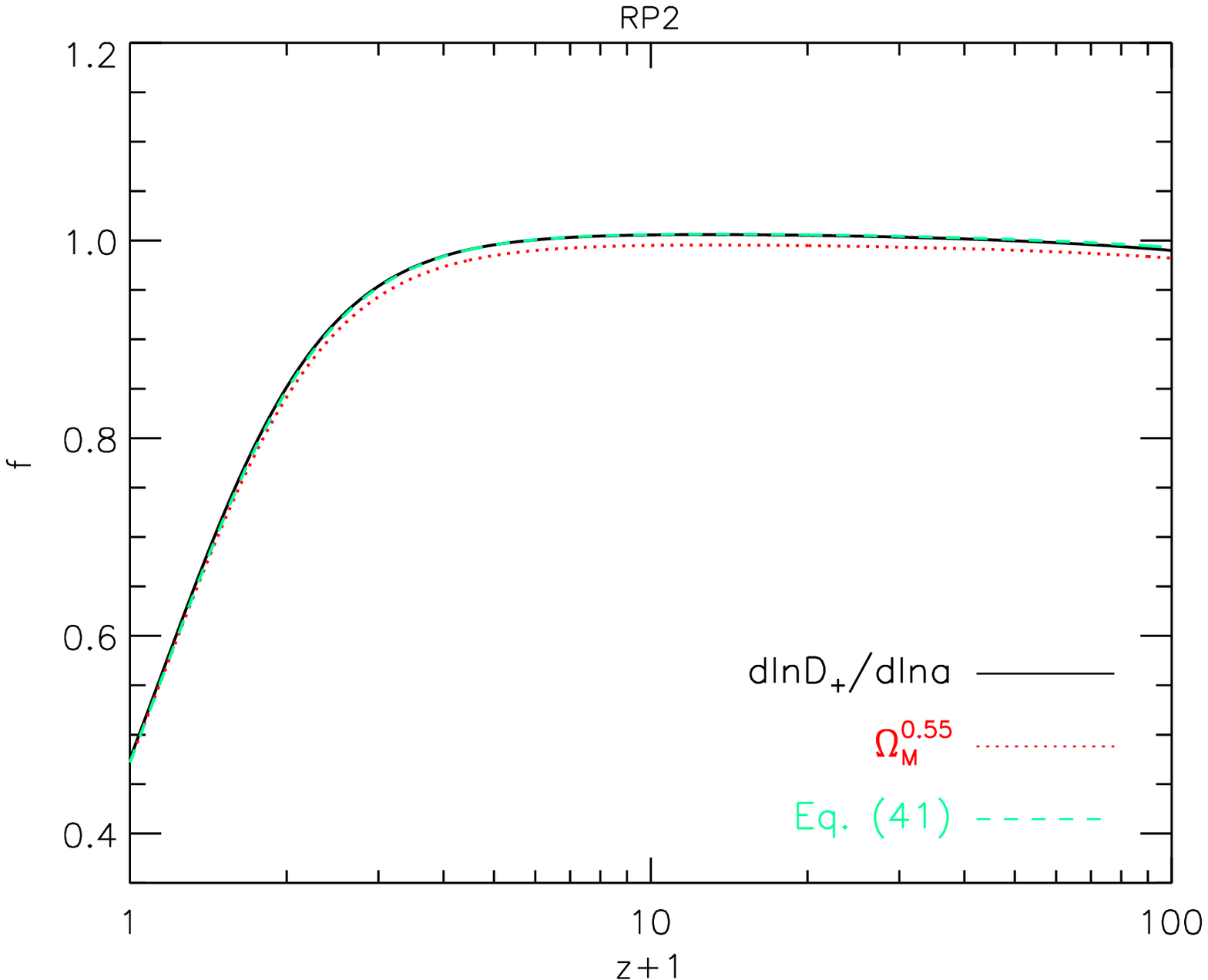}\\
\includegraphics[scale=0.32]{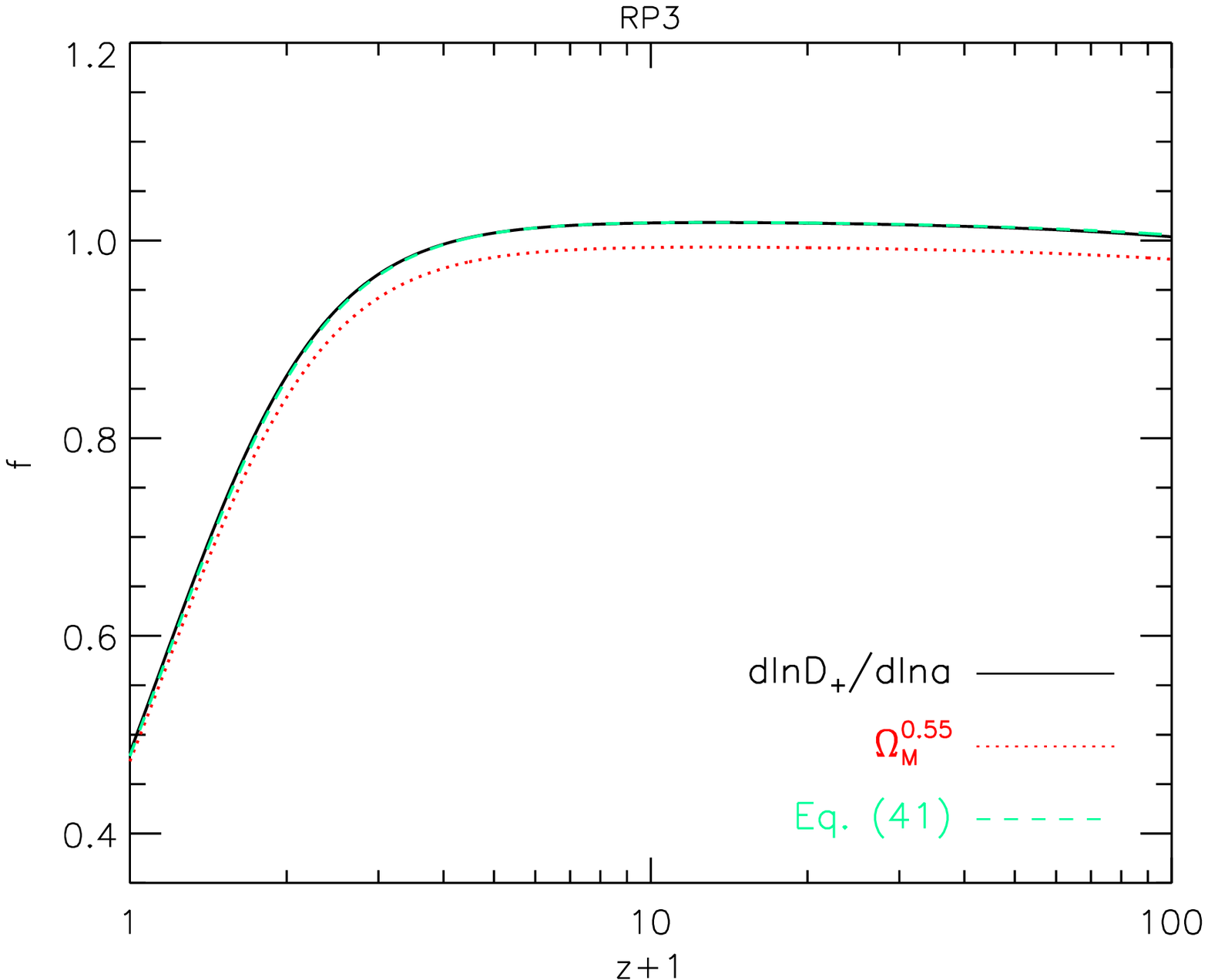}
\includegraphics[scale=0.32]{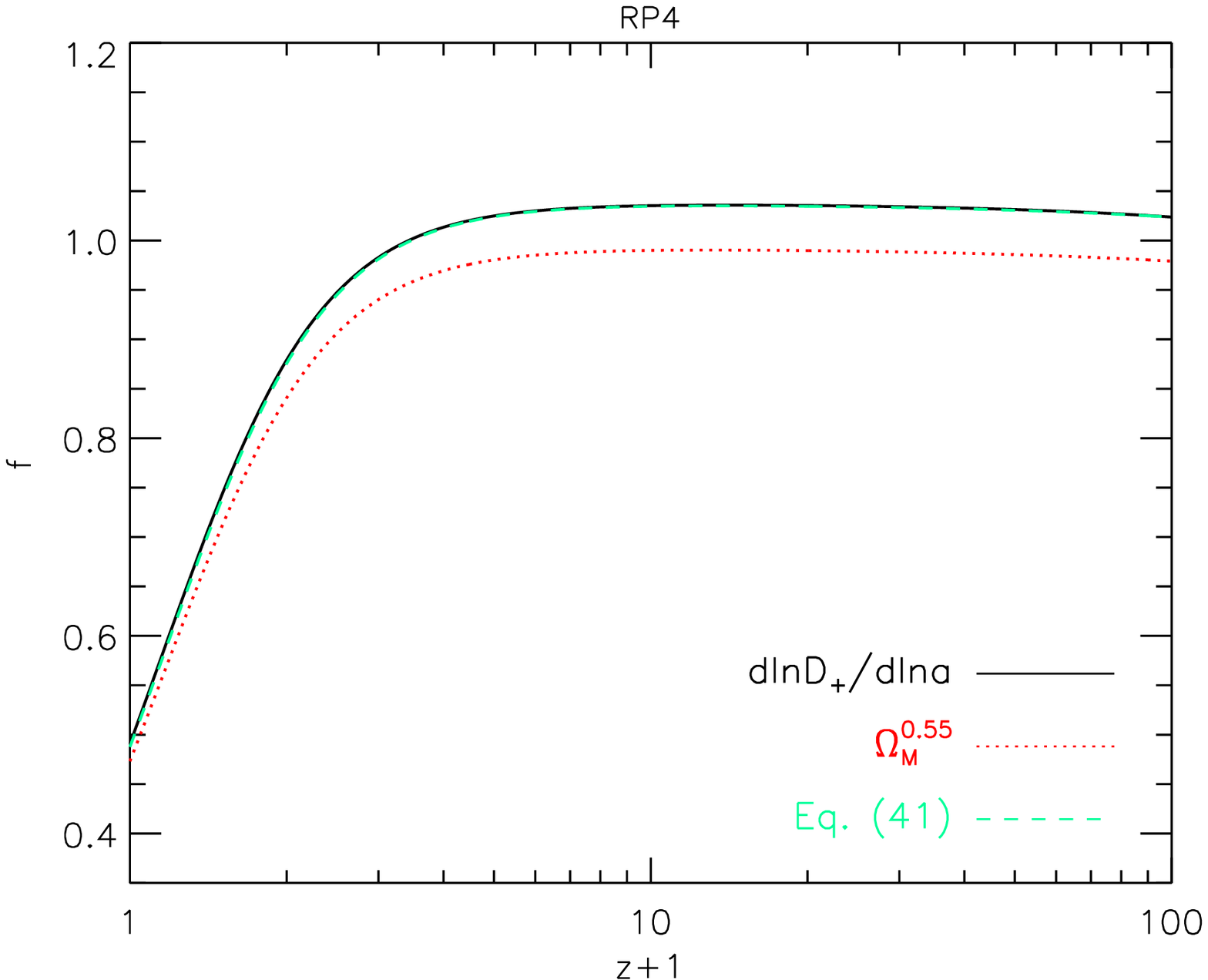}
\includegraphics[scale=0.32]{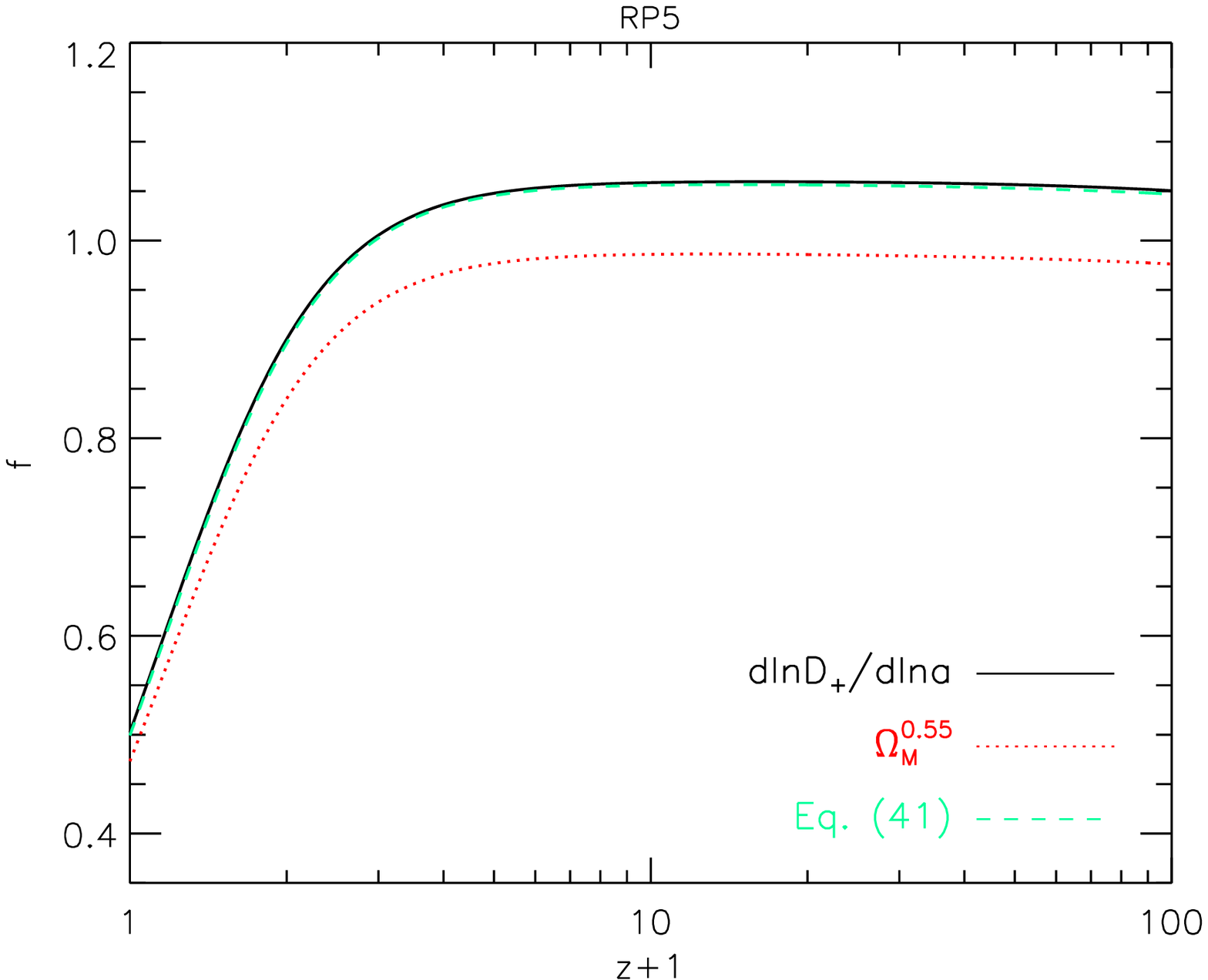}\\
\includegraphics[scale=0.32]{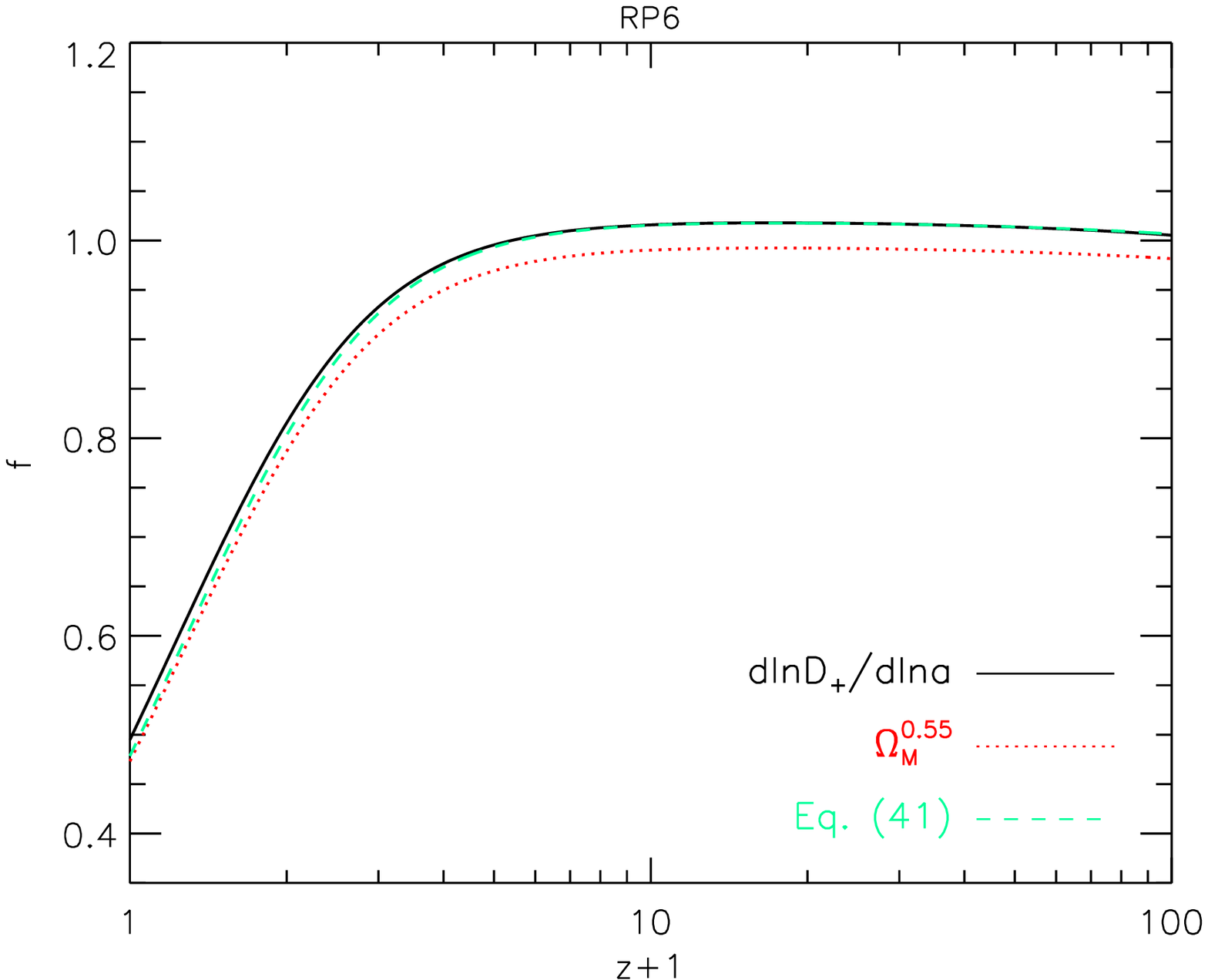}
\caption{Comparison of the function $f(a)$ with its usual 
approximation $f=\Omega _{m}^{0.55}$ and with the new fit of 
Eqn.~(\ref{gf_fit}) for a $\Lambda$CDM model and for a series of coupled DE models.}
\label{f_omega}
\end{figure*}

We note that a phenomenological parameterization of the growth function for
coupled DE models with constant coupling to dark matter has recently been made
available by \citet{DiPorto_Amendola_2008}. However, it is only valid for
models with no admixture of uncoupled matter, whereas in our case we also have
a baryonic component.  Also, in the $\Lambda$CDM cosmology, the total growth
rate is well approximated by a power of the total matter density $\Omega
_{M}^{\gamma }$, with $\gamma = 0.55$, roughly independently of the
cosmological constant density \citep{Peebles_1980}. This is however no longer
true in coupled cosmologies, as we show in Fig.~\ref{f_omega}.  We find that,
for our set of coupled DE models, a different phenomenological fit given
by
\begin{equation}
\label{gf_fit}
f(a) \sim \Omega _{M}^{\gamma }(1 + \gamma \frac{\Omega _{\rm CDM}}{\Omega _{M}} \epsilon _{c}\beta _{c}^2) \,,
\end{equation}
with $\gamma = 0.56$ \citep[as previously found
in][]{Amendola_Quercellini_2004} and $\epsilon _{c}=2.4$ works well.  The fit
(\ref{gf_fit}) reproduces the growth rate with a maximum error of $\sim 2\%$
over a range of coupling values between $0$ and $0.2$ and for a cosmic baryon
fraction $\Omega _{b}/\Omega _{m}$ at $z=0$ in the interval $0.0 - 0.1$ for
the case of the potential slope $\alpha = 0.143$ (corresponding to the slope
of the RP1-RP5 models).  For a value of $\alpha = 2.0$ (corresponding to the
slope assumed for RP6) the maximum error increases to $\sim 4\%$ in the same
range of coupling and baryon fraction.  In Fig.~\ref{f_omega}, we plot both
the fitting formulas together with the exact $f(a)$.  For our initial
conditions setup we in any case prefer to use the exact function $f(a)$
directly computed for each model with {\small CMBEASY}, rather than any of the
phenomenological approximations.

\subsection{Tests of the numerical implementation: the linear growth factor}

\begin{figure}
\includegraphics[scale=0.45]{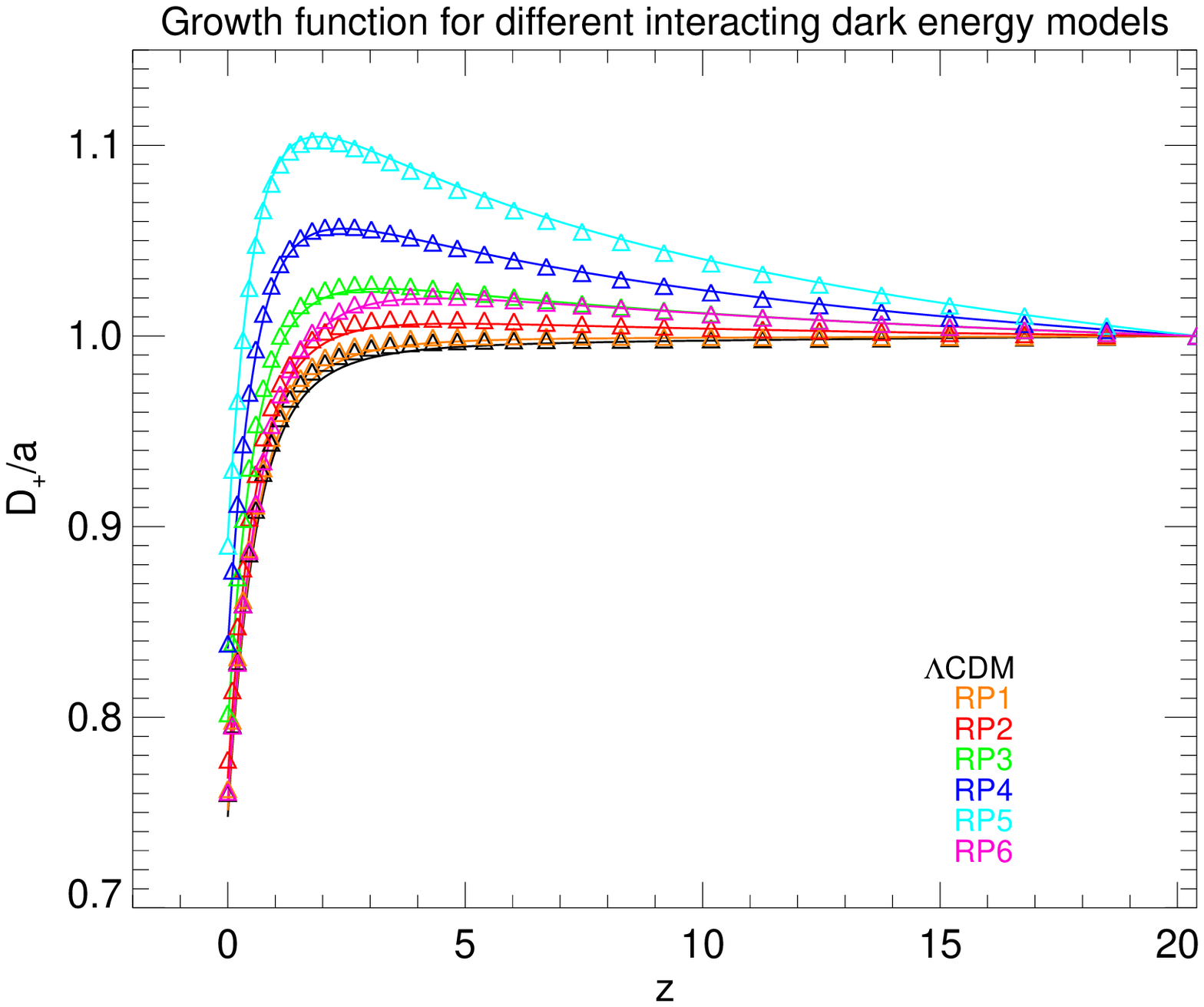}
  \caption{Evolution of the growth function with redshift for the seven models
    of coupled DE investigated with the low-resolution simulations ran with
    our modified version of {\small GADGET-2}. The solid lines are the total
    growth functions as evaluated numerically with {\small CMBEASY}, while the
    triangles are the growth function evaluated from the simulations. The
    relative accuracy in reproducing the theoretical prediction is of the
    order of a few percent irrespective of the coupling value $\beta _{c}$.}
\label{growth_factor}
\end{figure}

As a first test of our implementation we check whether the linear growth of
density fluctuations in the simulations is in agreement with the linear theory
prediction for each coupled DE model under investigation.  To do so, we
compute the growth factor from the simulation outputs of the low-resolution
simulations described in Table~\ref{Simulations_Table} by evaluating the
change in the amplitude of the matter power spectrum on very large scales, and
we compare it with the solution of the system of coupled equations for linear
perturbations (\ref{eq_newtonian_gauge}), numerically integrated with {\small
  CMBEASY}.  The comparison is shown in Fig.~\ref{growth_factor} for all the
constant coupling models.  The accuracy of the linear growth computed from the
simulations in fitting the theoretical prediction is of the same order for all
the values of the coupling, and the discrepancy with respect to the numerical
solution obtained with our modified version of {\small GADGET-2} never exceeds
a few percent.

\subsection{Our set of N-body simulations} \label{Simulations}

In our simulations, we are especially interested in the effects that the
presence of a coupling between DE and CDM induces in the properties of
collapsed structures, and we would like to understand which of these effects
are due to linear features of the coupled theory, and which due to the
modified gravitational interaction in the dark sector. This goal turns out to
be challenging due to the presence of several different sources of changes in
the simulation outcomes within our set of runs.  To summarize this, let us
briefly discuss in which respect, besides the different gravitational
interactions, the high-resolution simulations listed in
Table~\ref{Simulations_Table} are different from each other:

\begin{itemize}
\item the initial conditions of the simulations are generated using a
  different matter power spectrum for each model, i.e. the influence of the
  coupled DE on the initial power spectrum is taken into account and this
  means that every simulation will have a slightly different initial power
  spectrum shape;
\item the amplitude of density fluctuations is normalized at $z=0$ for all the
  simulations to $\sigma_{8}=0.796$, but due to the different shapes of the
  individual power spectra the amplitude of density fluctuations at the
  present time will not be the same in all simulations at all scales;
\item the initial displacement of particles is computed for each simulation by
  scaling down the individual power spectrum amplitudes as normalized at $z=0$
  to the initial redshift of the simulations ($z_{i}=60$) by using for each
  simulation the appropriate growth function. This results in a lower initial
  amplitude for more strongly coupled models;
\item hydrodynamical forces are acting on baryon particles in all the four
  fully self-consistent simulations ($\Lambda$CDM, RP1, RP2, RP5), and
  therefore differences in the evolution of the dark matter and baryon
  distributions might be due to a superposition of hydrodynamics and modified
  gravitational interaction; 
\item non-adiabatic processes like e.g.~radiative cooling, star formation, and feedback
  are not included in any of the simulations presented in this work.
\end{itemize}

In order to try to disentangle which of these differences cause significant
changes in our results, we decided to run three further test simulations
in which, in turn, some of the new physics has been disabled.
\begin{itemize}
\item In the two simulations labelled as ``NO-SPH'' ($\Lambda $CDM-NO-SPH,
  RP5-NO-SPH), we disabled hydrodynamical SPH (Smoothed Particle
  Hydrodynamics) forces in the code integration. We can then compare a
  $\Lambda $CDM model with a strongly coupled model treating both baryons and
  cold dark matter particles as collisionless particles. The differences in
  the dynamics will then be due only to the different gravitational
  interaction implemented in the RP5 model. However, the shape and amplitude
  of the initial power spectrum for the two simulations are still different;
\item In the simulation labelled RP5-NO-GF, we ran a RP5 cosmological model
  using as initial conditions the same file we used for the $\Lambda $CDM
  run. This means that no effect arising in this simulation compared to
  $\Lambda $CDM can be due to different initial conditions, i.e.~due to the
  differences in the shape and amplitude of the initial power spectra that are
  present in the other simulations.
\end{itemize}

\section{Results}\label{results}

We now describe our results for the effect of the coupling between DE and CDM
on nonlinear structures.  As first basic analysis steps we apply the
Friends-of-Friends (FoF) and {\small SUBFIND} algorithms \citep{Springel2001}
to identify groups and gravitationally bound subgroups in each of our
simulations.  Given that the seed used for the random realization of the power
spectrum in the initial conditions is the same for all the seven simulations,
structures will form roughly at the same positions in all simulations, and it
is hence possible to identify the same objects in all the simulations and to
compare their properties. Nevertheless, due to the different timestepping
induced by the different physics implemented in each run, and the slightly
different transfer functions, objects in the different simulations can be
slightly offset from each other. We therefore apply a selection criterion and
identify objects found in the different simulations as the same structure only
if the most bound particle of each of them lies within the virial radius of
the corresponding structure in the $\Lambda $CDM simulation.  If this
criterion is not fulfilled for all the different simulations we want to
compare, we do not consider the corresponding halo in any of the comparison
analysis described below. We restrict this matching procedure to the 200 most
massive halos identified by the FoF algorithm, which have virial masses
ranging from $4.64 \times 10^{12} h^{-1} M_{\odot}$ to $2.83 \times 10^{14}
h^{-1} M_{\odot}$. Note that depending on the specific set of simulations we
consider in our comparative analysis, this can result in small differences in
the number of halos included in each of the comparison samples.

\subsection{Halo mass function} 

\begin{figure*}
\includegraphics[scale=0.45]{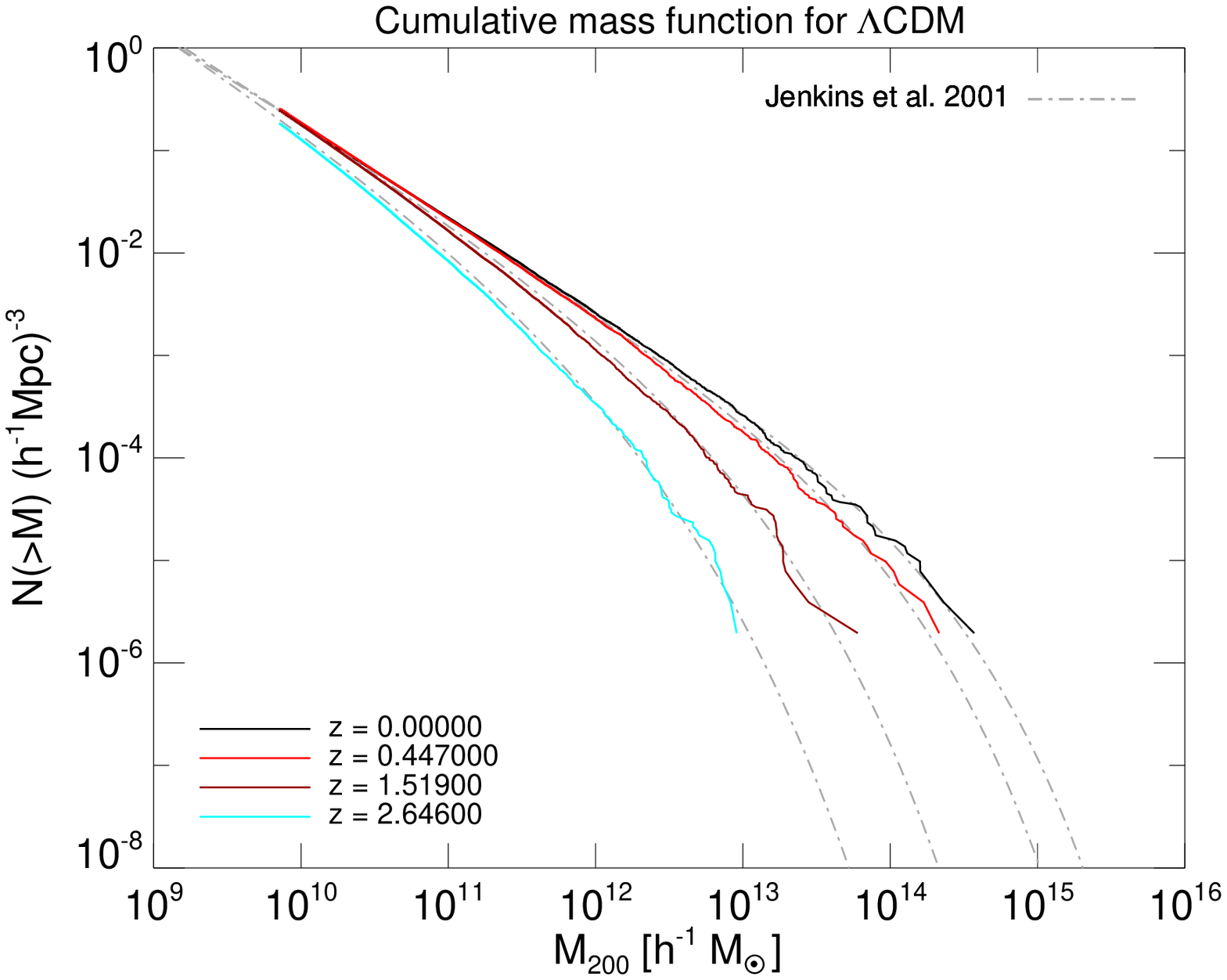}
\includegraphics[scale=0.45]{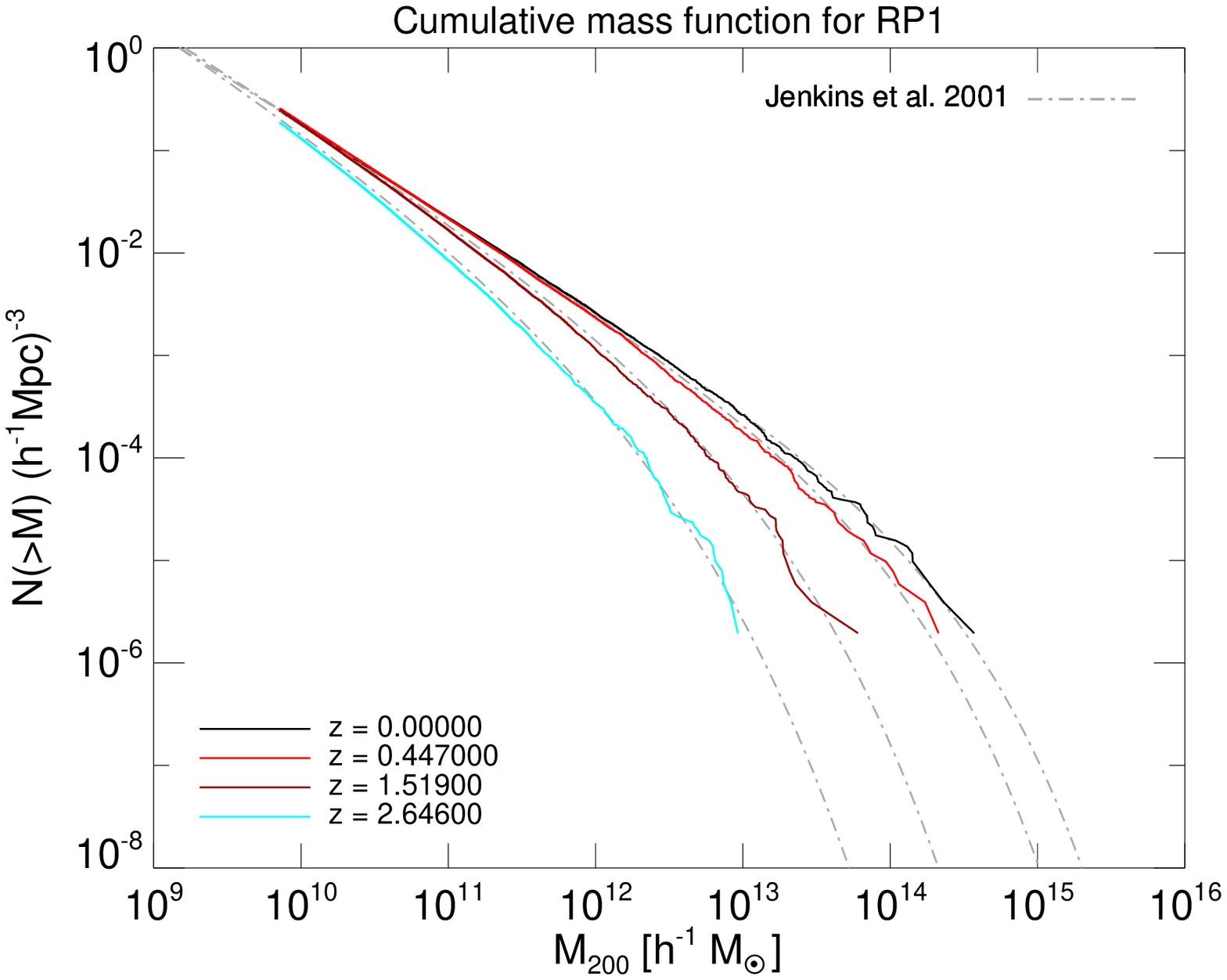}\\
\includegraphics[scale=0.45]{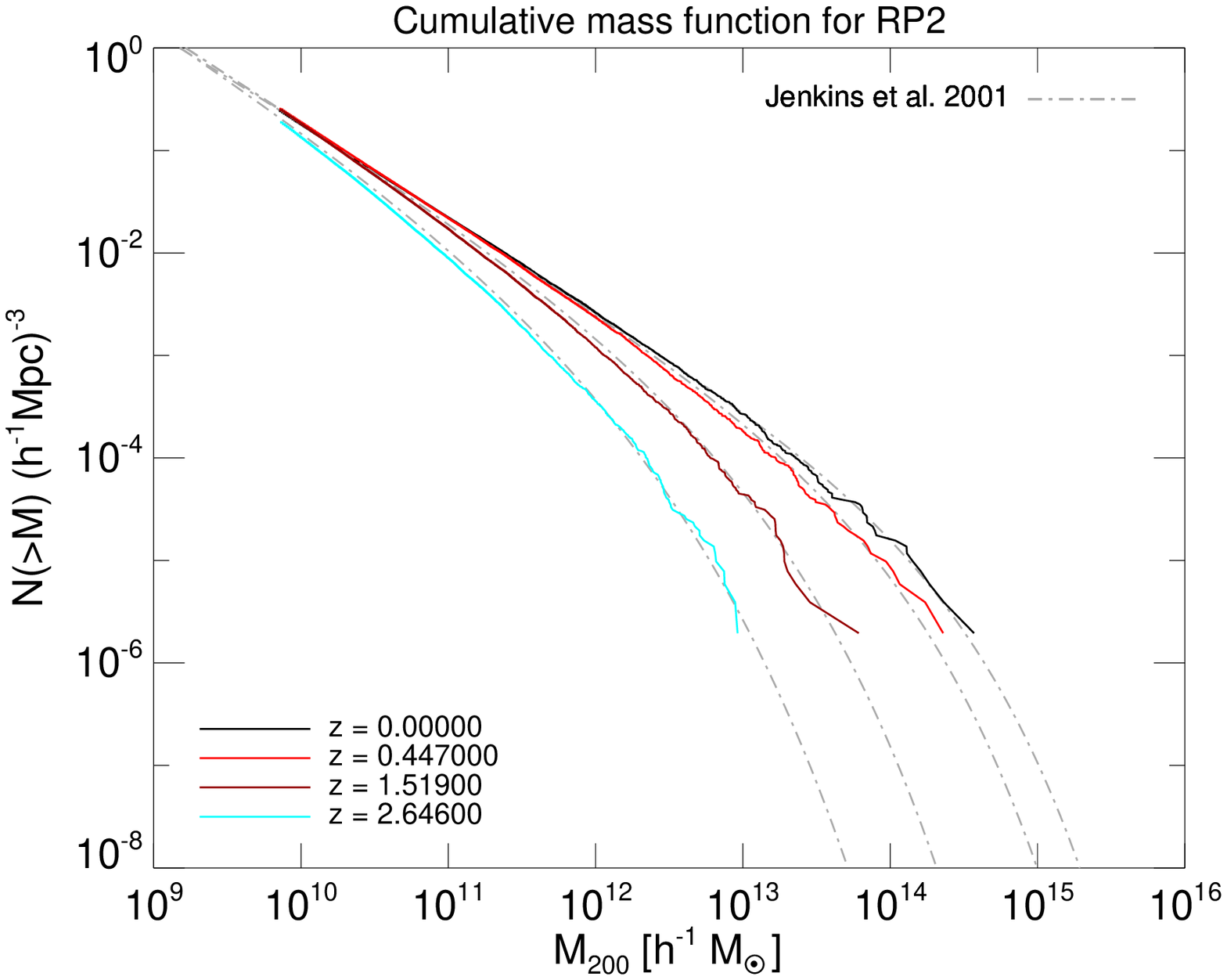}
\includegraphics[scale=0.45]{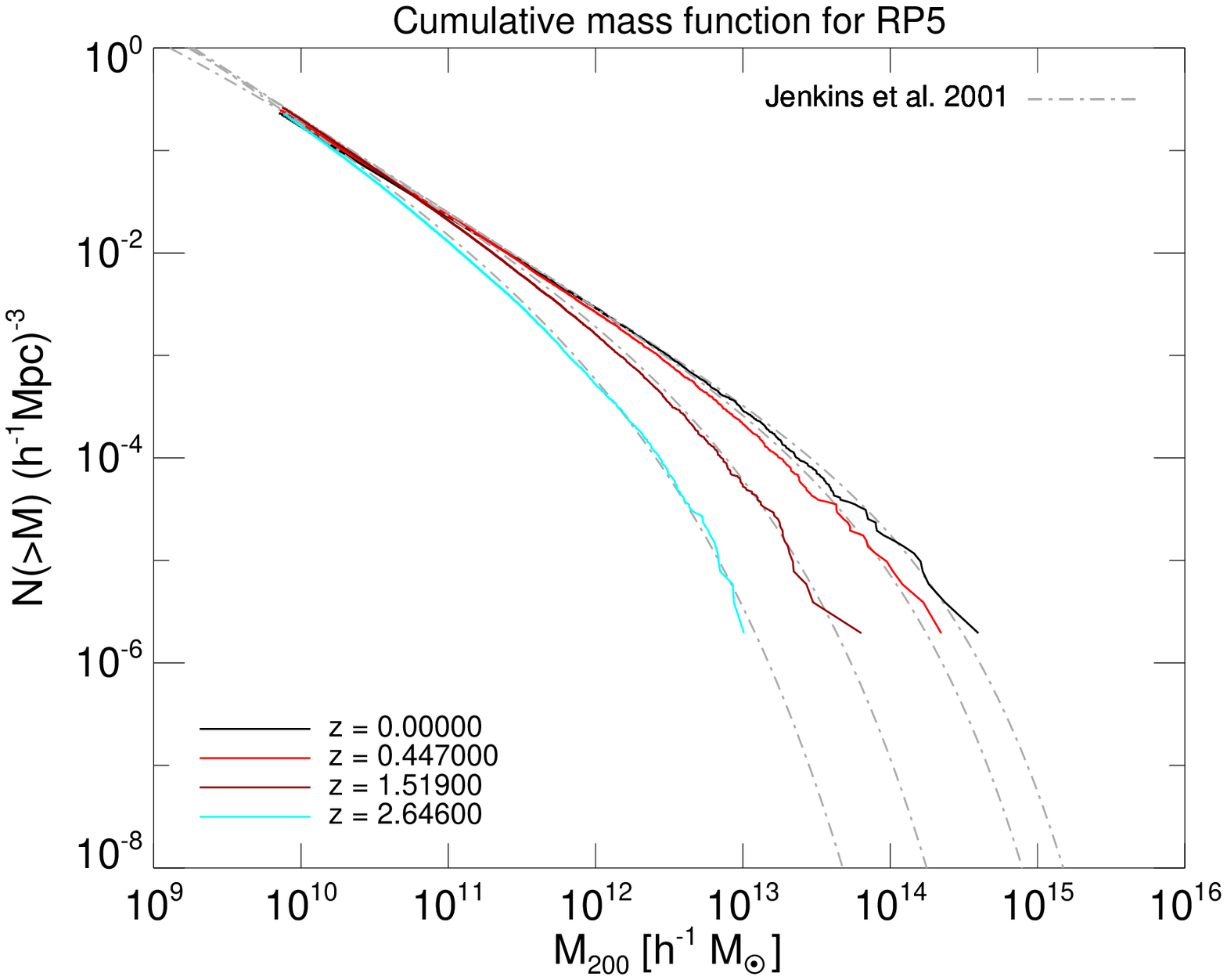}
  \caption{Cumulative mass functions for the four fully self-consistent
    high-resolution simulations of the coupled DE models. The four
    differently-coloured solid lines in each figure represent the cumulative
    mass function at four different redshifts in each of the investigated
    models. The dot-dashed lines are the corresponding predictions according
    to the \citet{Jenkins_etal_2000} formula, computed for each simulation
    with the appropriate growth function and power spectrum.}
\label{cumulative_massfunctions}
\end{figure*}

For the four fully self-consistent high-resolution simulations listed in
Table~\ref{Simulations_Table} ($\Lambda $CDM, RP1, RP2, RP5), we have computed
the halo mass function based on the groups identified with the
Friends-of-Friends (FoF) algorithm with a linking length of $\lambda = 0.2
\times \bar{d}$, where $\bar{d}$ is the mean particle spacing. It is important
to recall that our simulations have the same initial random phases and are
normalized to the same $\sigma_{8}$ today, but the shapes of the input power
spectra are slightly different for each simulation. In
Fig.~\ref{cumulative_massfunctions} we plot the cumulative mass functions for
the four simulations at different redshifts.  Remarkably, the mass functions
of all the cosmological models and at all the different redshifts considered
are well fit by the formula given in \citet{Jenkins_etal_2000}, provided it is
evaluated with the actual power spectrum and the correct linear growth factor
for the corresponding cosmological model. The usual mass function formalism
hence continues to work even for coupled DE cosmologies, a result in line with
recent findings for early dark energy cosmologies \citep{Grossi_2008, Francis_etal_2008, Francis_etal_2008b}.

\begin{figure*}
\includegraphics[scale=0.45]{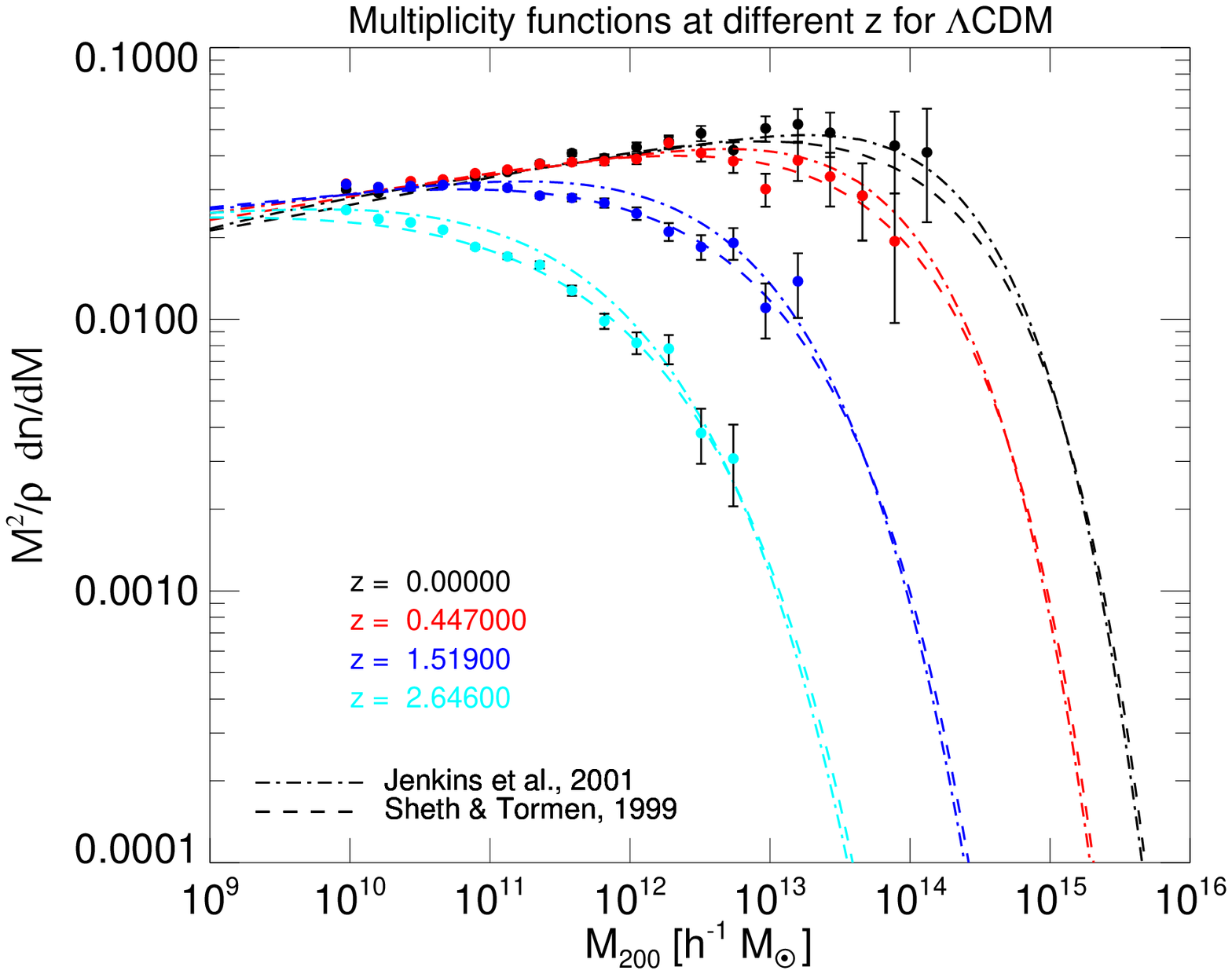}
\includegraphics[scale=0.45]{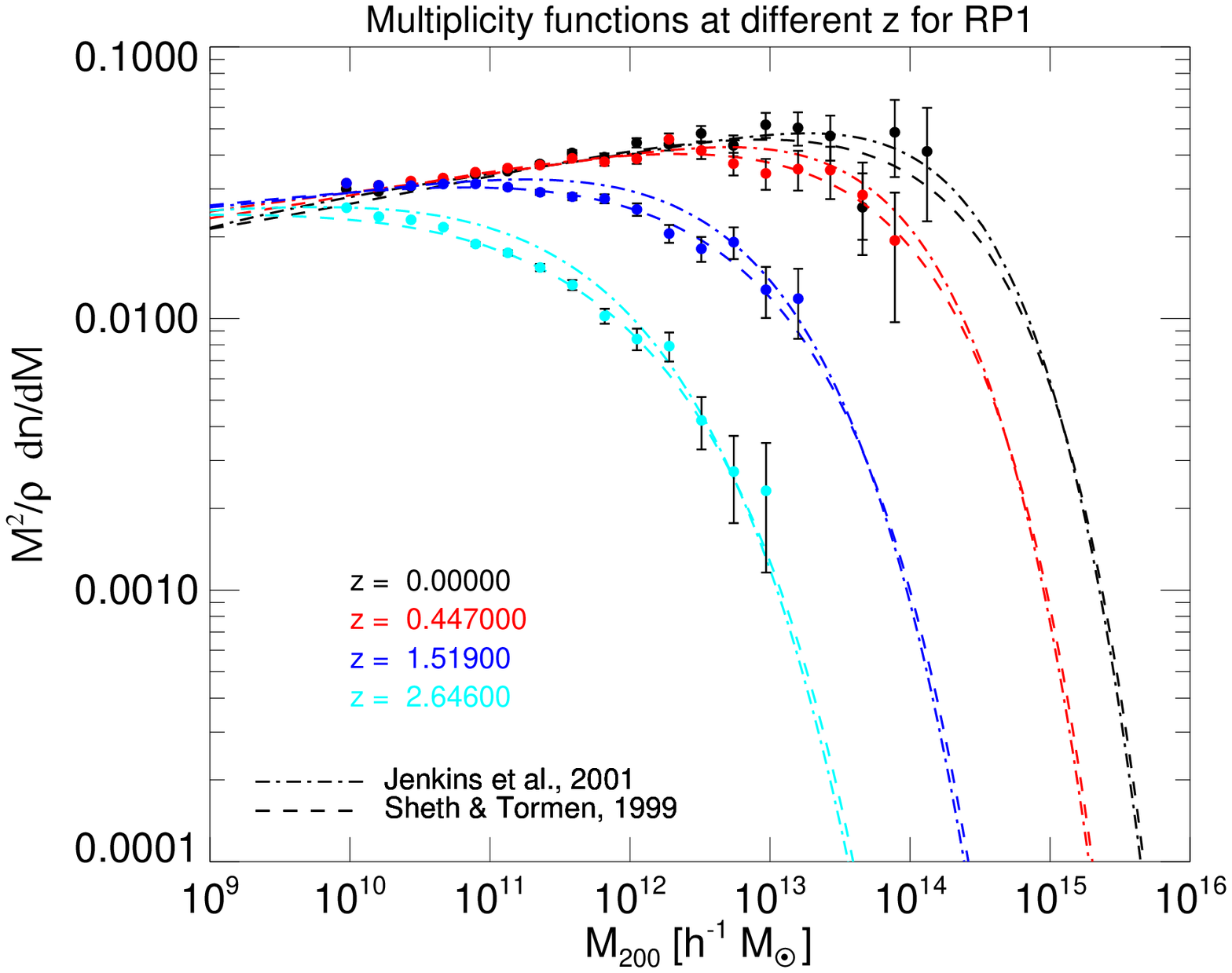}\\
\includegraphics[scale=0.45]{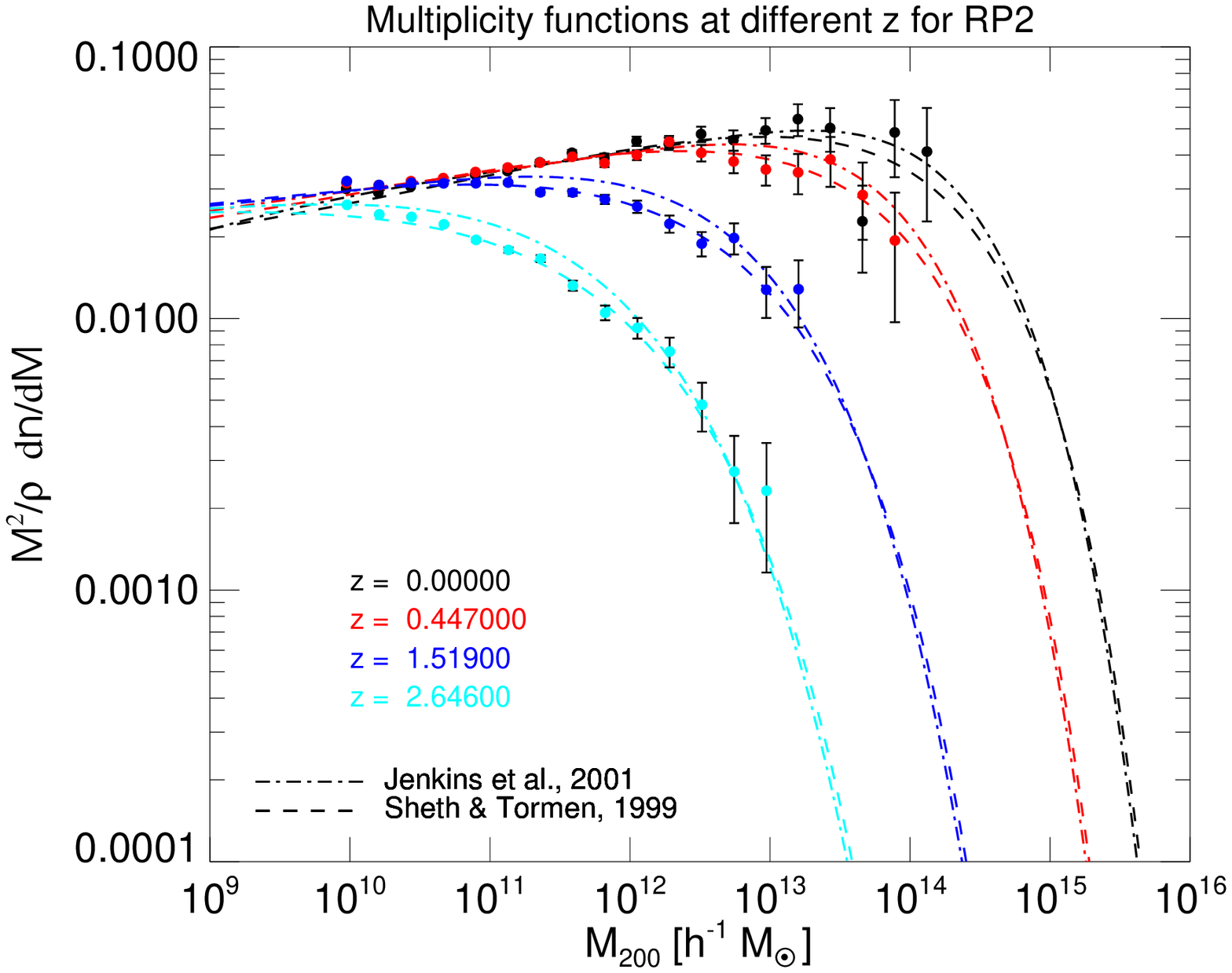}
\includegraphics[scale=0.45]{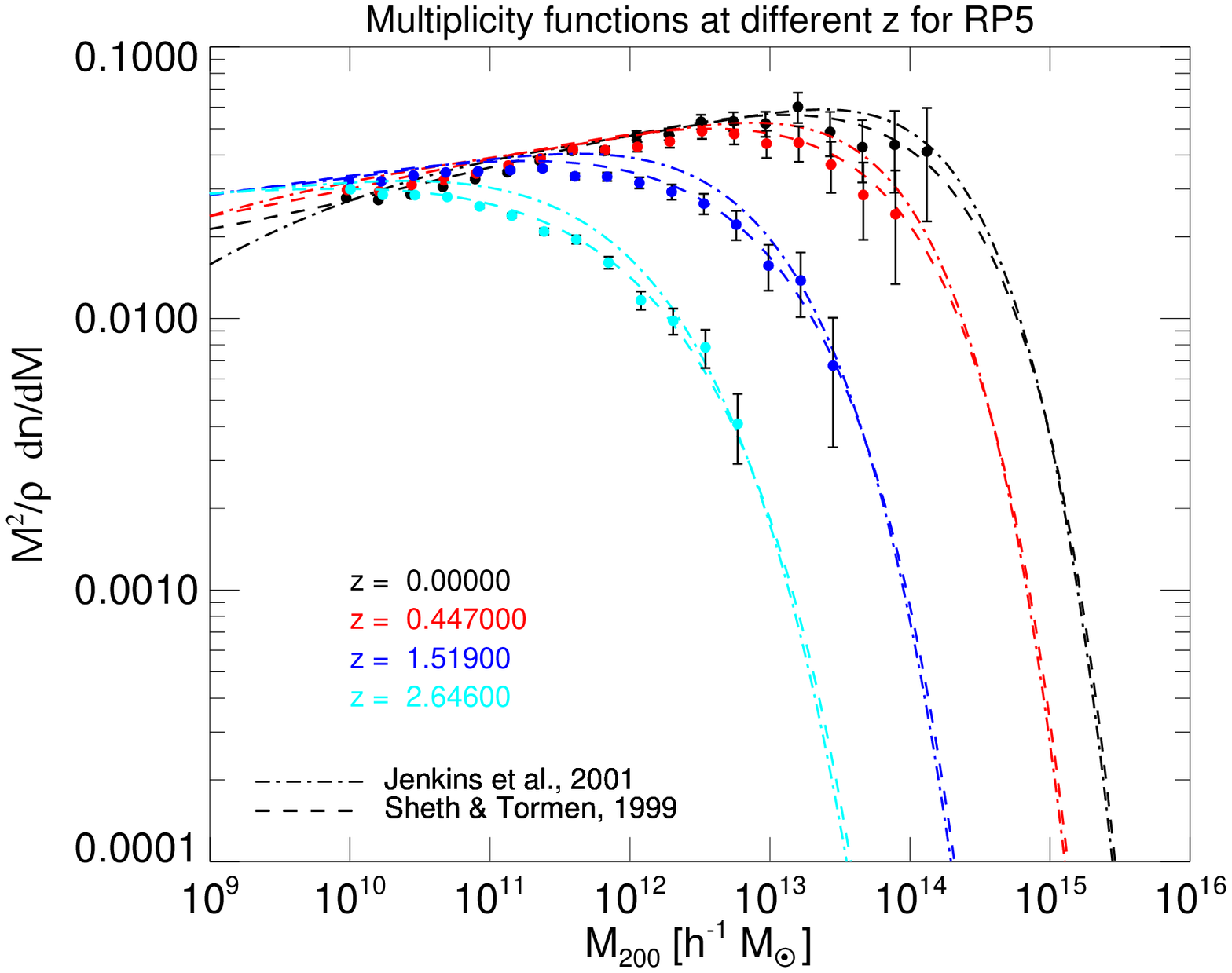}
  \caption{Multiplicity functions for the four high-resolution simulations of
    the interacting DE models. The four differently-coloured sets of data
    points are the multiplicity functions evaluated in equally spaced
    logarithmic mass bins at four different redshifts. The dot-dashed and
    dashed lines represent the predictions for the multiplicity function from
    \citet{Jenkins_etal_2000} and \citet{Sheth_Tormen_1999}, respectively,
    computed for each simulation with the appropriate growth function and
    power spectrum. The comparison clearly shows a slightly better agreement
    with the fitting function by \citet{Sheth_Tormen_1999}, in particular at
    high redshift.}
\label{multiplicity_function}
\end{figure*}

We also plot in Fig.~\ref{multiplicity_function} the multiplicity function,
defined as the derivative of the mass function with respect to the mass
\mbox{($M^{2}/\rho \cdot {\rm d}n(<M)/{\rm d}M$)}, for each simulation at
different redshifts. This more sensitive representation of the mass function
reveals a slightly better agreement with the formula by
\citet{Sheth_Tormen_1999} compared with that of \citet{Jenkins_etal_2000},
which are both overplotted for a direct comparison.

\subsection{Matter power spectrum}

\begin{figure*}
\includegraphics[scale=0.45]{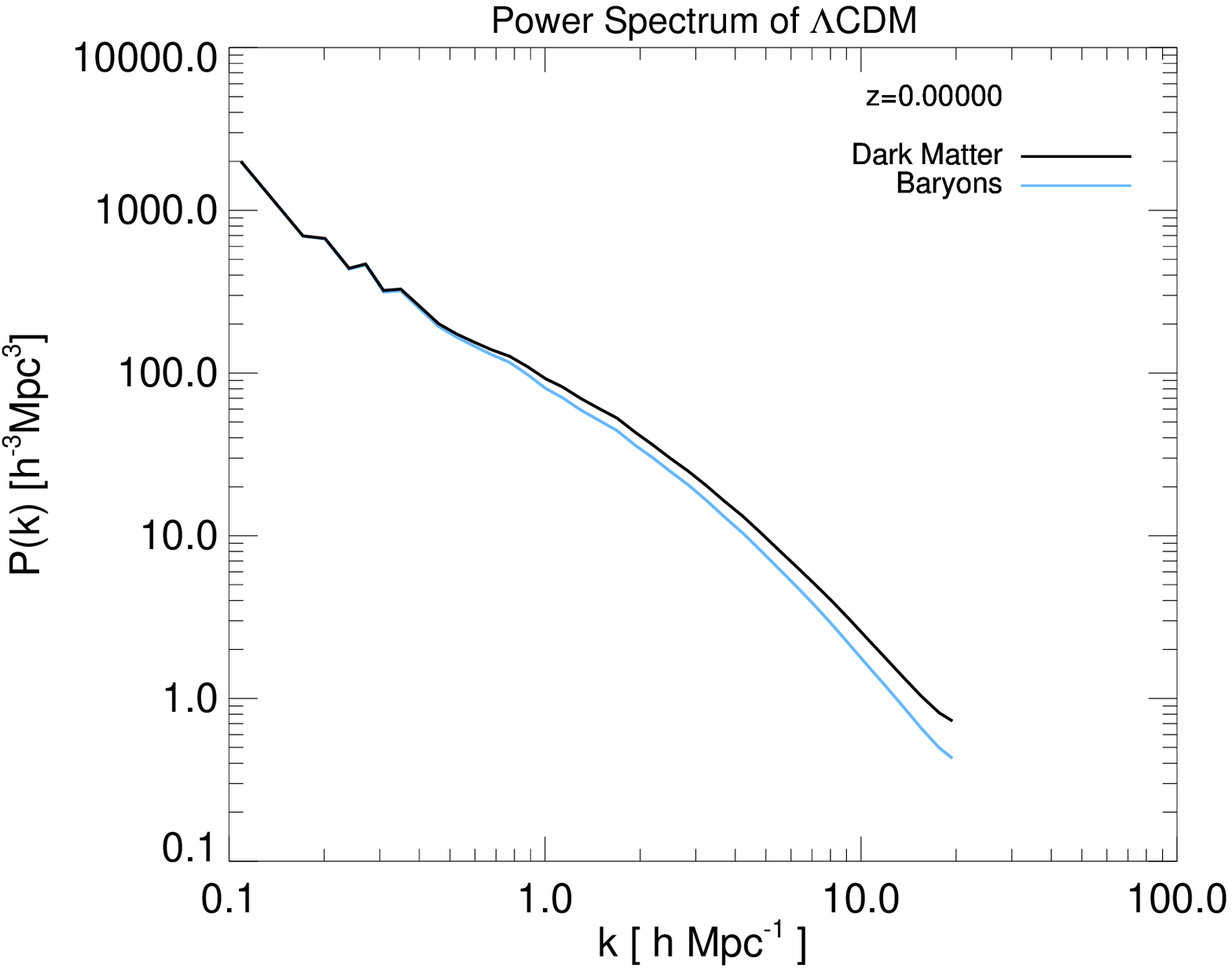}
\includegraphics[scale=0.45]{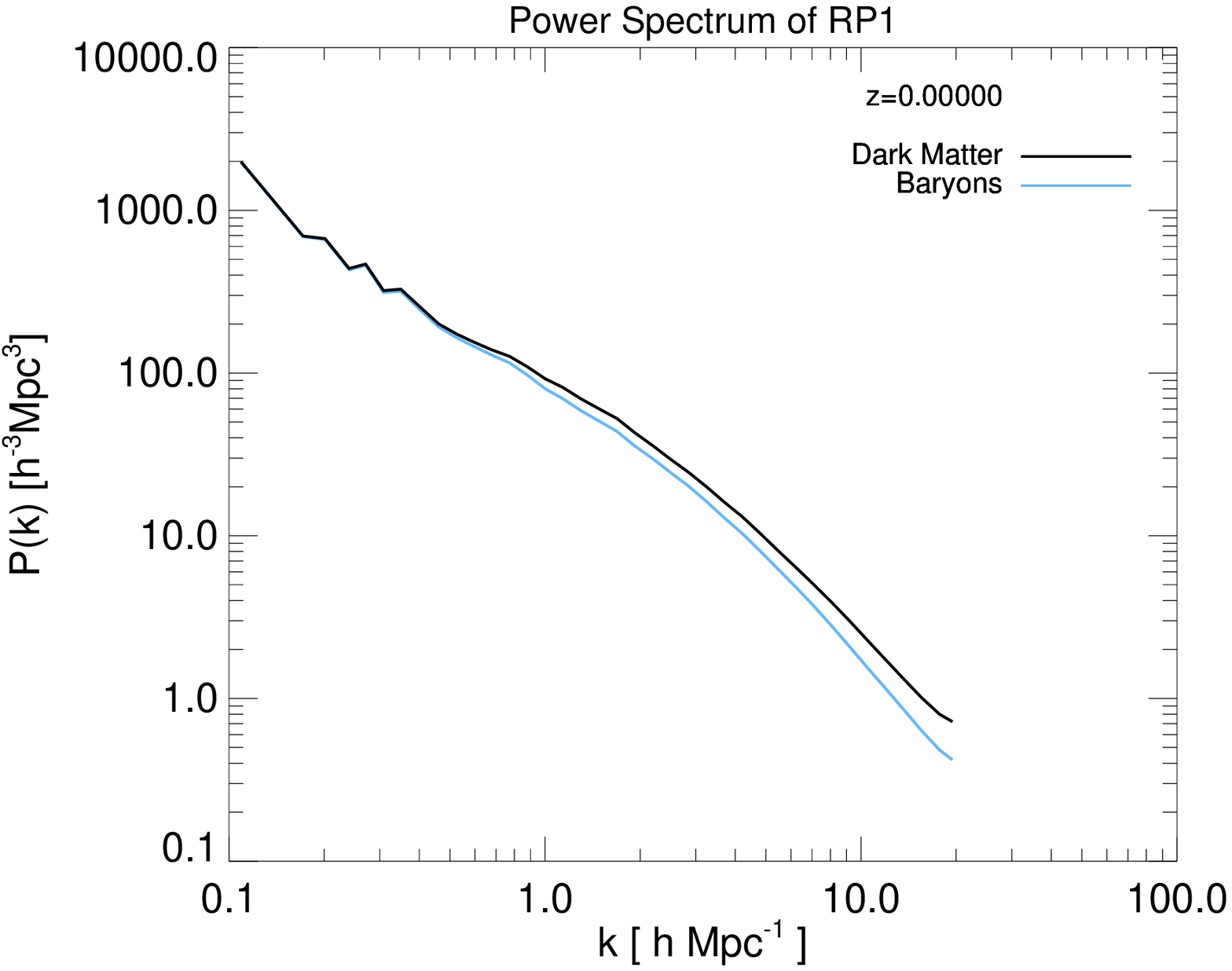}\\
\includegraphics[scale=0.45]{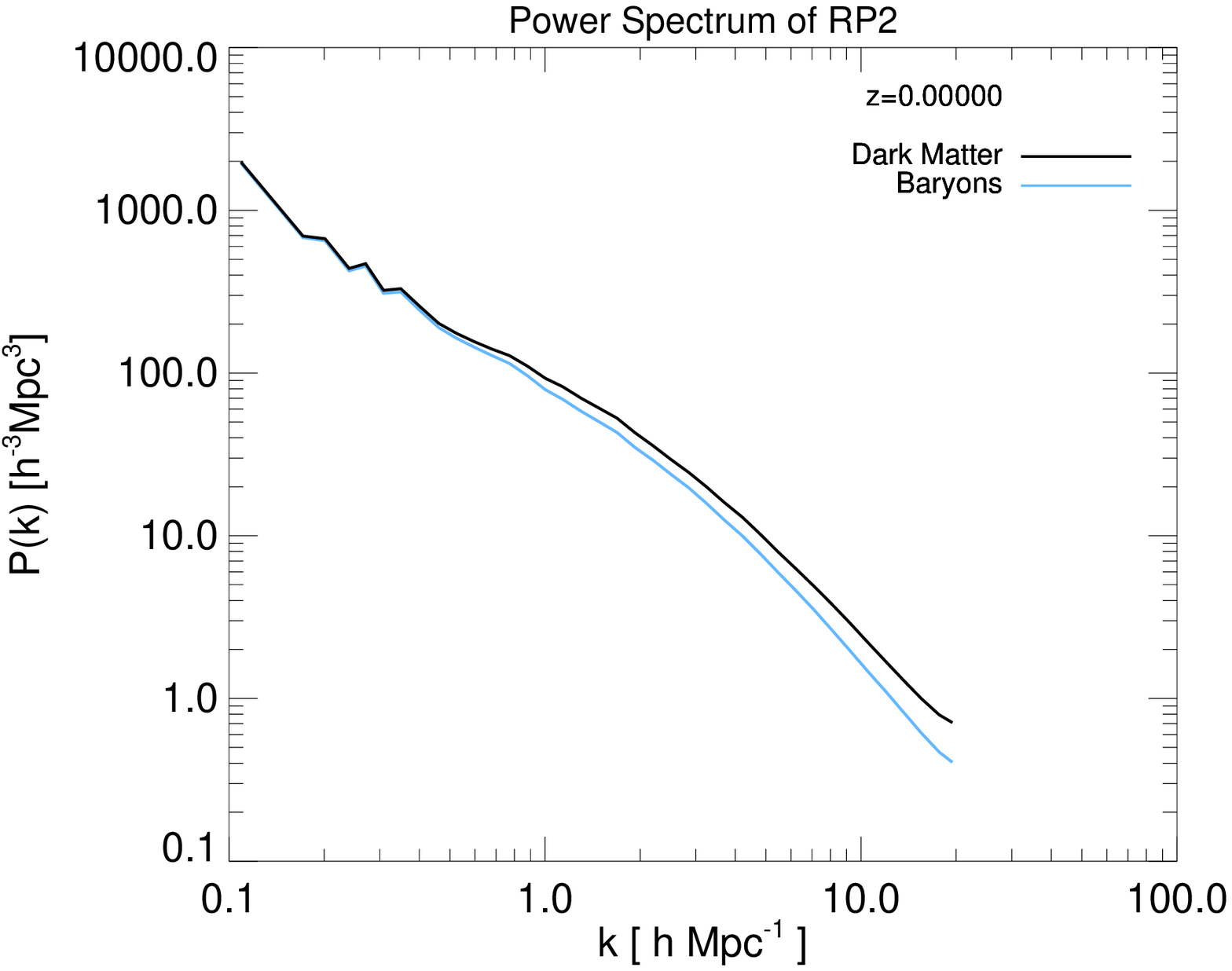}
\includegraphics[scale=0.45]{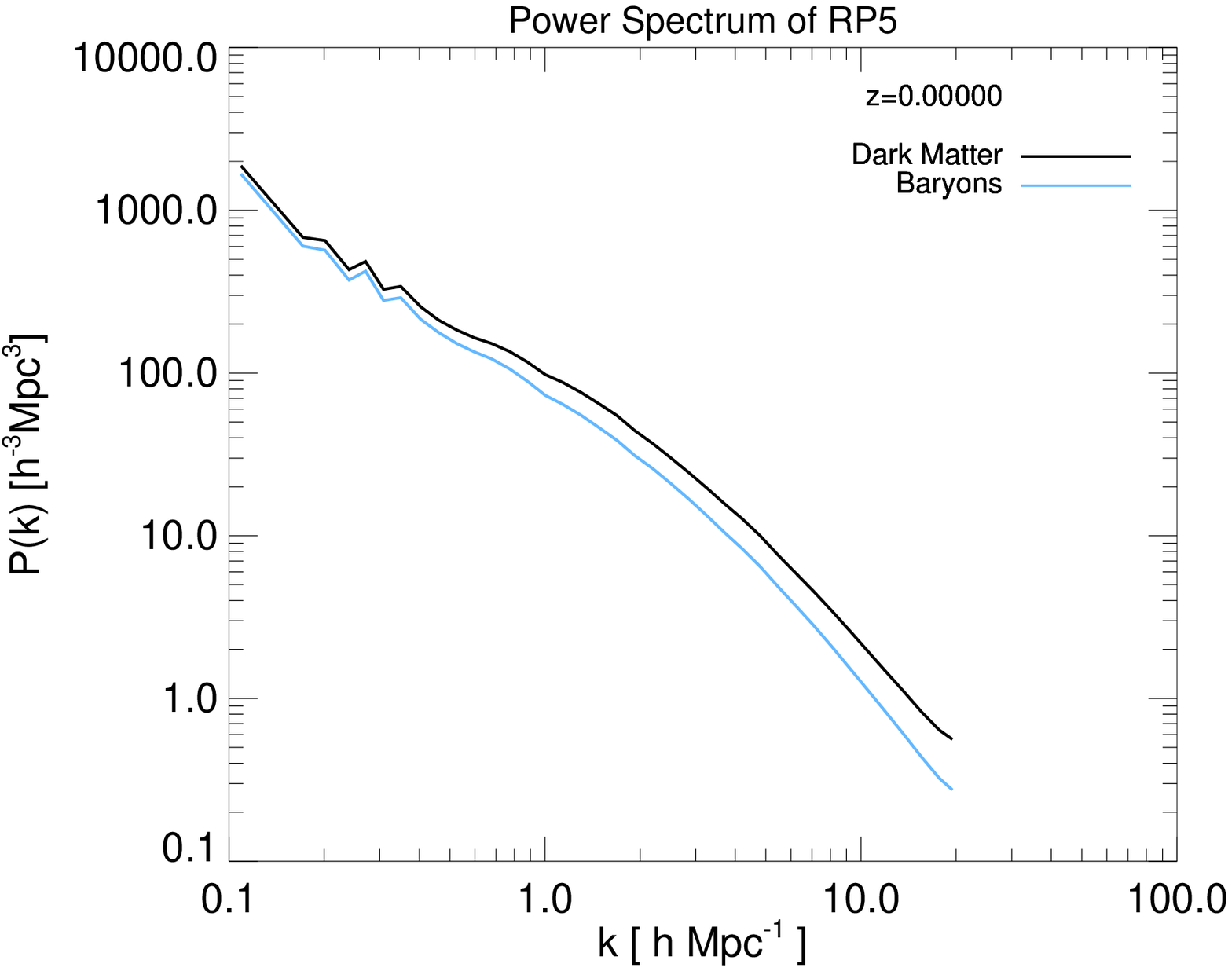}\\
\includegraphics[scale=0.45]{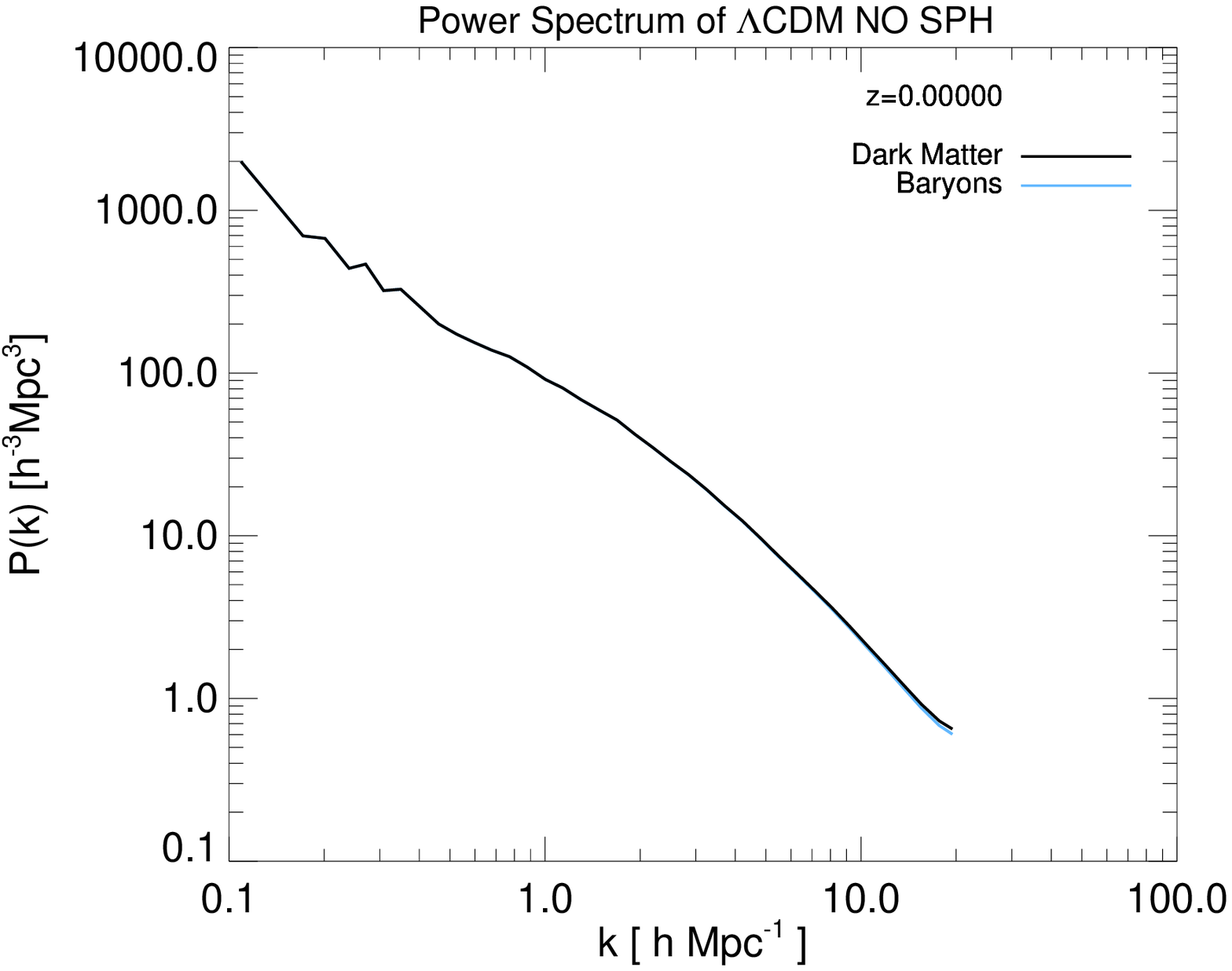}
\includegraphics[scale=0.45]{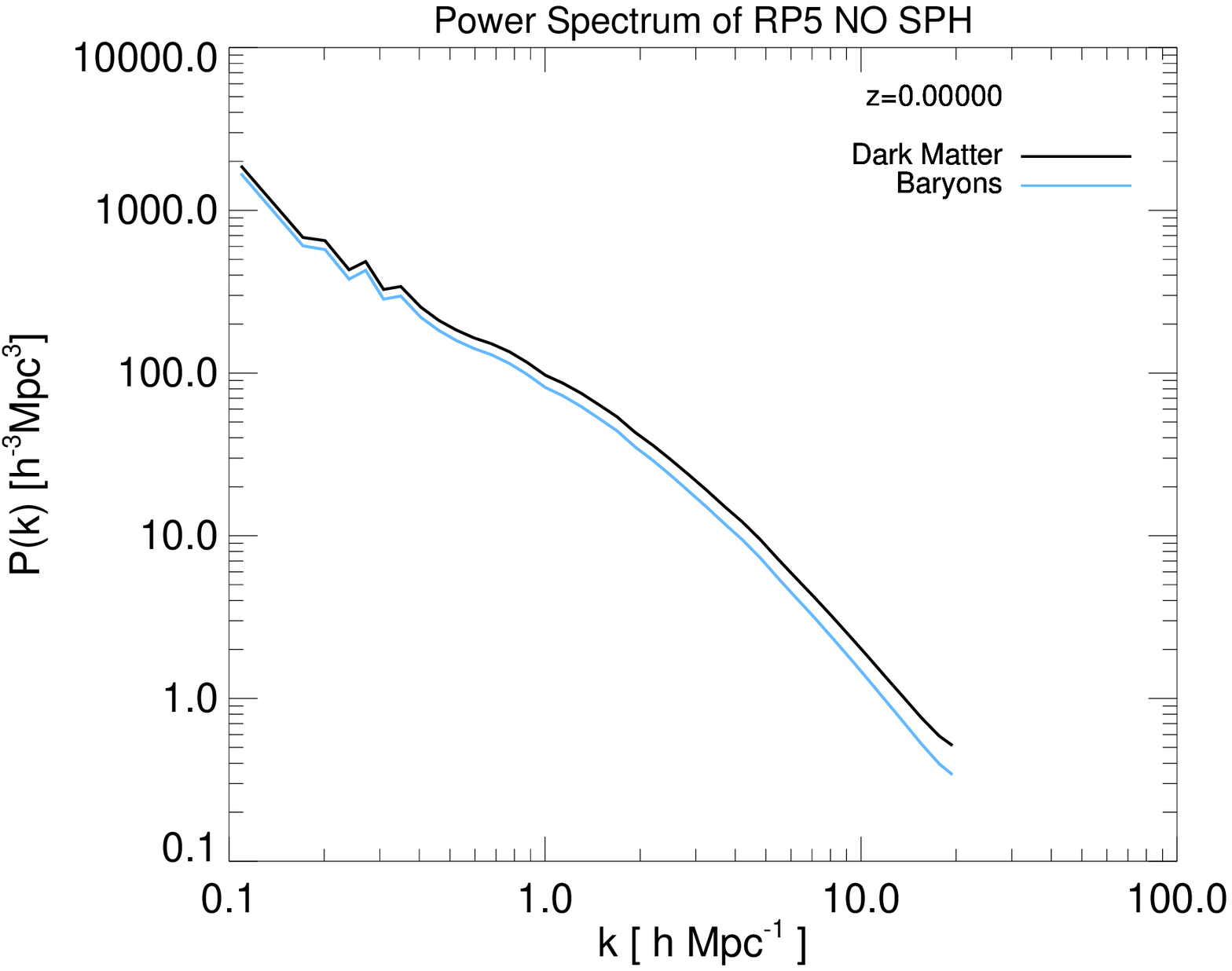}
  \caption{Power spectra of CDM (black line) and baryons (blue line) at $z=0$
    for the set of coupled DE models under investigation. The appearance of a
    bias between the two distributions, which grows with increasing coupling
    $\beta _{c}$, is clearly visible at the large scale end of the plots. The
    last two panels show the comparison of a $\Lambda $CDM and a coupled DE
    cosmology with $\beta _{c}=0.2$ in absence of hydrodynamic forces acting
    on baryons. In these two panels, the bias on all scales is purely due to
    the interaction of CDM with the DE scalar field $\phi $. }
\label{sim_power_spectra}
\end{figure*}

The presence of a long-range fifth-force acting only between CDM particles
induces a linear bias on all scales between the amplitude of density
fluctuations in baryons and CDM.  These density fluctuations, in fact, start
with the same relative amplitude in the initial conditions of all the
simulations, and then grow at a different rate due to the presence of the
extra force.  Such a bias is then easily distinguishable from the
hydrodynamical bias that develops only at small scales as structure evolves.
This effect is clearly visible in our simulations, as it can be seen from the
baryon and CDM power spectra at $z=0$ in the four different cosmologies we
analyze (Fig.~\ref{sim_power_spectra}): the density power spectra of baryons
and CDM end up with a different amplitude on all scales at $z=0$ in the
coupled DE models. CDM always has a larger amplitude, and the difference grows
with increasing coupling $\beta _{c}$.

In order to disentangle this effect from the small scale bias due to pressure
forces acting on baryons, we can make use of our ``NO-SPH'' simulations to
show the result if only the fifth-force effect is included. This is shown in
the last two panels of Fig.~\ref{sim_power_spectra}.
\normalsize

\begin{figure*}
\includegraphics[scale=0.45]{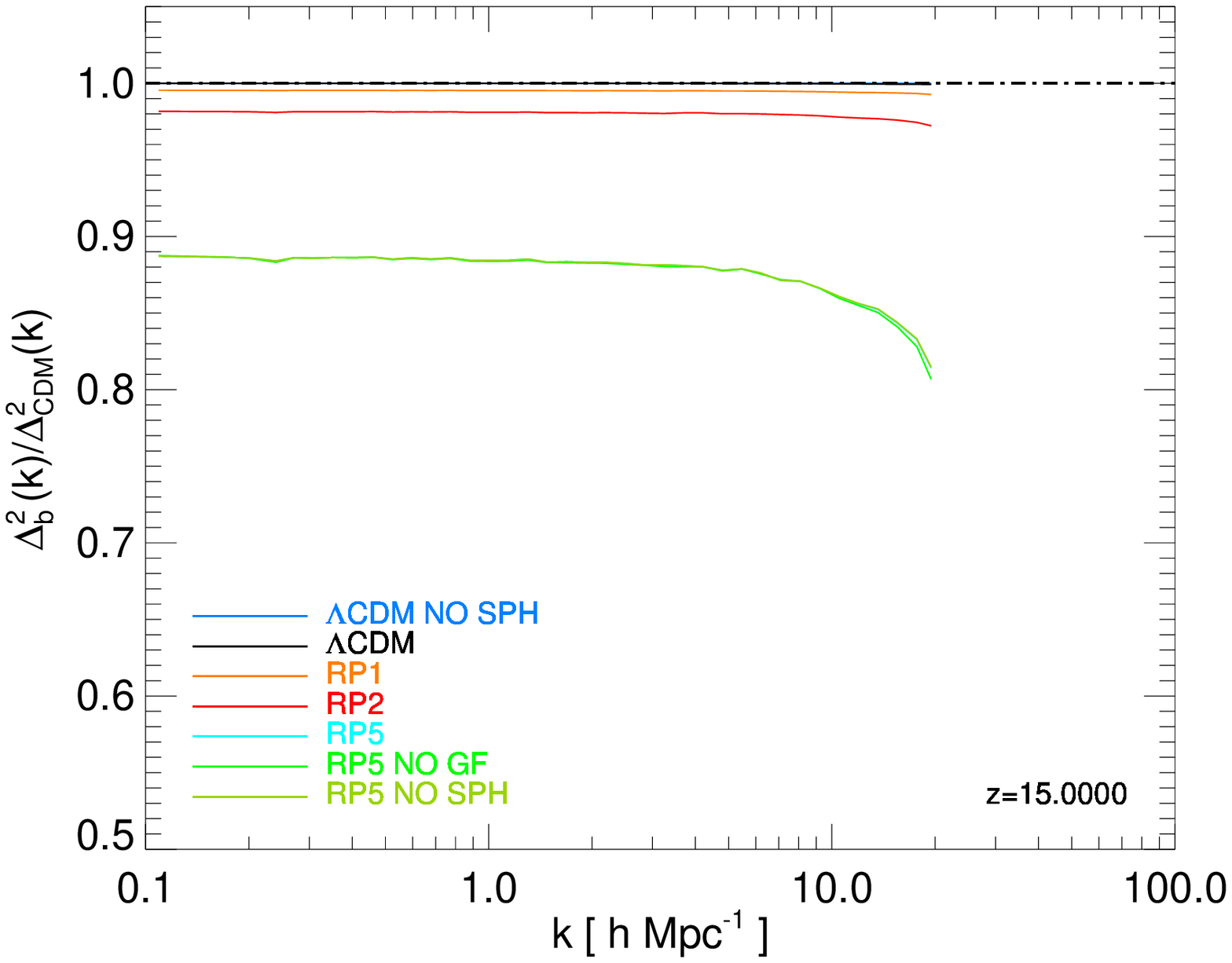}
\includegraphics[scale=0.45]{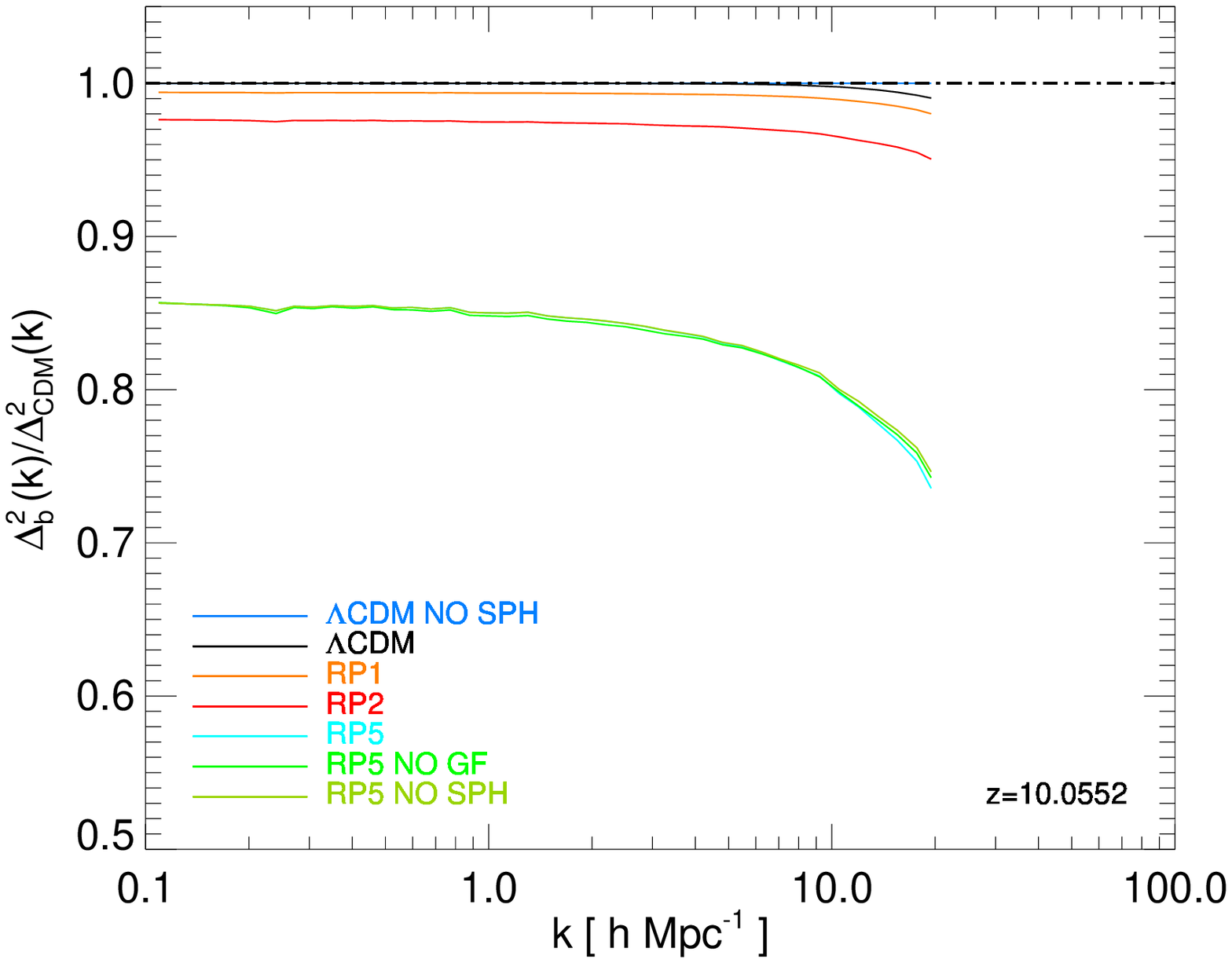}\\
\includegraphics[scale=0.45]{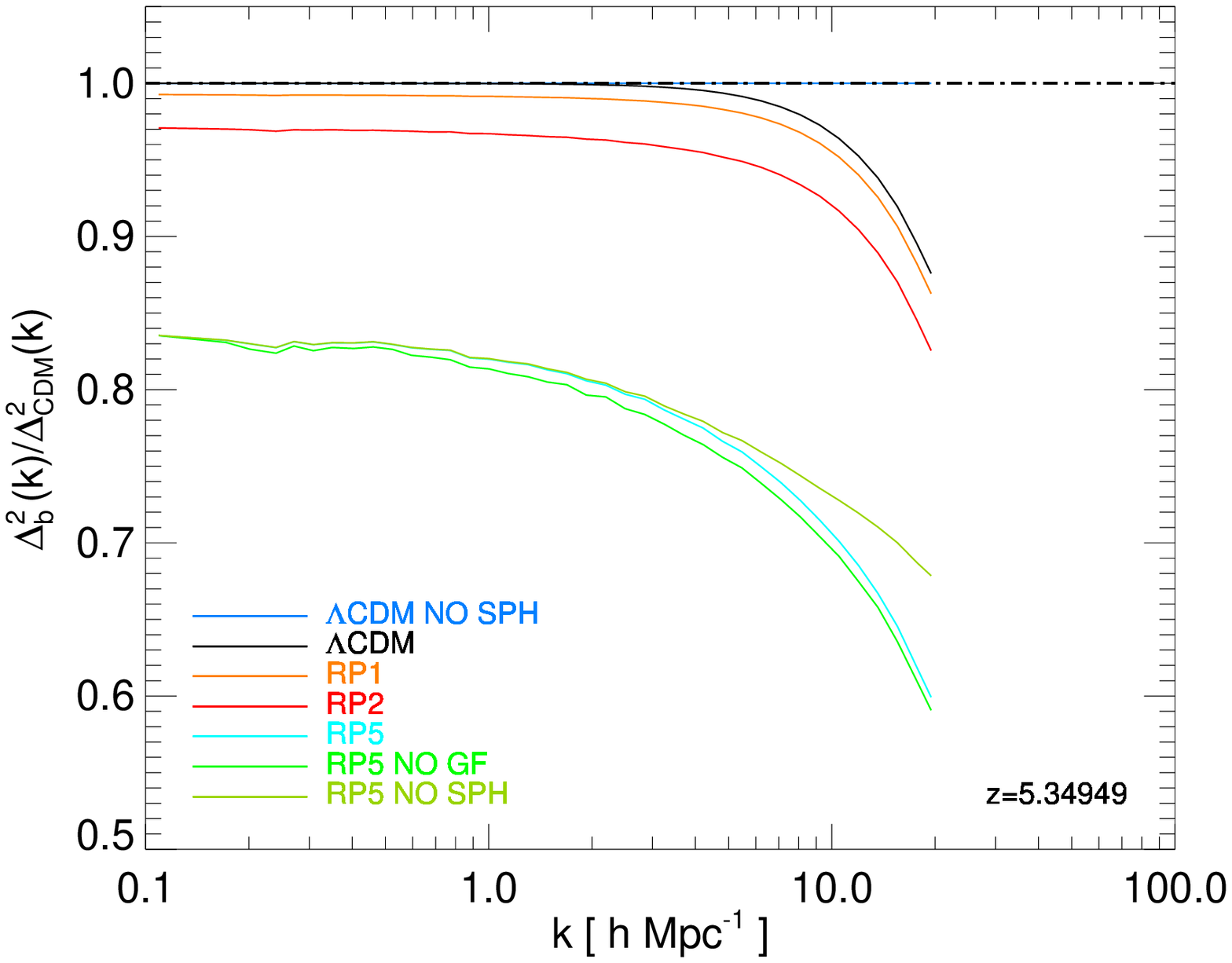}
\includegraphics[scale=0.45]{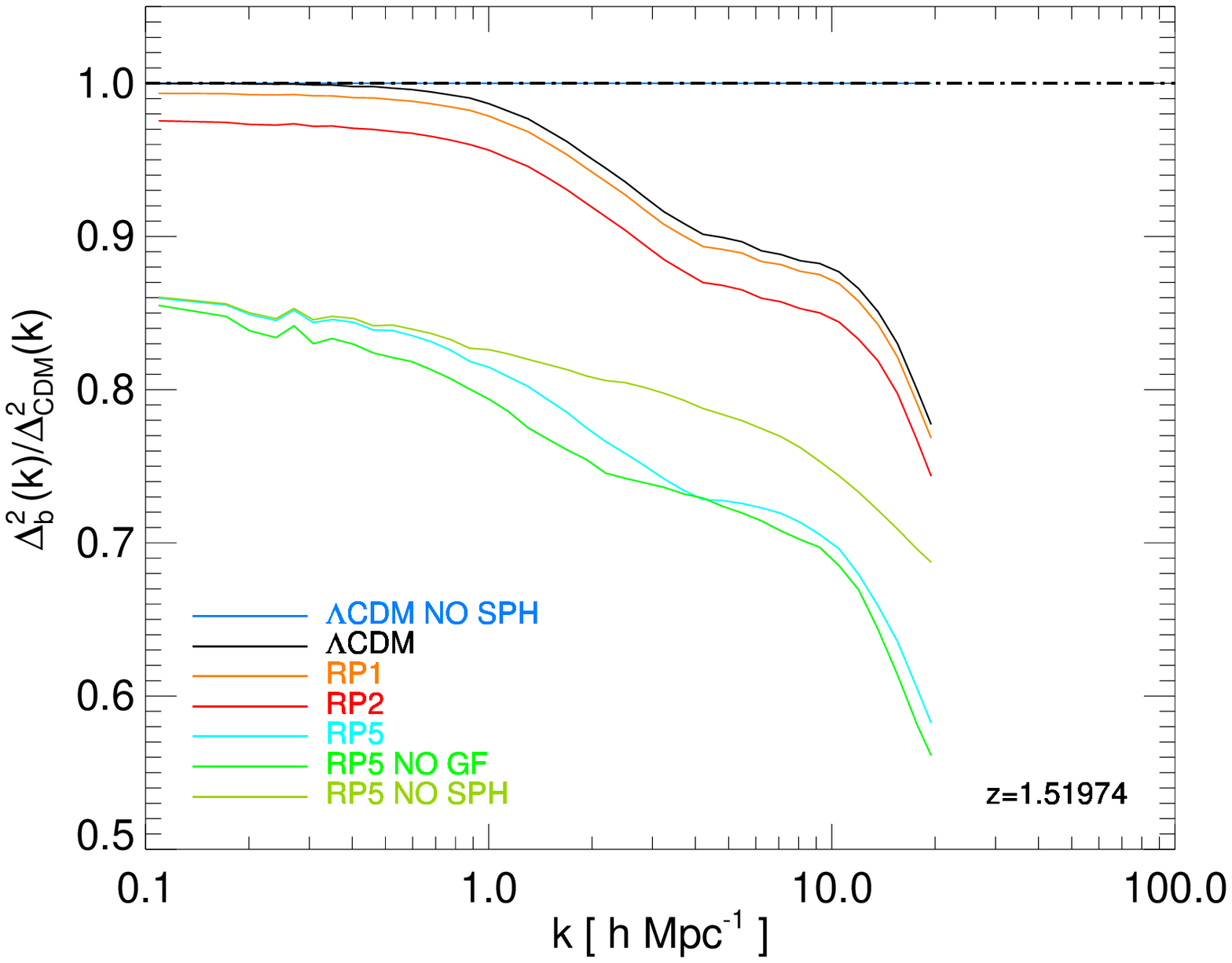}\\
\includegraphics[scale=0.45]{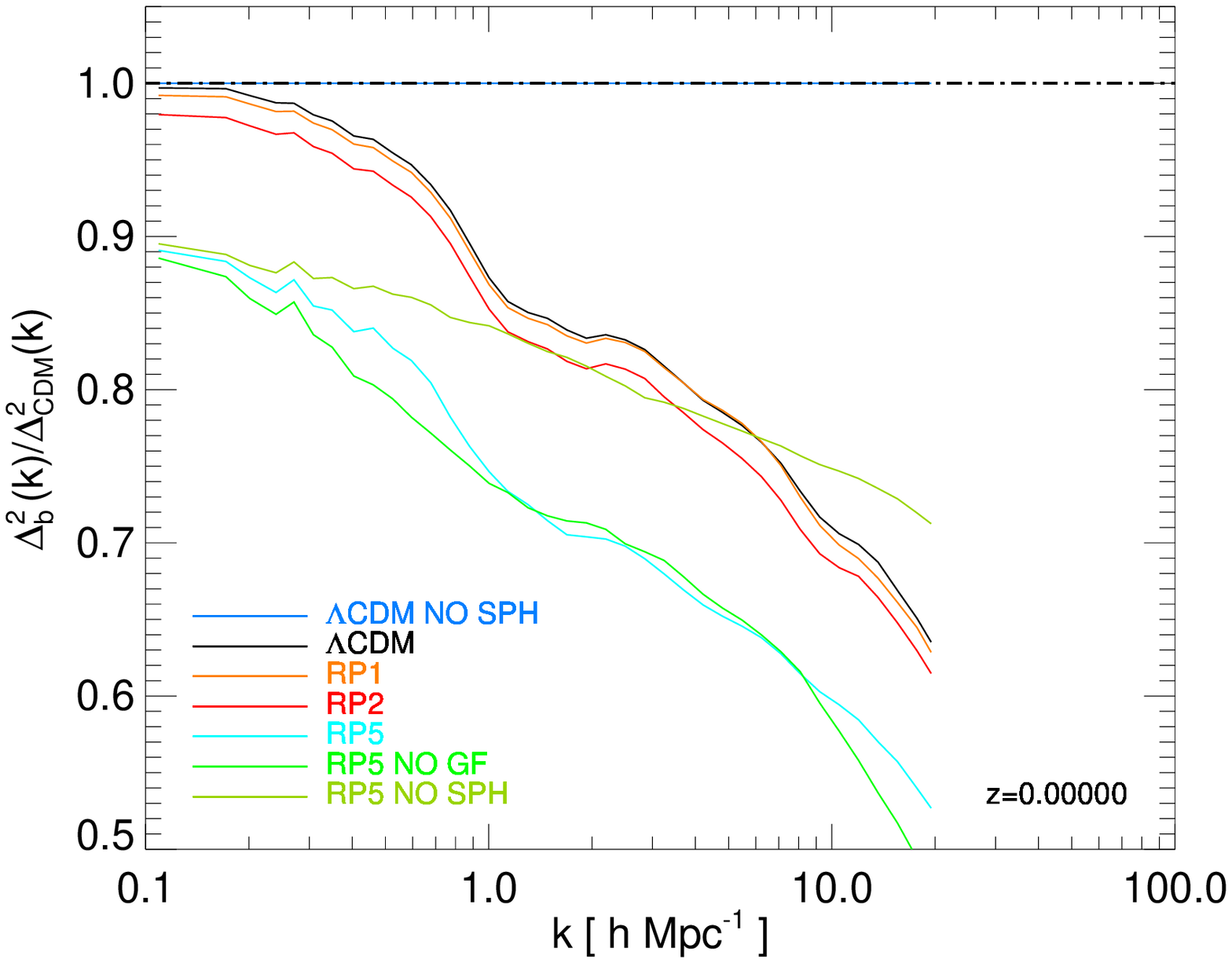}
  \caption{Ratio of the power spectra of baryons and CDM as a function of
    wavenumber for the set of high-resolution simulations ran with our
    modified version of {\small GADGET-2}, for five different redshifts. The
    linear large-scale bias appears already at high redshifts, while at lower
    redshifts the hydrodynamic forces start to suppress power in the baryon
    component at small scales. In absence of such hydrodynamic forces the
    progressive enhancement of the large scale bias at small scales for the
    RP5-NO-SPH run (light green curve) as compared to the completely flat
    behaviour of the $\Lambda $CDM-NO-SPH simulation (blue curve) -- where no
    bias is expected -- shows clearly that nonlinearities must increase the
    effect of the coupling on the different clustering rates of the two
    species. All the curves have been corrected for a spurious numerical drop
    of the baryonic power at small scales as described in the text.}
\label{power_ratio}
\end{figure*}

In order to make the effect described above even more visible, and to better
show the difference of the hydrodynamical bias from the gravitational one
induced by the coupled DE component, we also plot in Fig.~\ref{power_ratio}
the ratio of the power $\Delta ^{2}(k) = P(k)k^{3}/2\pi $ in baryons to that
in CDM at different redshifts. In these plots, we have corrected all the
curves for a spurious effect on small scales due to the mass difference
between baryon and CDM particles in the simulations that induces a small drop
in the baryon power. This effect is of purely numerical origin and could be
easily removed by using particles of equal mass for the two cosmological
matter species in the N-body runs.

Let us now briefly comment on the plots of Fig.~\ref{power_ratio}. The curves
represent the bias between the baryon and the CDM density fluctuation
amplitudes as a function of the wave number. A constant bias of 1.0 at all
redshifts is the expected result for two collisionless matter species
interacting with the same gravitational strength, and this is what we find for
our $\Lambda $CDM-NO-SPH simulation (dark blue curve). In the $\Lambda $CDM
simulation, we notice that this value of 1.0 is maintained at all redshifts
only for large scales, while on smaller scales, as structures evolve, the
collisional nature of baryon particles progressively induces a drop of this
ratio (black curve). The same behaviour is seen for all the other hydrodynamic
simulations (RP1, RP2, RP5). However, in the latter cases, the large-scale
bias is always smaller than 1.0, already at high redshifts, and it decreases
with increasing values of the CDM coupling $\beta _{c}$, as expected. This is
the gravitational bias that appears also at large scales in
Fig.~\ref{sim_power_spectra}.

Particularly interesting is then again the case of the RP5-NO-SPH simulation
(dark-green curve), as compared to the $\Lambda $CDM-NO-SPH one, because it
allows us to disentangle the hydrodynamic effects from the effects due to the
coupled DE extra physics. In this case we find that the bias of RP5-NO-SPH
perfectly overlaps with the one of the other two RP5 simulations at high
redshifts, while at lower and lower redshifts, the small scale behaviour is
progressively more and more different: as expected the absence of hydrodynamic
forces acting on baryon particles induces a larger value of the bias, which is
now solely due to the different gravitational strength felt by the two
particle species. It is however very interesting to notice that the bias does
not keep the large-scale linear value at all scales, as it is the case for the
$\Lambda $CDM-NO-SPH run, but evolves towards lower and lower values for
smaller and smaller scales. This clearly shows that nonlinearities enhance the
effect of the coupling on the growth of overdensities in the two differently
interacting matter species.

\normalsize

\subsection{Halo density profiles}

Applying the selection criterion described above to our four self-consistent
simulations ($\Lambda$CDM, RP1, RP2, RP5) we select among the 200 most massive
groups identified at $z=0$ for each run 74 objects that can be considered with
certainty to be the same structure in the different simulations.  For these 74
halos we compute the spherically averaged density profiles of CDM and baryons
as a function of radius around the position of the particle with the minimum
gravitational potential.

\begin{figure*}
\includegraphics[scale=0.45]{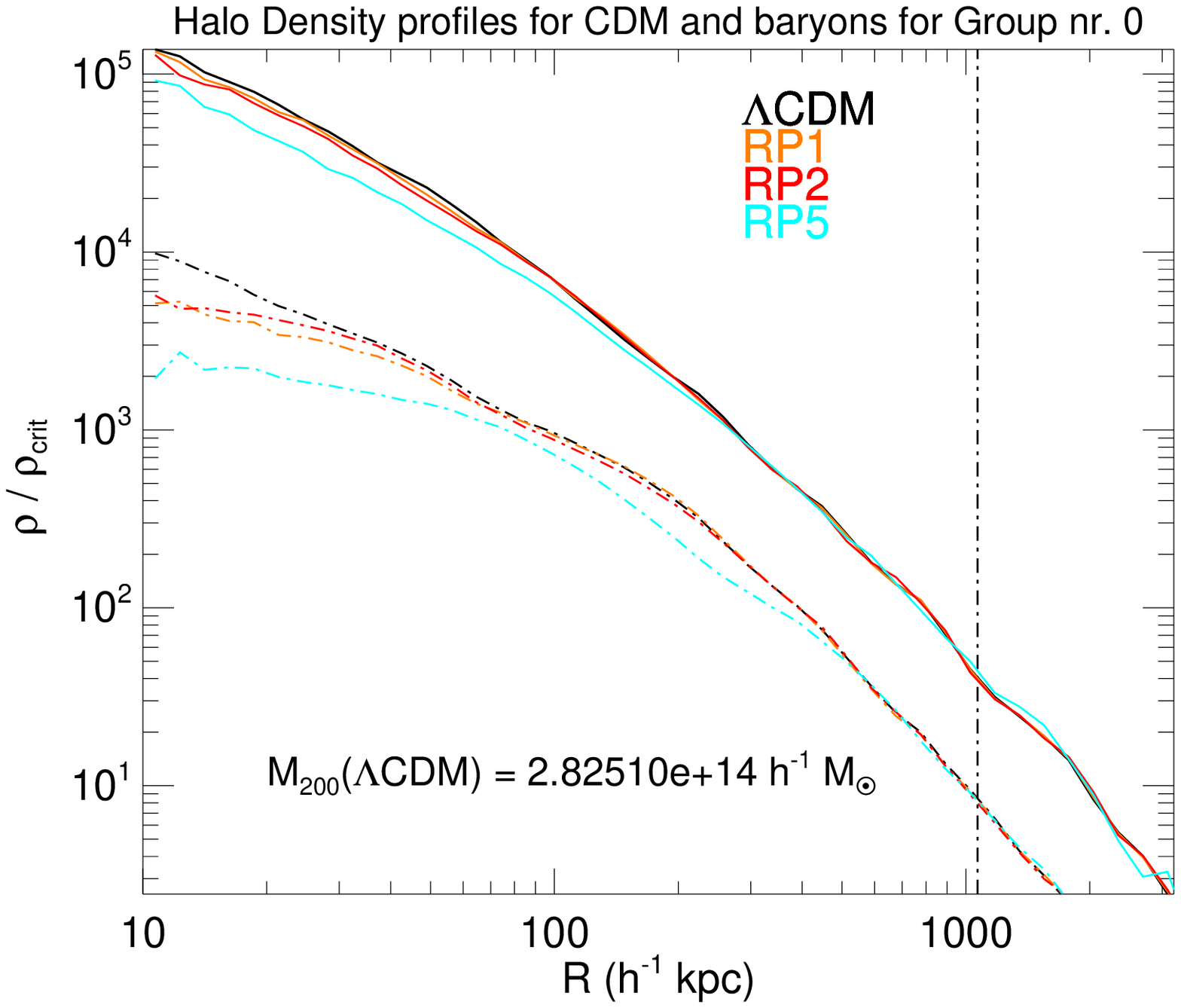}
\includegraphics[scale=0.45]{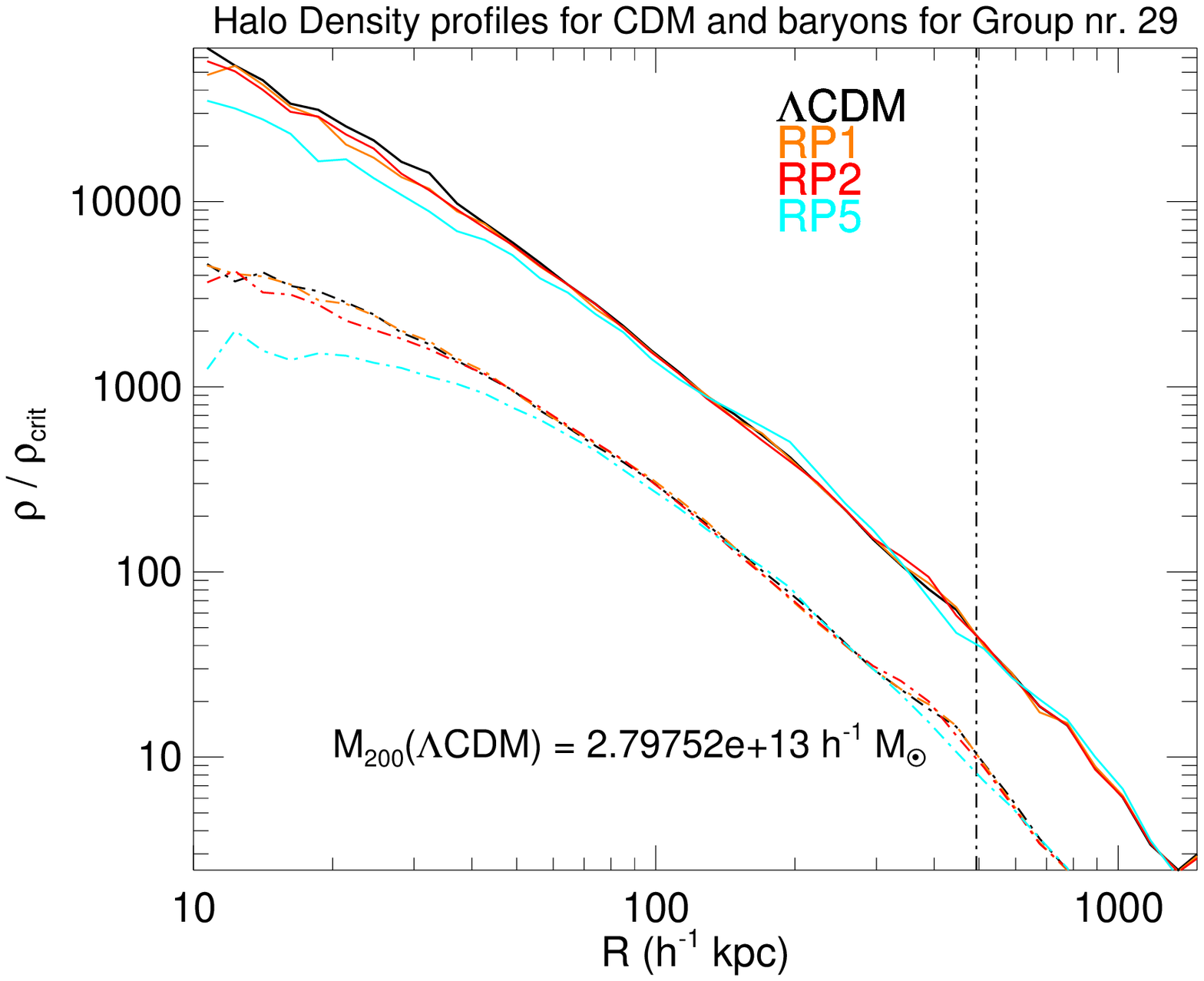}\\
\includegraphics[scale=0.45]{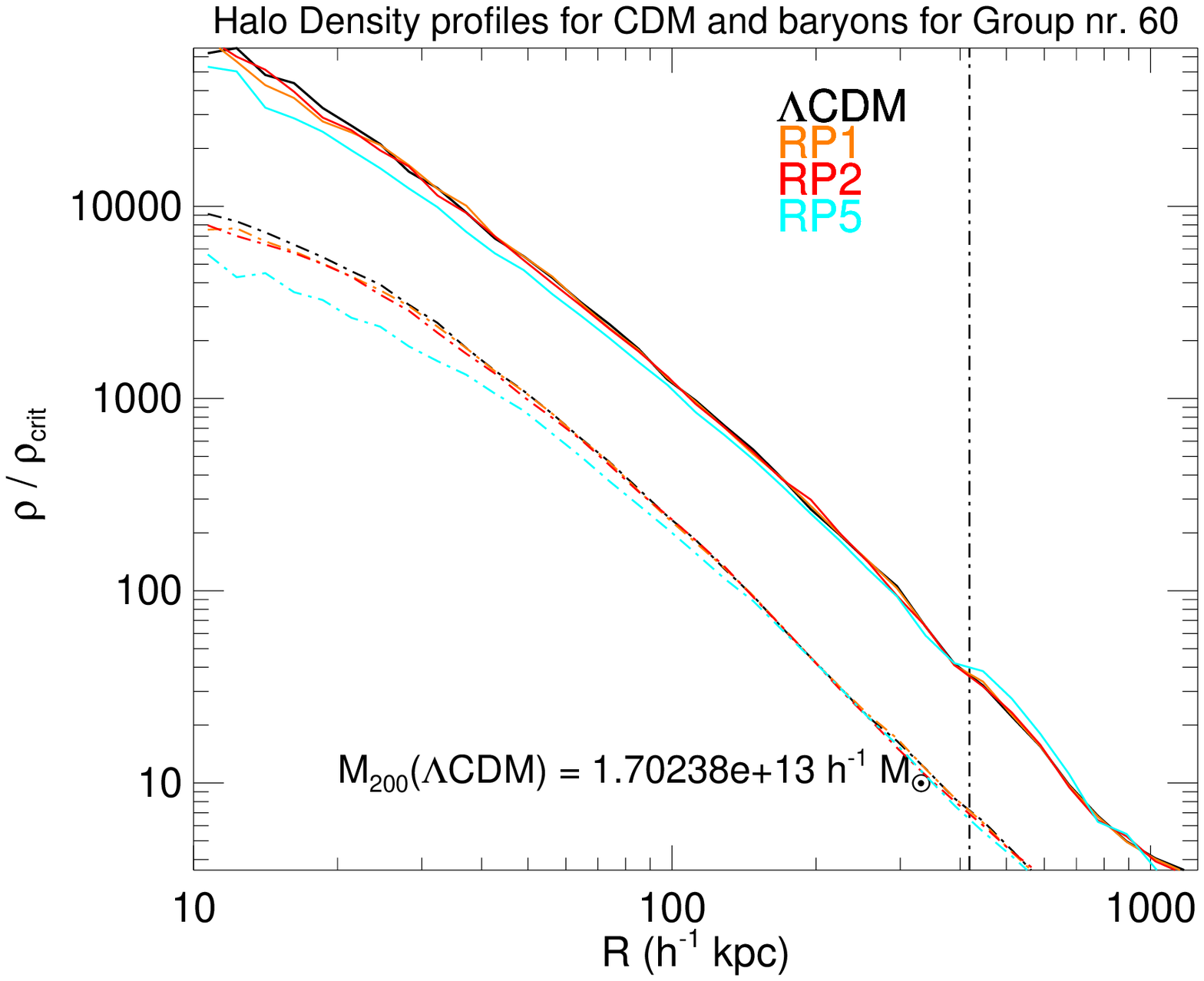}
\includegraphics[scale=0.45]{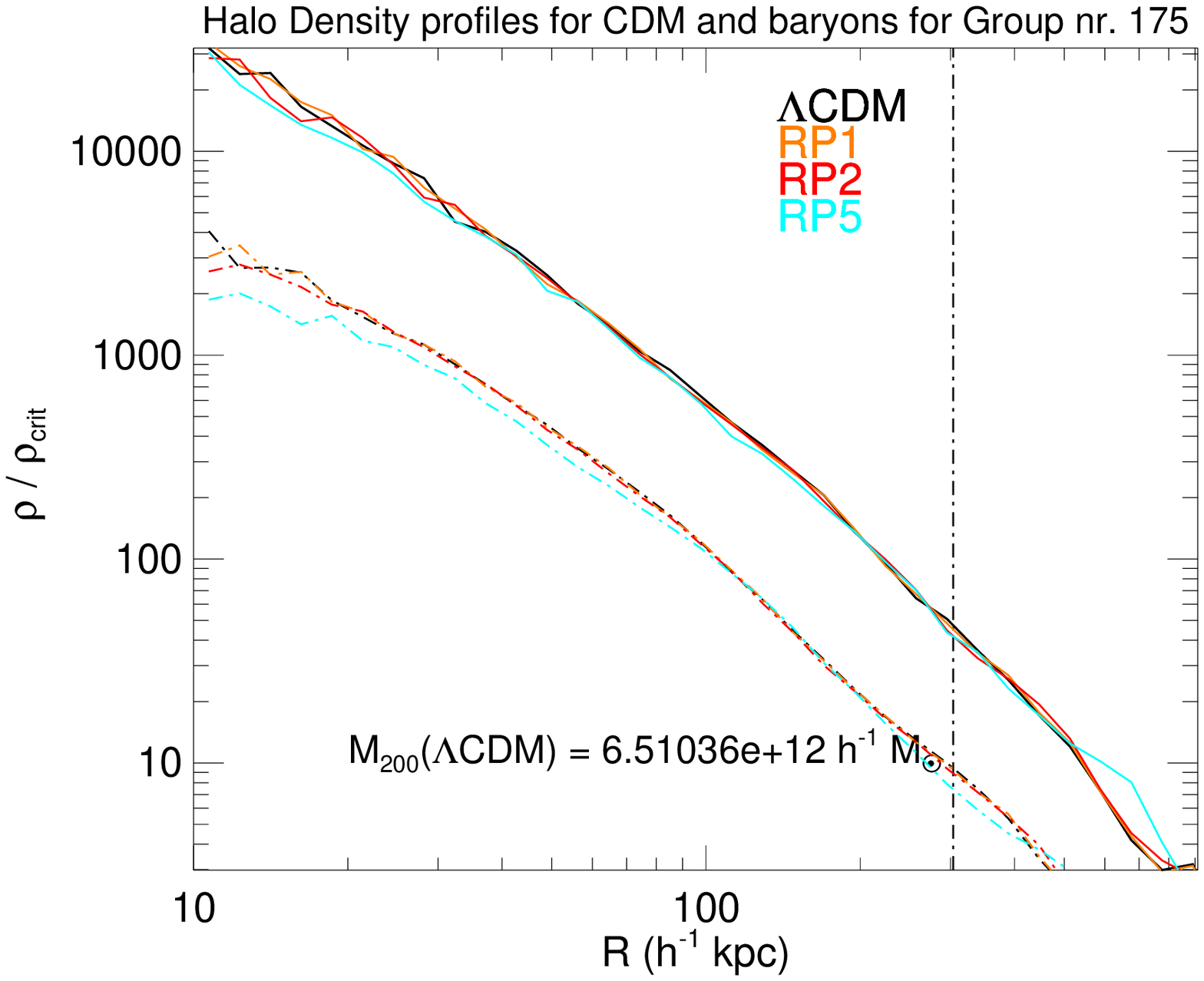}
  \caption{Density profiles of CDM (solid lines) and baryons (dot-dashed
    lines) for four halos of different mass in the simulation box at
    $z=0$. The vertical dot-dashed line indicates the location of the virial
    radius for the $\Lambda$CDM halo. The decrease of the inner overdensity of
    the profiles with increasing coupling is clearly visible in all the four
    plots.}
\label{density_profiles}
\end{figure*}

Interestingly, the halos formed in the coupled DE cosmologies show
systematically a lower inner overdensity with respect to $\Lambda $CDM, and
this effect grows with increasing coupling. This is clearly visible in
Fig.~\ref{density_profiles} where we show the density profiles of CDM and
baryons in the four different cosmologies for a few selected halos of
different virial mass in our sample.  We remark that this result is clearly
incompatible with the essentially opposite behaviour previously reported by
\citet{Maccio_etal_2004}, and deserves a more detailed discussion.

Unlike found in \citet{Maccio_etal_2004},
all the 74 halos in our comparison sample 
have density profiles
that are well fitted  by the  NFW  fitting function \citep{NFW}
\begin{equation}
\label{NaFrWh}
\frac{\rho(r) }{\rho _{\rm crit}} = \frac{\delta ^{*}}{({r}/{r_{s})}(1+{r}/{r_{s}})^2}\,,
\end{equation}
independent of the value of the coupling.  Here $\delta^{*}$ is a parameter
that sets the characteristic halo density contrast relative to the critical
density $\rho_{\rm crit}$. The scale radius $r_{s}$ increases for each halo
with increasing coupling $\beta _{c}$, and becomes larger than that found in
$\Lambda $CDM. In other words, the halos become {\em less concentrated} with
increasing coupling.

\begin{figure}
\includegraphics[scale=0.45]{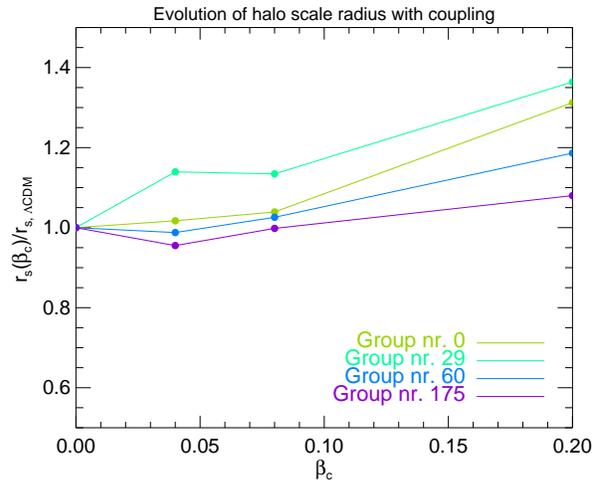}
  \caption{Relative evolution with respect to $\Lambda $CDM of the scale
    radius $r_{s}$ for the four halos plotted in Fig.~\ref{density_profiles}
    as a function of coupling $\beta _{c}$.}
  \label{plot_scale_radii}
\end{figure}

For example, for the four halos shown in Fig.~\ref{density_profiles}, the
scale radius grows with increasing coupling by roughly 10\% to 35\% going from
$\Lambda $CDM to RP5, as listed in Table~\ref{Table_scale_radii}, and shown in
Fig.~\ref{plot_scale_radii}.
The effect is not strong enough to fully address the ``cusp-core'' problem, but clearly shows how the interaction between DE and CDM can produce shallower halo density profiles with respect to a $\Lambda$CDM cosmology with the same cosmological parameters. This result goes then in the direction of alleviating the problem, and therefore opens up new room for the study of dark interactions as opposite to previous claims. In fact, although the models presented here span over the full observationally allowed parameter space for constant-coupling models (\citet{Bean:2008ac}, but see also e.g. \citet{LaVacca_etal_2009}), it is well conceivable that more realistic scenarios with a variable coupling strength could produce larger effects in the nonlinear regime without running into conflict with present observational bounds from linear probes. We defer to future work the investigation of such models, but we stress here that our present results constitute the first evidence that the nonlinear dynamics of generalized coupled cosmologies might provide a solution to the ``cusp-core'' problem.
\begin{center}
\begin{table*}
\begin{tabular}{ccccccccc}
\hline
& \begin{minipage}{45pt}
Group 0 \\ $r_{s}$ (h$^{-1}$ kpc)
\end{minipage} &  
\begin{minipage}{45pt}
Group 0 \\ $\frac{r_{s}}{r_{s}(\Lambda \mathrm{CDM})}$
\end{minipage} &
\begin{minipage}{45pt}
Group 29 \\ $r_{s}$ (h$^{-1}$ kpc)
\end{minipage} &
\begin{minipage}{45pt}
Group 29 \\ $\frac{r_{s}}{r_{s}(\Lambda \mathrm{CDM})}$
\end{minipage} &
\begin{minipage}{45pt}
Group 60 \\ $r_{s}$ (h$^{-1}$ kpc)
\end{minipage} &
\begin{minipage}{45pt}
Group 60 \\ $\frac{r_{s}}{r_{s}(\Lambda \mathrm{CDM})}$
\end{minipage} &
\begin{minipage}{45pt}
Group 175 \\ $r_{s}$ (h$^{-1}$ kpc)
\end{minipage} &
\begin{minipage}{45pt}
Group 175 \\ $\frac{r_{s}}{r_{s}(\Lambda \mathrm{CDM})}$
\end{minipage} \\
\hline
$\Lambda $CDM & 225.14  & 1.0 & 105.51 & 1.0 & 61.92 & 1.0 & 70.61 & 1.0\\
RP1 & 229.00 & 1.02 & 120.21 & 1.14 &  61.16 & 0.99 & 67.45 & 0.96\\
RP2 & 233.96 & 1.04 & 119.68 & 1.13 &  63.52 & 1.03 & 70.48 & 1.0\\
RP5 & 295.47 & 1.31 & 143.92 & 1.36 &  73.46 & 1.19 & 76.26 & 1.08\\
\hline
\end{tabular}
\caption{Evolution of the scale radius $r_{s}$ for the four halos shown in
  Fig.~\ref{density_profiles} with respect to the corresponding $\Lambda $CDM
  value. The trend is towards larger values of $r_{s}$ with increasing
  coupling $\beta _{c}$,   with a relative growth of up to 36\% for the 
 largest coupling value $\beta _{c} = 0.2$.}
\label{Table_scale_radii}
\end{table*}
\end{center}

\subsection{Halo concentrations}

For all the 200 most massive halos found in each of our four fully self
consistent simulations we compute halo concentrations as
\begin{equation}
  c = \frac{r_{\rm 200}}{r_{s}}\,,
\end{equation}
based on our NFW fits to the halo density profiles.  Here $r_{\rm 200}$ is the
radius enclosing a mean overdensity 200 times the critical density. Note that
here no further selection criterion is applied, and the concentration is
computed for all the 200 most massive halos in each simulation.

\begin{figure*}
\includegraphics[scale=0.45]{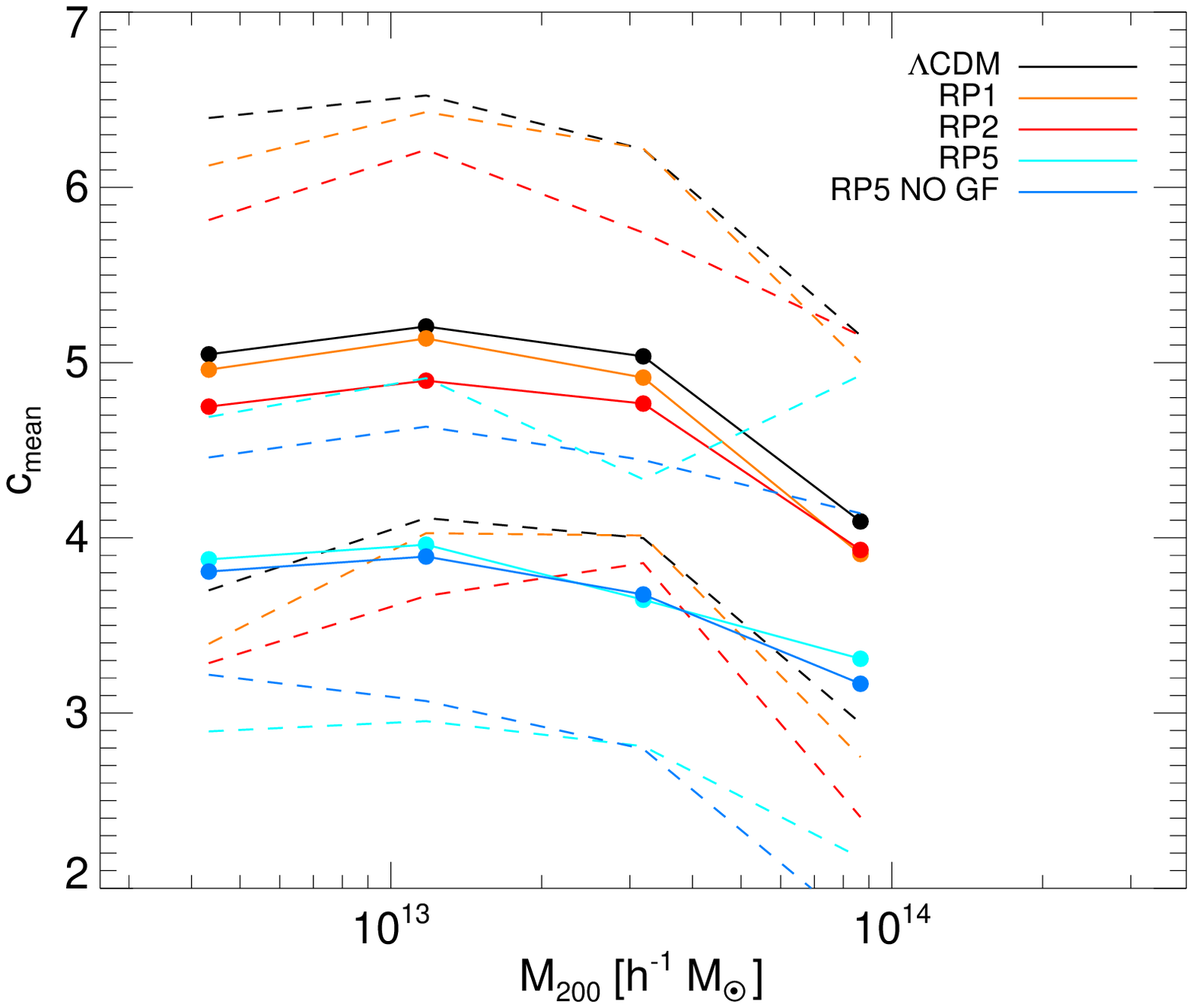}
\includegraphics[scale=0.45]{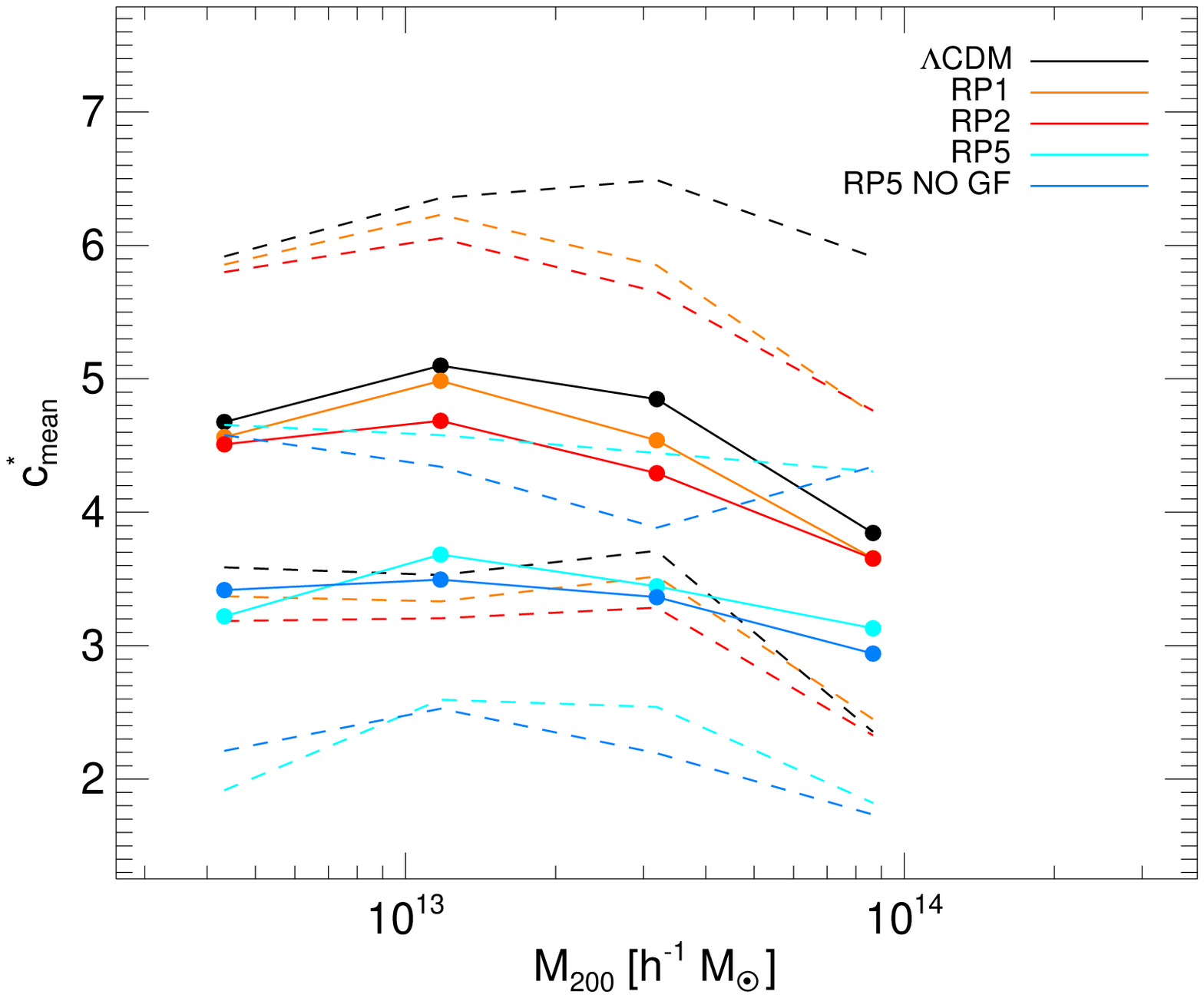}
  \caption{Evolution of the mean halo concentration as a function of mass for
    the 200 most massive halos in our simulations and for the different
    cosmological models under investigation. The concentrations have been
    computed by directly fitting the halo density profile of each halo with an
    NFW model ({\em left panel}) or by using the method introduced by
    \citet{Aquarius} and described in
    Eqs.~(\ref{conc_aquarius},\ref{deltav_aquarius}) ({\em right panel}). The
    halos have been binned by mass, and the mean concentration in each bin is
    plotted as a filled circle. The coloured dashed lines indicate for each
    simulation  the spread of 68\% of the halos in each mass
    bin. The highest mass bin is not plotted because of its very low number of
    halos. The decrease of the mean concentration with increasing coupling
    appears in the same way in both plots.}
\label{concentrations}
\end{figure*}

Consistently with the trend found for the inner overdensity in the halo
density profiles and for the evolution of the scale radius with coupling, we
find that halo concentrations are on average significantly lower for coupled
DE models with respect to $\Lambda $CDM, and the effect again increases with
increasing coupling $\beta_{c}$.  This behaviour is shown explicitly in the
left panel of Fig.~\ref{concentrations}, where we plot halo concentrations as
a function of the halo virial mass $M_{200}$ for a series of our
high-resolution simulations.  In the standard interpretation, the halo
concentrations are thought to reflect the cosmic matter density at the time of
formation of the halo, leading to the association of a larger value of the
concentration with an earlier formation epoch, and vice versa. In the context
of this standard picture, the effect we found for the concentrations could be
interpreted as a sign of a later formation time of massive halos in the
coupled DE models as compared to the $\Lambda $CDM model. Such a later
formation time could be possibly due to the fact that matter density
fluctuations start with a lower amplitude in the initial conditions of the
coupled cosmologies with respect to $\Lambda $CDM, and this would make them
forming massive structures later, despite their faster linear growth (as shown
in Fig.~\ref{growth_factor}).

However, we can demonstrate that this is not the case, just making use of our
RP5-NO-GF simulation, in which the Universe evolves according to the same
physics as RP5, but starting with the identical initial conditions as used for
the $\Lambda$CDM run. Therefore, any difference between these two simulations
can not be due to the initial amplitude of fluctuations. The evolution of halo
concentrations with mass for this run is also plotted in
Fig.~\ref{concentrations} (dark blue curve), and shows a very similar behaviour
to the RP5 curve.

As a cross check of this result, we have repeated the same analysis by
computing halo concentrations with an independent method that circumvents the
profile fitting. The concentration can be related to two other basic
structural properties of a halo, namely its maximum rotational velocity
$V_{\rm max}$, and the radius at which this velocity peak is located, $r_{\rm
  max}$.  According to \citet{Aquarius}, the concentration can then be related
to these two quantities by the relation:
\begin{equation}
\label{conc_aquarius}
\frac{200}{3}\frac{c^{3}}{\ln (1 + c) - c/(1 + c)} = 7.213\, \delta _{V}\,,
\end{equation}
where $\delta _{V}$ is a simple function of $V_{\rm max}$ and $r_{\rm max}$:
\begin{equation}
\label{deltav_aquarius}
\delta _{V} = 2\left( \frac{V_{\rm max}}{H_{0}r_{\rm max}}\right) ^{2}\,.
\end{equation}
We denote the concentrations evaluated in this way as $c^{*}$, and include our
results as a function of halo mass in the right panel of
Fig.~\ref{concentrations}.  Although not identical in detail, as expected due
to the different methods used to measure concentrations, the two plots of
Fig.~\ref{concentrations} show the same trend for the evolution of halo
concentrations with coupling, and the same independence of this effect from
the initial conditions of the simulations.
The slight down-turn of the concentrations curve at the low-mass end that is visible in both the two panels of Fig.~\ref{concentrations} is an expected effect due to numerical resolution, as was discussed in detail in \citet{Neto_etal_2007} and in \citet{Aquarius}.
Although the concentration decrease is found to have the same trend and amplitude for the poorly resolved low-mass halos as for the larger structures for which no resolution problem is expected, one should keep in mind as an element of caution that higher resolution simulations would be required for a definitive confirmation of this effect in the low-mass range of our comparison sample, i.e. for masses below $10^{13} $ M$_{\odot }$/h.
\normalsize

In order to directly verify that the lower concentrations cannot be a
consequence of a later formation time we have also  computed the
average formation redshift of the halos in our sample for all the four
self-consistent simulations, by building merger trees for all the halos in our
sample, and by following backwards in time the main progenitor of each halo
until the redshift at which its virial mass is only half of the final virial
mass of the halo at $z=0$. We define the corresponding time as the formation
redshift $z_{f}$ of the halo.  

\begin{figure}
\includegraphics[scale=0.45]{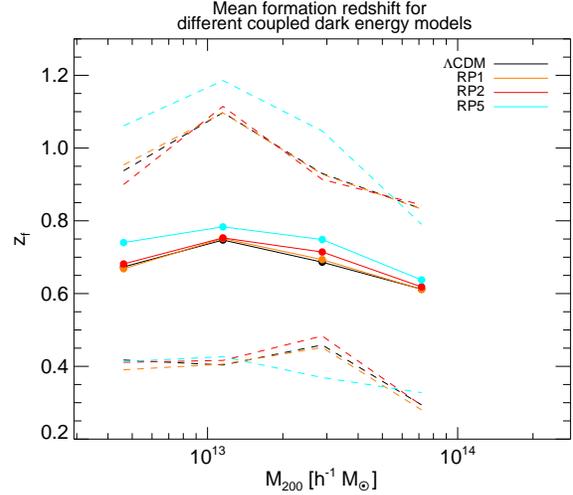}
  \caption{Evolution of the mean halo formation redshift $z_{f}$ as a function
    of halo mass for the 200 most massive halos in our simulations and for the
    different cosmological models under investigation. The formation redshift
    $z_{f}$ is defined as the redshift at which the main progenitor of the
    halo has a virial mass equal to half the virial mass of the halo at
    $z=0$. The halos have been binned by mass, and the mean formation redshift
    in each bin is plotted as a filled circle. The coloured dashed lines
    indicate for each simulation the spread of 68\% of the halos in each mass
    bin. The highest mass bin is not plotted because of its too low number of
    halos.}
\label{formation_redshift}
\end{figure}

In Figure~\ref{formation_redshift}, we show the evolution of $z_{f}$ as a
function of halo mass for all our cosmological models.  It is evident that
massive halos in the different cosmologies form approximately at the same
time, with a slightly earlier formation for the RP5 cosmology, which one might
have expected to translate into slightly larger values of the concentrations.
Therefore we conclude that the unambiguous trend of lower halo concentrations
for larger coupling values must be a peculiar feature that arises from the
extra physics that characterizes the coupled DE cosmologies.  A more detailed
investigation of how this peculiar behaviour arises is hence required in order
to understand this phenomenology of the dynamics in coupled DE cosmologies.

We perform such an investigation by switching off individually the two main
effects which could be responsible of the concentrations drop, which are the
variation of particle mass and the extra velocity-dependent term.  To save computational
time, we do this only for the most strongly coupled model RP5 and only for the
late stages of cosmic evolution. More specifically, we take one of our RP5
simulation snapshots at a given redshift $z^{*}$, and use it as initial
conditions file for a new run starting at $z=z^{*}$ down to $z=0$ in which one
of these two effects is switched off. We label these simulations as
``RP5-NO-MASS'' and ``RP5-NO-FRIC'' for the cases where the mass decrease or
the velocity-dependent term are dropped, respectively.  We set $z^{*}=1.5$ as a
conservative choice based on the consideration that, according to our
definition of formation redshift of a halo, all the halos in our sample have a
formation redshift $z<z^{*}$, as shown in Fig.~\ref{formation_redshift}.  

\begin{figure*}
\includegraphics[scale=0.45]{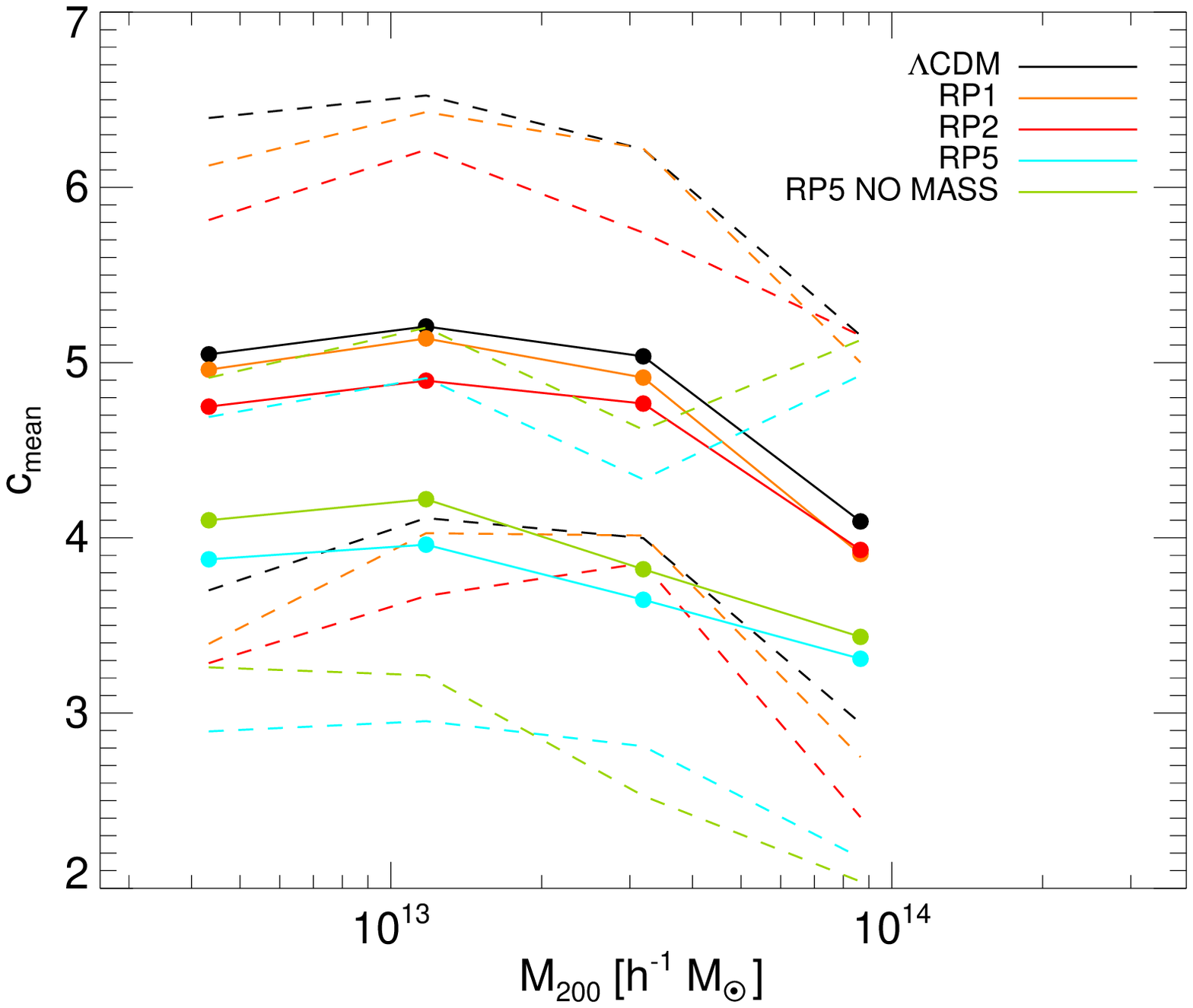}
\includegraphics[scale=0.45]{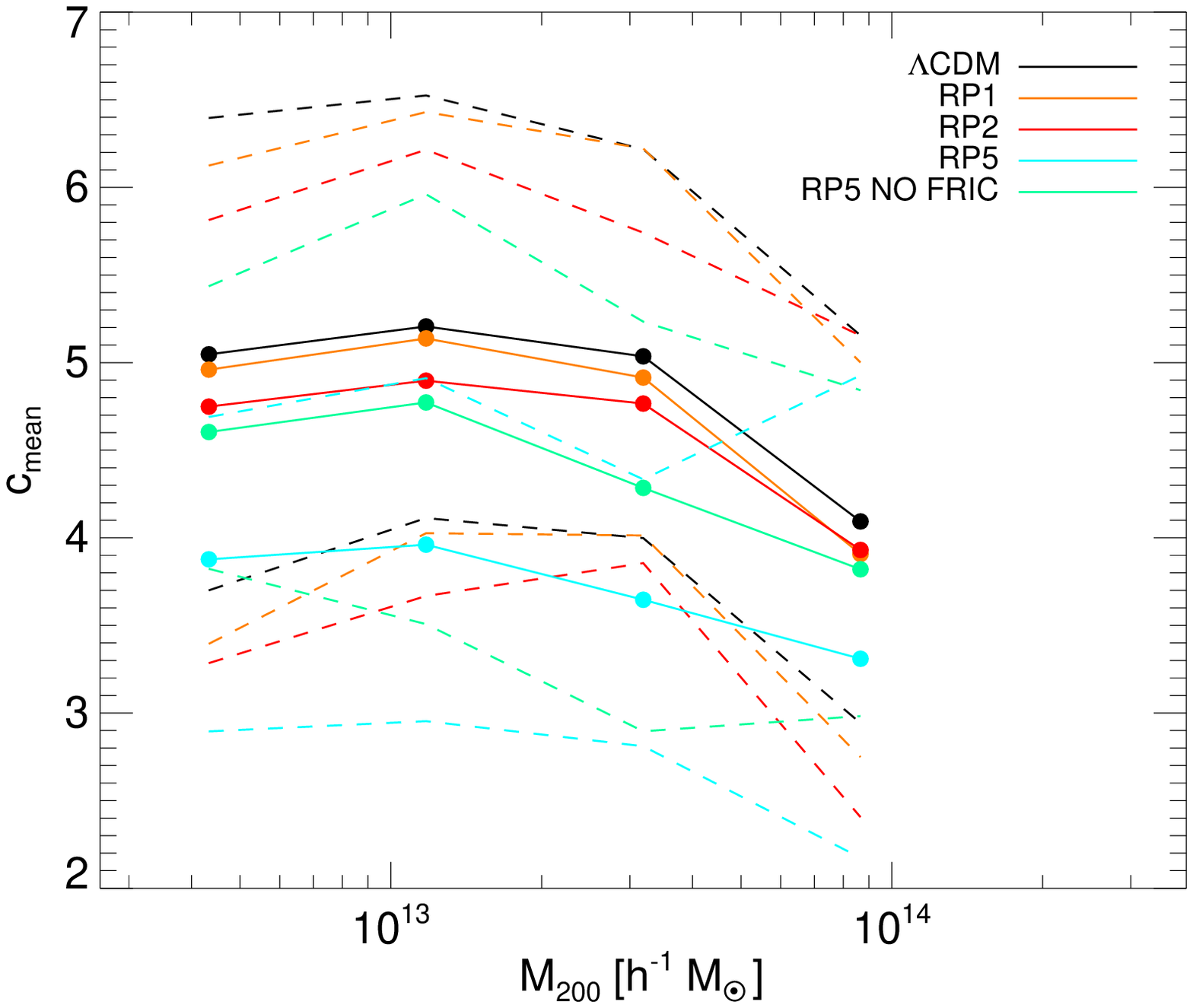}
  \caption{Evolution of halo concentrations for the same models and the same
    halo sample as in Fig.~\ref{concentrations}, and for an additional test
    simulation in each of the two panels. In the {\em left panel}, the
    simulation RP5-NO-MASS shows the effect of switching off the mass
    correction for $z<z^{*}\sim 1.5$: there is a small but systematic increase
    of average halo concentrations over the whole mass range. In the {\em
      right panel}, the simulation RP5-NO-FRIC shows the effect of switching
    off in the same redshift interval the velocity-dependent term. The increase of
    concentrations is in this case much more consistent and accounts for a
    large fraction of the total concentration reduction of RP5.}
\label{concentrations_nomass_nofric}
\end{figure*}

By switching off the mass variation for $z<z^{*}$, we find that the halo
concentrations at $z=0$ show a slight increase over the whole mass range of
the sample with respect to the fully self-consistent RP5 simulation. This
effect is shown in the left panel of
Fig.~\ref{concentrations_nomass_nofric}. We interpret this as a sign of the
fact that the mass decrease reduces the total gravitational potential energy
of halos, resulting in a modification of their virial equilibrium
configuration.

In fact, if the potential well of a halo gets shallower as time goes by as a
consequence of the decrease of the mass of its CDM content, the system will
find itself with an excess of kinetic energy, and will therefore expand in
order to restore virial equilibrium. Such an expansion is expected to cause a
drop of the halo concentrations, which we confirm here because switching off
this mechanisms yields consistently higher concentrations at $z=0$.  However,
it is clear from Fig.~\ref{concentrations_nomass_nofric} that this mechanism
cannot account for the total effect of concentration decrease, but only for a
small fraction of it.  

We therefore now investigate the other possible origin of this effect,
i.e.~the impact of the velocity-dependent term (\ref{friction_term}) on the dynamics of
CDM particles.  To this end we switch off for $z<z^{*}$ the additional
acceleration arising from the velocity-dependent term for coupled particles described by
Eqn.~(\ref{extra_friction}). The outcome of this test simulation is shown in
the right panel of Fig.~\ref{concentrations_nomass_nofric}: the increase
of the concentrations with respect to the fully self-consistent RP5 simulation
is now much more substantial than in the case of RP5-NO-MASS, and shows that
the velocity-dependent term is actually the dominant mechanism in determining the
decrease of halo concentrations and the decrease of the inner overdensity of
CDM halos discussed above.  The interpretation of this effect seems quite
unambiguous: the velocity-dependent term induces an extra acceleration on coupled
particles in the direction of their velocity, and this produces an increase of
the kinetic energy of the particles, moving the system out of its virial
equilibrium configuration. The system responds by a small expansion and a
lowering of the concentration.

\begin{figure}
\includegraphics[scale=0.45]{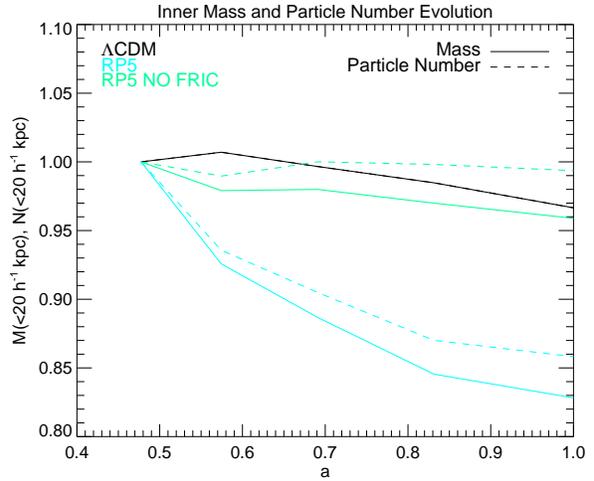}
  \caption{Evolution with respect to the scale factor $a$ of the average mass
    (solid) and of the average number of particles (dashed) enclosed in a
    sphere of physical radius $r=20 \mathrm{h}^{-1}\mathrm{kpc}$ centred on
    the potential minimum of each halo in our sample. The curves are
    normalized at $a=0.48$ ($z\sim 1$) and show the expected flat behaviour for
    the $\Lambda $CDM case (black line) for which the solid and the dashed
    curves coincide due to the constancy of the mass of particles. For the RP5
    case (light blue curves), there is a strong decrease in time of both mass
    and particle number, which clearly illustrates the expansion of RP5 halos
    with respect to the $\Lambda $CDM case. By switching off the extra
    velocity-dependent force acting on CDM particles (RP5-NO-FRIC, light green curves), an
    almost flat behaviour is recovered again for the particle number, while the
    decrease of mass is now due to the particle mass variation -- which is
    still in place for this simulation -- on top of the particle number
    evolution. This plot therefore clearly shows that the extra physics of
    coupled DE cosmologies induces an overall expansion of CDM halos at low
    redshifts, and clearly identifies in the velocity-dependent} term the leading
    mechanism that produces this expansion.
\label{halo_expansion}
\end{figure}

As a further check of this interpretation of our results, we also test
directly the dynamic evolution of halos to check whether they really slightly
expand in the presence of DE-CDM coupling. To this end, we compute for all the
halos in our sample the time evolution of the mass and the number of particles
contained in a sphere of physical radius $r = 20\,h^{-1}\mathrm{kpc}$
centred on the potential minimum of each halo. This sphere represents the
very innermost part of all the halos in our sample at any redshift between
$z^{*}$ and 0, and we refer to it as the halo ``core''; its mass content is
expected to be roughly constant for $\Lambda $CDM cosmologies at low
redshifts. Indeed, we can recover this behaviour for our $\Lambda $CDM
simulation by averaging the evolution of core masses and particle numbers over
the whole halo sample. On the other hand, for RP5, as expected according to
our interpretation, both the mass and the number of particles in the halo
cores strongly decrease with time. This is shown in Fig.~\ref{halo_expansion},
where the solid lines represent the average evolution of mass, and the dashed
lines represent the average evolution of the particle number. Evidently, for
$\Lambda$CDM the two curves coincide because the mass of the particles is
constant and any change of the enclosed mass in the core must be due to a
change of the number of enclosed particles. On the other hand, for RP5, the
mass and the particle number behave differently due to the mass variation of
CDM particles. The decrease of the number of particles contained in the core
can be interpreted as a manifestation of an expansion of the halos.  Moreover,
if we compute the same evolution for our RP5-NO-FRIC simulation, we find an
almost constant evolution of the core particle number and a very weak decrease
of the core mass due to the variation of CDM particle mass. This result also
confirms our interpretation concerning the origin of the decrease of
concentrations: the velocity-dependent term is the most relevant mechanism for inducing
halos expansion at low redshifts, and as a consequence the decrease of the
inner overdensity of CDM halos and of their concentration.

While further investigations of these effects are certainly required in order
to understand all the potential phenomenological features of interacting DE
models, our conclusion that a coupling between DE and CDM produces less peaked
halo density profiles and lower halo concentrations seems to be quite robust
based on the analysis of our simulations that we have discussed here.  We note
again that our findings are in stark contrast with the results of previous
work by \citet{Maccio_etal_2004} who found for coupled DE models a strong
increase in concentration and density profiles in the centre that more steeply
rise than $\Lambda$CDM.  The effects we find go in the direction of less
``cuspyness'' of halo density profiles, which is preferred by observations and
thus in fact opens up new room for the phenomenology of interacting DE models.

\subsection{Integrated bias and halo baryon fraction}

The extra force felt by CDM particles induces, as we have already seen for the
evolution of the matter power spectrum, a bias in the evolution of density
fluctuations of baryons and CDM \citep{Mainini:2005fe, Mainini:2006zj, Manera_Mota_2006}. We can
then use our selected halo sample to test the evolution of this bias from the
linear regime already probed by the power spectrum on large scales to the
highly nonlinear regime in the centre of massive collapsed structures.  We
then test the evolution of the integrated bias 
\begin{equation}
B(<r) \equiv \frac{\rho _{b}(<r) - \bar{\rho}_{b}}{\bar{\rho }_{b}} \cdot  \frac{\bar{\rho}_{c}}{ \rho _{c}(<r) - \bar{\rho }_{c}}\,,
\label{integrated_bias}
\end{equation}
as defined in \citet{Maccio_etal_2004}, where $\rho _{b}(<r)$ and $\rho
_{c}(<r)$ are the densities within a sphere of radius $r$ around the potential
minimum of a halo, for baryons and CDM, respectively.  Following
\citet{Maccio_etal_2004}, we have not used the innermost part of the halos ($r
< 10 h^{-1}{\rm kpc} \sim 3 \times \epsilon _{s}$) in order to avoid potential
resolution problems.  

\begin{figure*}
\includegraphics[scale=0.4]{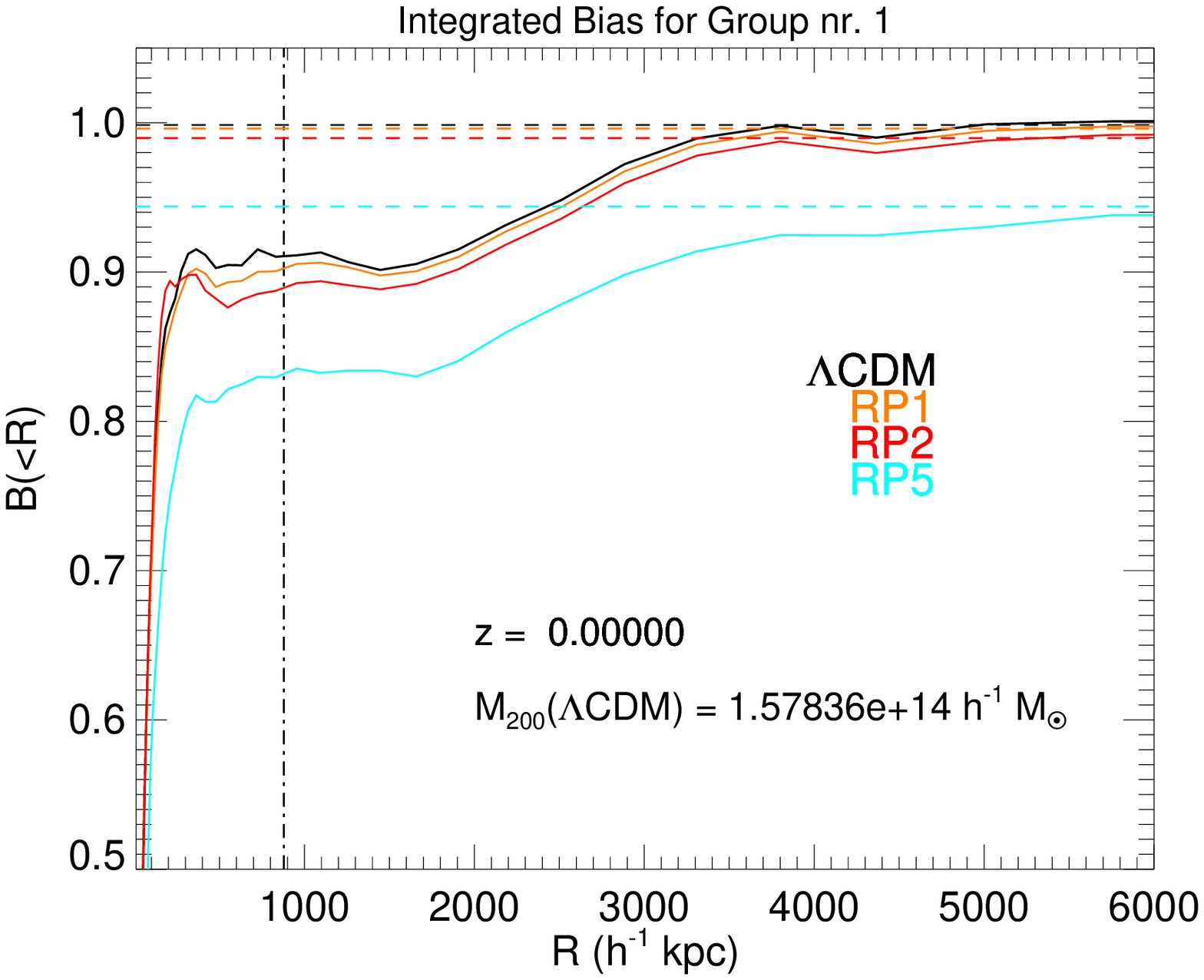}
\includegraphics[scale=0.4]{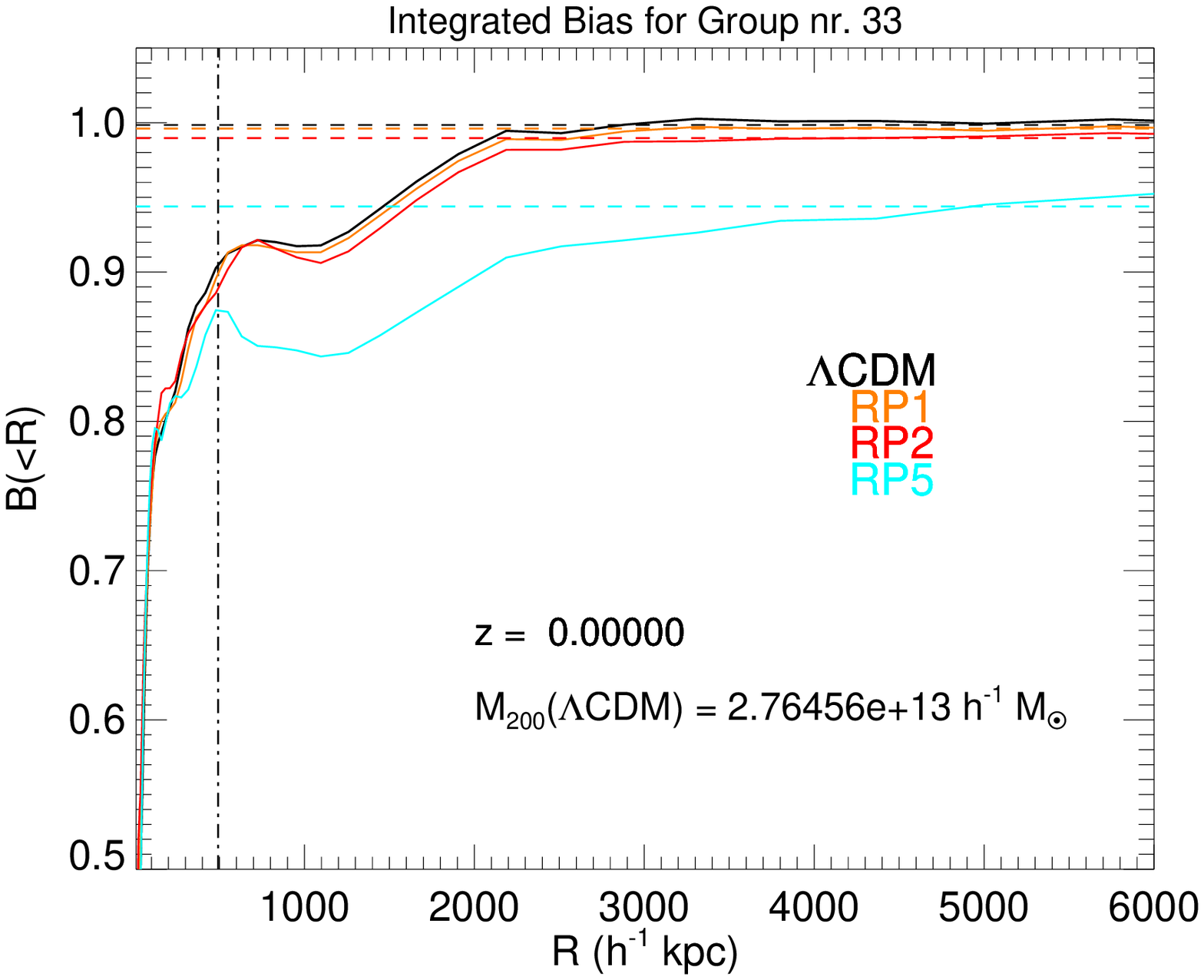}\\
\includegraphics[scale=0.4]{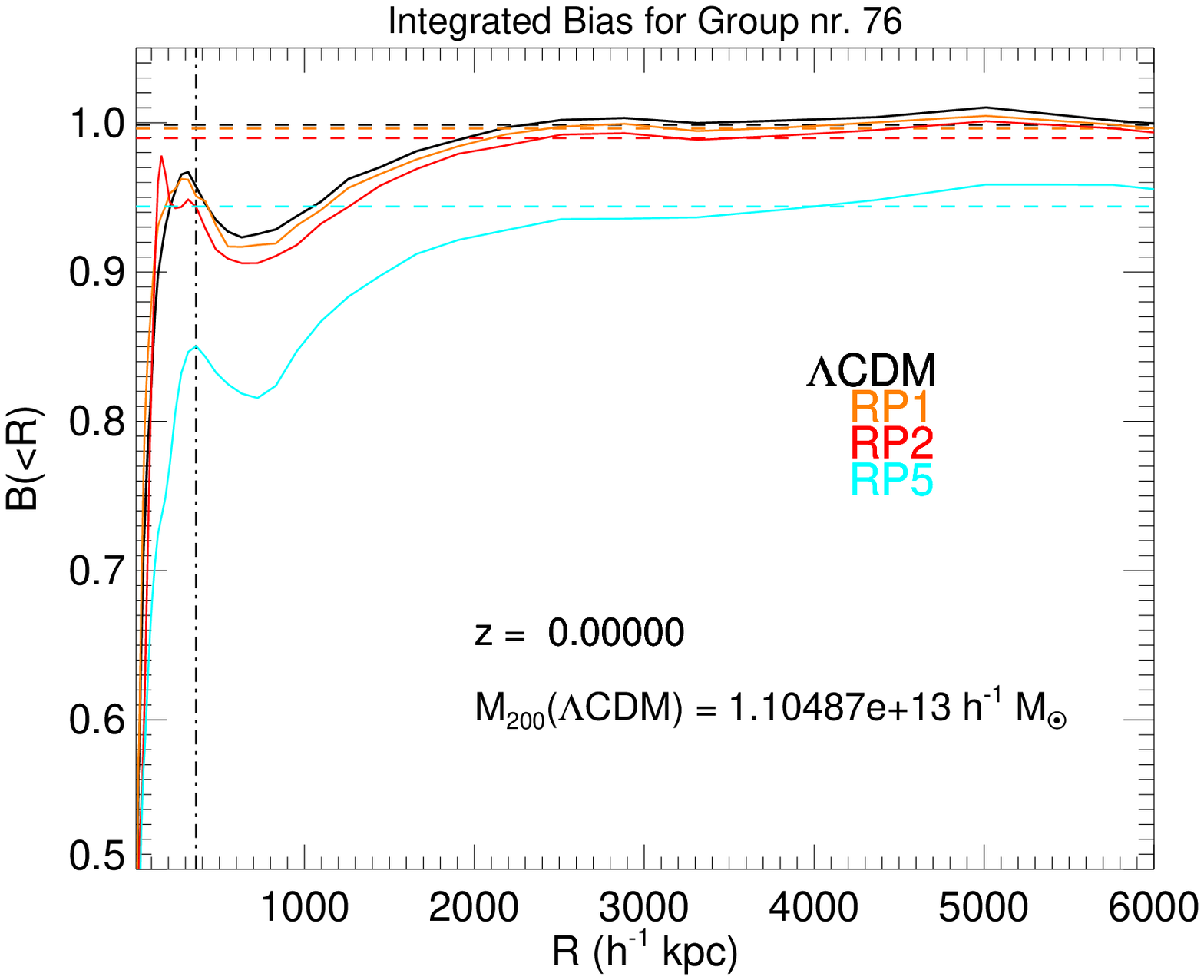}
\includegraphics[scale=0.4]{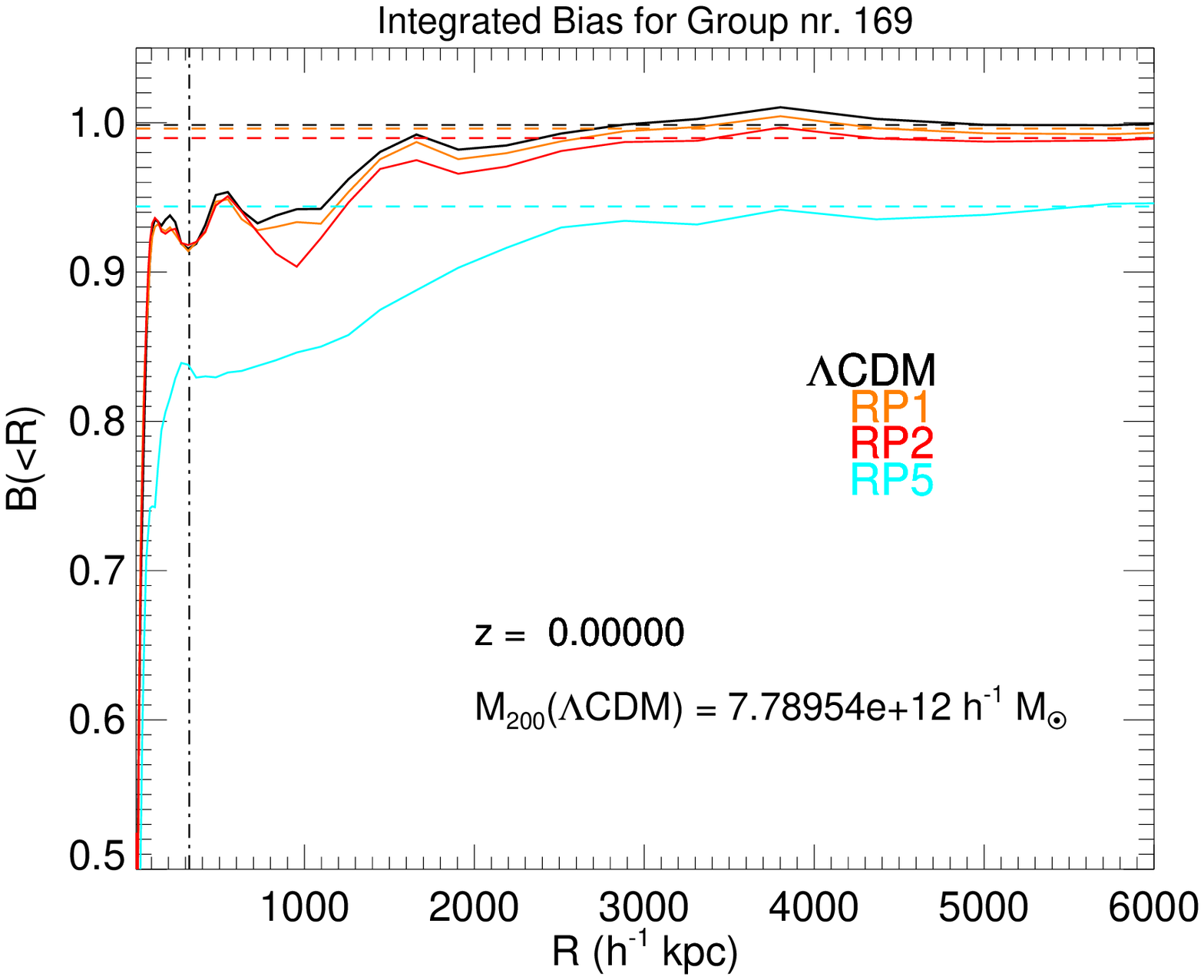}
  \caption{Evolution of the integrated bias $B(<r)$ for the four fully
    self-consistent high-resolution simulations and for four selected halos of
    different mass in our sample. The horizontal dashed lines indicate the
    value of the large scale linear bias as evaluated from the power spectrum
    amplitudes of baryons and CDM. The vertical black dot-dashed line shows
    the position of the virial radius for the $\Lambda$CDM halo in the
    sample. The drop of the value of $B(<r)$ in the innermost regions of the
    halos is evident but in these runs is given by a superposition of effects
    due to hydrodynamical forces and to the modified gravitational
    interaction. On large scales, the bias tends to converge to the linear
    value, as expected.}
\label{plot_bias}
\end{figure*}

In Figure~\ref{plot_bias}, we show the evolution of the bias for four selected
halos of our sample with similar masses as the ones shown in
Fig.~\ref{density_profiles}. It clearly appears that the bias is considerably
enhanced in the nonlinear regime, while at large scales it converges to the
linear value evaluated from the power spectrum amplitude on large scales,
represented in Fig.~\ref{plot_bias} by the horizontal dashed lines.

\begin{figure*}
\includegraphics[scale=0.4]{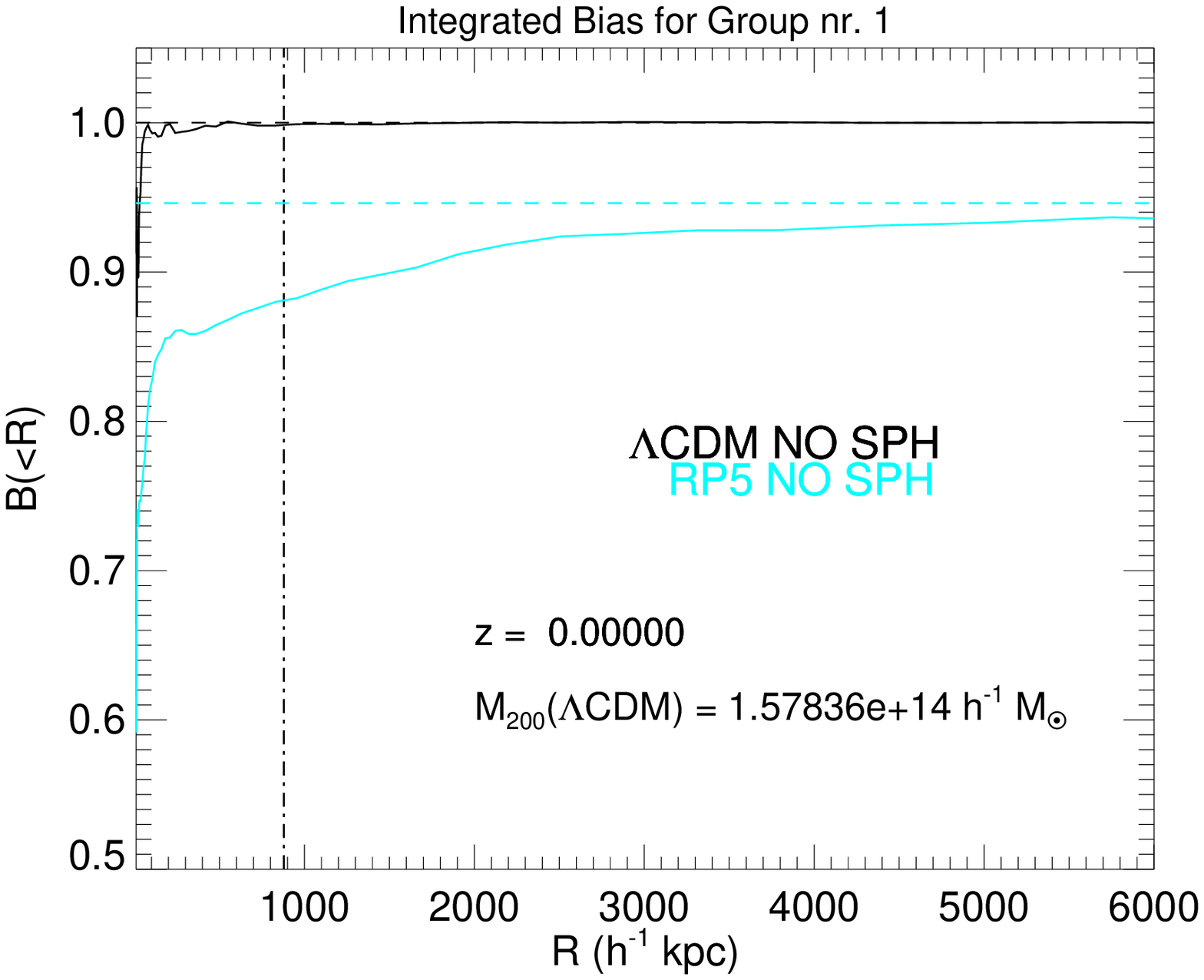}
\includegraphics[scale=0.4]{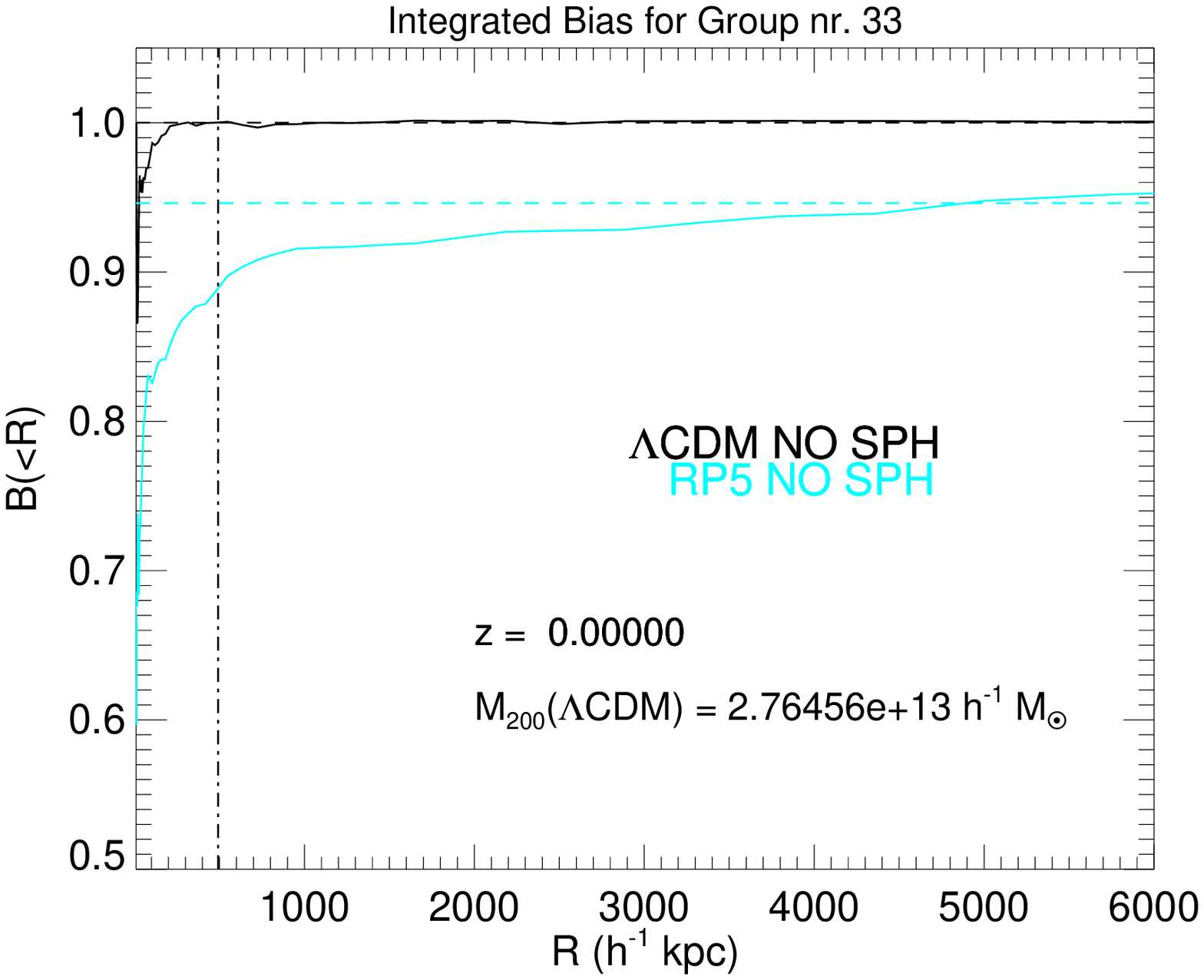}\\
\includegraphics[scale=0.4]{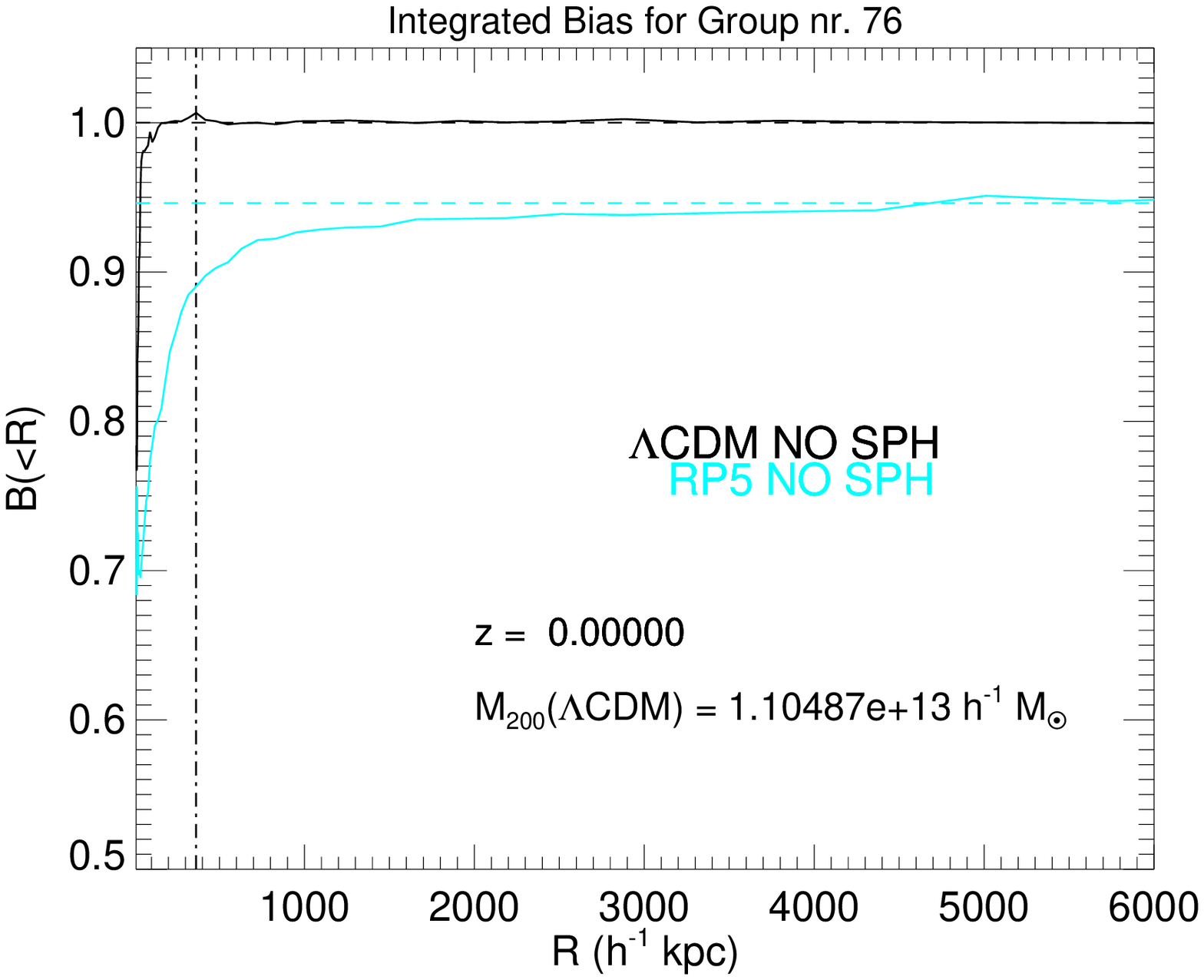}
\includegraphics[scale=0.4]{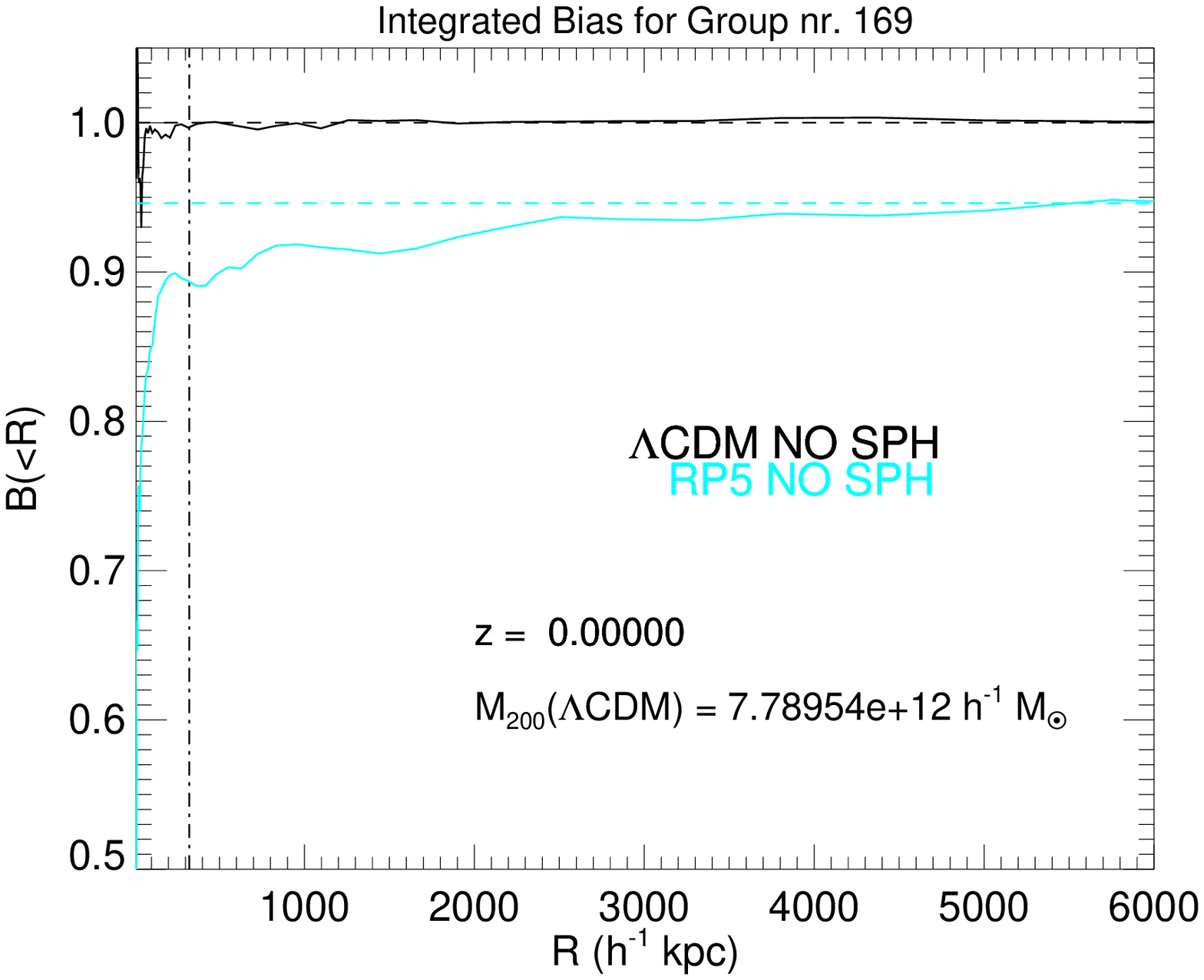}
  \caption{Evolution of the integrated bias $B(<r)$ for the two
    high-resolution simulations without hydrodynamical forces on baryon
    particles for the same four halos shown in Fig.~\ref{plot_bias}. The
    enhancement of the bias due to the extra scalar force in the core of
    highly nonlinear structures appears here clearly.}
\label{plot_bias_nosph}
\end{figure*}

However, also in this case this effect could be due only to the presence of
hydrodynamical forces acting on the baryons, and may not really be caused by
the fifth-forces from the coupled DE scalar field, as we can infer from the
fact that also the $\Lambda $CDM curve, where no coupled DE is present, shows
a departure from the large scale value of 1.0 when approaching the centre of
the halos.  Once again we make use of our additional test simulations
$\Lambda$CDM-NO-SPH and RP5-NO-SPH in order to disentangle the two effects.
In Fig.~\ref{plot_bias_nosph}, we show the same four plots as in
Fig.~\ref{plot_bias} for the two simulations without hydrodynamic forces, and
the appearance of a nonlinear bias imprinted only by the coupled DE scalar
field acting on CDM particles is then absolutely evident.  On the other hand,
the absence of any bias, as expected, in the $\Lambda$CDM-NO-SPH run shows
clearly that no major numerical problems can be responsible for the effect in
the RP5-NO-SPH simulations.
This evolution of the bias $B(<r)$ with radius for massive halos is in good agreement with that found by \citet{Maccio_etal_2004} despite the starkly different behavior of the individual overdensities in baryons and CDM found in the two works.

\begin{figure}
\includegraphics[scale=0.45]{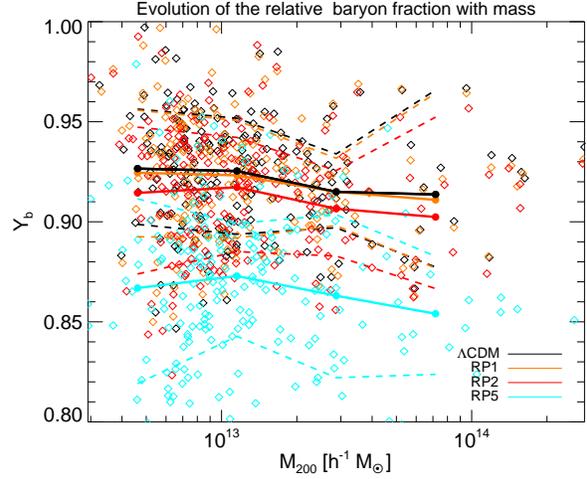}
  \caption{Evolution with virial mass $M_{200}$ of the relative baryon
    fraction $Y_{b}$ within the virial radius $r_{200}$ of all the halos in
    our sample. The coloured diamonds represent the relative baryon fraction of
    each single halo, while the filled circles and the coloured curves show the
    behaviour of the mean relative baryon fraction in each mass bin for the
    four fully self-consistent high-resolution simulations. A decrease of
    $Y_{b}$ with increasing coupling is clearly visible both in the
    distribution of the individual halos and in the averaged curves.}
\label{baryon_fraction}
\end{figure}

It is interesting that the above effect produces a baryon deficit in
virialized halos, i.e.~they contain fewer baryons than expected based on their
mass and the universal cosmological baryon fraction. 
In particular, this means
that one can not expect that baryon fractions determined through X-ray
measurements in clusters would yield the cosmological value.  
It is important to stress again here that our approach does not include non-adiabatic processes (like e.g.~star formation, cooling, and possible feedback mechanisms) to the physics of the baryonic component. Therefore our outcomes cannot be expected to provide a prediction of baryon fractions to be directly confronted with observations. However, this approach has the advantage to clearly show in which direction the DE coupling affects the baryon budget of collapsed structures which would then be available for such radiative processes.
In order to give
a rough estimate of the magnitude of this effect we compute the baryon
fraction within the virial radius $r_{200}$ of all the halos in our sample
defined as
\begin{equation}
f_{b} \equiv \frac{M_{b}(<r_{200})}{M_{\rm tot}(<r_{200})}
\label{b_frac}
\end{equation}
for our four fully self-consistent simulations.  We plot in
Fig.~\ref{baryon_fraction} as a function of halo virial mass the relative
baryon fraction defined as:
\begin{equation}
Y_{b}\equiv \frac{f_{b}}{\Omega _{b}/\Omega _{m}}\,.
\end{equation}
For the $\Lambda $CDM case, our results for the evolution of $Y_{b}$ are
consistent with the value of $Y_{b} \sim 0.92$ found by the {\em Santa Barbara
  Cluster Comparison Project} \citep{SBCCP}, and with the more recent results
of \citet{Ettori_etal_2006} and \citet{Gottloeber_Yepes_2007}, while for the
coupled models the relative baryon fraction shows a progressive decrease with
increasing coupling, down to a value of $Y_{b} \sim 0.86-0.87$ for the RP5
case.

It is also important to notice that this effect is always towards lower baryon
fractions in clusters with respect to the cosmological value. This could in
fact alleviate tensions between the high baryon abundance estimated from CMB
observations, and the somewhat lower values inferred from detailed X-ray
observations of galaxy clusters \citep{Vikhlinin_etal_2006,
  McCarthy_etal_2007, LaRoque_etal_2006, Afshordi_etal_2007}.

\section{Conclusions}
\label{concl}

We have investigated coupled DE cosmologies, both with respect to their
expected background and linear perturbation evolution, as well as in their
predictions for the nonlinear regime of structure formation.  To do so we have
developed and tested in this work a modified version of the cosmological
N-body code {\small GADGET-2} suitable for evolving these kinds of
cosmological models.  The numerical implementation we have developed is in
fact quite general and not restricted to the simple specific models of coupled
quintessence that we have investigated in this paper. Instead, it should be
well suited for a much wider range of DE models. We also note that the ability
to selectively enable or disable each of the modifications discussed above,
makes the code suitable for cosmological models that are unrelated to the coupled
DE scenario but require similar new degrees of freedom that our implementation
allows. These are:
\begin{enumerate}
\item the expansion history of the Universe can be specified according to any
  desired evolution of the Hubble rate as a function of the scale factor $a$;
\item a global variation in time of the gravitational constant and/or a
  variation in time of the gravitational strength for each individual matter
  species. This includes the possibility to have a long range repulsive
  interaction between different particle species;
\item variation in time of the particle mass of each individual matter
  species;
\item extra velocity-dependent terms in the equation of motion for each
  individual matter species.
\end{enumerate}

With this implementation we have investigated the effects of coupled DE models
with a constant coupling $\beta _{c}$ to the CDM fluid on structure
formation. We have shown that the halo mass function is modified in
coupled DE
models, but can still be well fitted at different redshifts by the {\em
  Jenkins et al.} \citep{Jenkins_etal_2000} fitting formula, or by the {\em
  Sheth \& Tormen} \citep{Sheth_Tormen_1999} formula, which yields a
moderately better agreement, especially at $z > 0$.

We have confirmed the analytic prediction that density fluctuations in baryons
and CDM will develop a bias on all scales due to the presence of a fifth-force
acting only between CDM particles.  We have also shown that in addition to
this the  bias is enhanced when moving from the linear regime of very
large scales to smaller and progressively more non-linear scales.

We have investigated the evolution of the bias between baryons and CDM
overdensities down to the very nonlinear regime found in the inner part of
collapsed objects, in the same fashion as described in
\citet{Maccio_etal_2004}.  We found here similar results with this previous
work, namely an enhancement of the bias in the nonlinear region within and
around massive halos. We also recover from this analysis the large scale value
of the linear bias computed from the power spectrum when integrating the bias
function up to very large radii from the centre of CDM halos.  The enhancement
of the bias in highly nonlinear structures has an impact on the determination
of the baryon fraction from cluster measurements, and we have computed for all our halos, without including non-adiabatic processes, the evolution of this fraction within the virial radius $r_{200}$
with coupling, finding that the baryon fraction is reduced with increasing
coupling by up to $\sim 8-10\%$ with respect to $\Lambda $CDM for the largest
coupling value.

We have also investigated the effect of the coupling on the halo density
profiles. We find that they are remarkably well fit over the resolved range by
the NFW formula for any value of the coupling. There is a clear trend of a
decrease of the inner halo overdensity with respect to $\Lambda $CDM with
increasing coupling (or, equivalently, an increase of the scale radius $r_{s}$
for increasing coupling). This result conflicts with previous claims for the
same class of coupled DE models \citep{Maccio_etal_2004}.

Using a number of special test simulations, we have identified the origin of
this effect of reduced halo concentrations for increasing coupling.  It
actually arises from a combination of two peculiar features that the coupling
introduces in the Newtonian limit of gravitational dynamics.  The first of
these is the decrease of CDM particle mass with time, which causes the total
potential energy of a halo to decrease, and hence effectively moves the system
to a configuration where an excess of kinetic energy is present relative to
virial equilibrium.  The second one is the additional velocity-dependent term,
which directly raises the total kinetic energy density of halos by
accelerating their particles in the direction of their peculiar velocity. Both
of these effects cause a halo to slightly expand in order to restore virial
equilibrium, and this reduces the halo concentration.

In conclusion, we have developed a general numerical implementation of coupled
dark energy models in the {\small GADGET-2} code. We have then performed the
first fully self-consistent high-resolution hydrodynamic N-body simulations of
interacting dark energy models with constant coupling, and carried out a basic
analysis of the non-linear structures that formed.  Interestingly, we found
that a larger coupling leads to a lower average halo concentration.
Furthermore, both the baryon fraction in massive halos and the inner
overdensity of CDM halos decrease with increasing coupling.  These effects
alleviate the present tensions between observations and the $\Lambda$CDM model
on small scales, implying that the coupled DE models are viable alternatives
to the cosmological constant included in standard $\Lambda$CDM.

\section*{Acknowledgments}

We are deeply thankful to L.~Amendola, C.~Baccigalupi, K.~Dolag,
A.~V.~Macci\`{o}, C.~Wetterich and S.~D.~M.~White for useful discussions and
suggestions on the physical models and on the numerical techniques.  MB wants
to acknowledge also M.~Boylan-Kolchin, M.~Grossi, D.~Sijacki and
M.~Vogelsberger for providing collaboration on numerical issues and for
sharing some pieces of codes for the post processing.  VP is supported by the
Alexander von Humboldt Foundation.  VP thanks C.~van~de~Bruck for helpful
discussion. VP and GR acknowledge M.~Frommert for collaborating to the modification
of {\small CMBEASY}.

This work has been supported by the TRR33 Transregio Collaborative Research
Network on the ``Dark Universe'', and by the DFG Cluster of Excellence
``Origin and Structure of the Universe''.


\bibliographystyle{mnras}
\bibliography{baldi_etal_2008_bibliography}

\begin{thebibliography}{82}
\expandafter\ifx\csname natexlab\endcsname\relax\def\natexlab#1{#1}\fi

\bibitem[Afshordi et~al.(2007)Afshordi, Lin, Nagai \&
  Sanderson]{Afshordi_etal_2007}
Afshordi N., Lin Y.-T., Nagai D., Sanderson A. J.~R., 2007, Mon. Not. Roy.
  Astron. Soc., 378, 293

\bibitem[Allen et~al.(2004)Allen, Schmidt, Ebeling, Fabian \& van
  Speybroeck]{Allen_etal_2004}
Allen S.~W., Schmidt R.~W., Ebeling H., Fabian A.~C., van Speybroeck L., 2004,
  Mon. Not. Roy. Astron. Soc., 353, 457

\bibitem[Amendola(2000)]{Amendola_2000}
Amendola L., 2000, Phys. Rev., D62, 043511

\bibitem[Amendola(2004)]{Amendola_2004}
Amendola L., 2004, Phys. Rev., D69, 103524

\bibitem[Amendola et~al.(2008)Amendola, Baldi \&
  Wetterich]{Amendola_Baldi_Wetterich_2008}
Amendola L., Baldi M., Wetterich C., 2008, Phys. Rev., D78, 023015

\bibitem[Amendola \& Quercellini(2004)]{Amendola_Quercellini_2004}
Amendola L., Quercellini C., 2004, Phys. Rev. Lett., 92, 181102

\bibitem[Anderson \& Carroll(1997)]{Anderson:1997un}
Anderson G.~W., Carroll S.~M., 1997

\bibitem[Astier et~al.(2006)]{SNLS}
Astier P., et~al., 2006, Astron. Astrophys., 447, 31

\bibitem[Baldi \& Macci\`{o}(in prep.)]{Baldi_Maccio_inprep}
Baldi M., Macci\`{o} A.~V., in prep.

\bibitem[Bardeen et~al.(1986)Bardeen, Bond, Kaiser \& Szalay]{BBKS}
Bardeen J.~M., Bond J.~R., Kaiser N., Szalay A.~S., 1986, Astrophys. J., 304,
  15

\bibitem[Bean et~al.(2008)Bean, Flanagan, Laszlo \& Trodden]{Bean:2008ac}
Bean R., Flanagan E.~E., Laszlo I., Trodden M., 2008

\bibitem[Bertolami et~al.(2007)Bertolami, Gil~Pedro \&
  Le~Delliou]{Bertolami:2007zm}
Bertolami O., Gil~Pedro F., Le~Delliou M., 2007, Phys. Lett., B654, 165

\bibitem[Binney \& Evans(2001)]{Binney_Evans_2001}
Binney J.~J., Evans N.~W., 2001, Mon. Not. Roy. Astron. Soc., 327, L27

\bibitem[Brookfield et~al.(2008)Brookfield, van~de Bruck \&
  Hall]{Brookfield:2007au}
Brookfield A.~W., van~de Bruck C., Hall L. M.~H., 2008, Phys. Rev., D77, 043006

\bibitem[Damour et~al.(1990)Damour, Gibbons \&
  Gundlach]{Damour_Gibbons_Gundlach_1990}
Damour T., Gibbons G.~W., Gundlach C., 1990, Phys. Rev. Lett., 64, 123

\bibitem[Di~Porto \& Amendola(2008)]{DiPorto_Amendola_2008}
Di~Porto C., Amendola L., 2008, Phys. Rev., D77, 083508

\bibitem[Doran(2005)]{CMBEASY}
Doran M., 2005, JCAP, 0510, 011

\bibitem[Doran \& Robbers(2006)]{EDE2}
Doran M., Robbers G., 2006, JCAP, 0606, 026

\bibitem[Doran et~al.(2001)Doran, Schwindt \& Wetterich]{EDE1}
Doran M., Schwindt J.-M., Wetterich C., 2001, Phys. Rev., D64, 123520

\bibitem[Eisenstein \& Hu(1998)]{Eisenstein_Hu_1997}
Eisenstein D.~J., Hu W., 1998, Astrophys. J., 496, 605

\bibitem[Ettori et~al.(2006)Ettori, Dolag, Borgani \&
  Murante]{Ettori_etal_2006}
Ettori S., Dolag K., Borgani S., Murante G., 2006, Mon. Not. Roy. Astron. Soc.,
  365, 1021

\bibitem[{Farrar} \& {Peebles}(2004)]{Farrar2004}
{Farrar} G.~R., {Peebles} P.~J.~E., 2004, \apj, 604, 1

\bibitem[{Farrar} \& {Rosen}(2007)]{Farrar2007}
{Farrar} G.~R., {Rosen} R.~A., 2007, Physical Review Letters, 98, 17, 171302

\bibitem[Flores \& Primack(1994)]{Flores_Primack_1994}
Flores R.~A., Primack J.~R., 1994, Astrophys. J., 427, L1

\bibitem[Francis et~al.(2008{\natexlab{a}})Francis, Lewis \&
  Linder]{Francis_etal_2008}
Francis M.~J., Lewis G.~F., Linder E.~V., 2008{\natexlab{a}}

\bibitem[Francis et~al.(2008{\natexlab{b}})Francis, Lewis \&
  Linder]{Francis_etal_2008b}
Francis M.~J., Lewis G.~F., Linder E.~V., 2008{\natexlab{b}}

\bibitem[{Frenk} et~al.(1999){Frenk}, {White}, {Bode} et~al.]{SBCCP}
{Frenk} C.~S., {White} S.~D.~M., {Bode} P., et~al., 1999, \apj, 525, 554

\bibitem[Gottloeber \& Yepes(2007)]{Gottloeber_Yepes_2007}
Gottloeber S., Yepes G., 2007, Astrophys. J., 664, 117

\bibitem[Gromov et~al.(2004)Gromov, Baryshev \& Teerikorpi]{Gromov:2002ek}
Gromov A., Baryshev Y., Teerikorpi P., 2004, Astron. Astrophys., 415, 813

\bibitem[Grossi \& Springel(2008)]{Grossi_2008}
Grossi M., Springel V., 2008

\bibitem[{Gubser} \& {Peebles}(2004)]{Gubser2004}
{Gubser} S.~S., {Peebles} P.~J.~E., 2004, \prd, 70, 12, 123511

\bibitem[Guo et~al.(2007)Guo, Ohta \& Tsujikawa]{Guo:2007zk}
Guo Z.-K., Ohta N., Tsujikawa S., 2007, Phys. Rev., D76, 023508

\bibitem[Huey \& Wandelt(2006)]{Huey_Wandelt_2006}
Huey G., Wandelt B.~D., 2006, Phys. Rev., D74, 023519

\bibitem[Jenkins et~al.(2001)]{Jenkins_etal_2000}
Jenkins A., et~al., 2001, Mon. Not. Roy. Astron. Soc., 321, 372

\bibitem[Kesden \& Kamionkowski(2006)]{Kesden_Kamionkowski_2006}
Kesden M., Kamionkowski M., 2006, Phys. Rev., D74, 083007

\bibitem[Keselman et~al.(2009)Keselman, Nusser \&
  Peebles]{Keselman_Nusser_Peebles_2009}
Keselman J.~A., Nusser A., Peebles P. J.~E., 2009

\bibitem[Kodama \& Sasaki(1984)]{Kodama_Sasaki_1984}
Kodama H., Sasaki M., 1984, Prog. Theor. Phys. Suppl., 78, 1

\bibitem[Komatsu et~al.(2008)]{wmap5}
Komatsu E., et~al., 2008

\bibitem[La~Vacca et~al.(2009)La~Vacca, Kristiansen, Colombo, Mainini \&
  Bonometto]{LaVacca_etal_2009}
La~Vacca G., Kristiansen J.~R., Colombo L. P.~L., Mainini R., Bonometto S.~A.,
  2009, JCAP, 0904, 007

\bibitem[LaRoque et~al.(2006)]{LaRoque_etal_2006}
LaRoque S., et~al., 2006, Astrophys. J., 652, 917

\bibitem[Laszlo \& Bean(2008)]{Laszlo_Bean_2008}
Laszlo I., Bean R., 2008, Phys. Rev., D77, 024048

\bibitem[Lee et~al.(2006)Lee, Liu \& Ng]{Lee:2006za}
Lee S., Liu G.-C., Ng K.-W., 2006, Phys. Rev., D73, 083516

\bibitem[Ma \& Bertschinger(1995)]{Ma_Bertschinger_1995}
Ma C.-P., Bertschinger E., 1995, Astrophys. J., 455, 7

\bibitem[Macci\`{o} et~al.(2004)Macci\`{o}, Quercellini, Mainini, Amendola \&
  Bonometto]{Maccio_etal_2004}
Macci\`{o} A.~V., Quercellini C., Mainini R., Amendola L., Bonometto S.~A.,
  2004, Phys. Rev., D69, 123516

\bibitem[Mainini(2005)]{Mainini:2005fe}
Mainini R., 2005, Phys. Rev., D72, 083514

\bibitem[Mainini \& Bonometto(2006)]{Mainini:2006zj}
Mainini R., Bonometto S., 2006, Phys. Rev., D74, 043504

\bibitem[Mainini \& Bonometto(2007{\natexlab{a}})]{Mainini:2007ft}
Mainini R., Bonometto S., 2007{\natexlab{a}}, JCAP, 0706, 020

\bibitem[Mainini \& Bonometto(2007{\natexlab{b}})]{Mainini_Bonometto_2007}
Mainini R., Bonometto S., 2007{\natexlab{b}}, JCAP, 0706, 020

\bibitem[Manera \& Mota(2006)]{Manera_Mota_2006}
Manera M., Mota D.~F., 2006, Mon. Not. Roy. Astron. Soc., 371, 1373

\bibitem[Mangano et~al.(2003)Mangano, Miele \& Pettorino]{Mangano:2002gg}
Mangano G., Miele G., Pettorino V., 2003, Mod. Phys. Lett., A18, 831

\bibitem[Matarrese et~al.(2003)Matarrese, Pietroni \&
  Schimd]{Matarrese_etal_2003}
Matarrese S., Pietroni M., Schimd C., 2003, JCAP, 0308, 005

\bibitem[McCarthy et~al.(2007)McCarthy, Bower \& Balogh]{McCarthy_etal_2007}
McCarthy I.~G., Bower R.~G., Balogh M.~L., 2007, Mon. Not. Roy. Astron. Soc.,
  377, 1457

\bibitem[Moore(1994)]{Moore_1994}
Moore B., 1994, Nature, 370, 629

\bibitem[Navarro et~al.(1996)Navarro, Frenk \& White]{Navarro_Frenk_White_1995}
Navarro J.~F., Frenk C.~S., White S. D.~M., 1996, Astrophys. J., 462, 563

\bibitem[Navarro et~al.(1997)Navarro, Frenk \& White]{NFW}
Navarro J.~F., Frenk C.~S., White S. D.~M., 1997, Astrophys. J., 490, 493

\bibitem[Navarro \& Steinmetz(2000)]{Navarro_Steinmetz_2000}
Navarro J.~F., Steinmetz M., 2000, Astrophys. J., 528, 607

\bibitem[Neto et~al.(2007)]{Neto_etal_2007}
Neto A.~F., et~al., 2007

\bibitem[Newman et~al.(2009)]{Newman_etal_2009}
Newman A.~B., et~al., 2009

\bibitem[Nusser et~al.(2005)Nusser, Gubser \&
  Peebles]{Nusser_Gubser_Peebles_2005}
Nusser A., Gubser S.~S., Peebles P. J.~E., 2005, Phys. Rev., D71, 083505

\bibitem[Oyaizu(2008)]{Oyaizu_2008}
Oyaizu H., 2008

\bibitem[Peebles(1980)]{Peebles_1980}
Peebles P. J.~E., 1980, {The Large-Scale Structure of the Universe}, Princeton
  University Press

\bibitem[Percival et~al.(2001)]{Percival_etal_2001}
Percival W.~J., et~al., 2001, Mon. Not. Roy. Astron. Soc., 327, 1297

\bibitem[Perlmutter et~al.(1999)]{Perlmutter_etal_1999}
Perlmutter S., et~al., 1999, Astrophys. J., 517, 565

\bibitem[Pettorino \& Baccigalupi(2008)]{Pettorino_Baccigalupi_2008}
Pettorino V., Baccigalupi C., 2008, Phys. Rev., D77, 103003

\bibitem[Quartin et~al.(2008)Quartin, Calvao, Joras, Reis \&
  Waga]{Quartin:2008px}
Quartin M., Calvao M.~O., Joras S.~E., Reis R. R.~R., Waga I., 2008, JCAP,
  0805, 007

\bibitem[Ratra \& Peebles(1988)]{Ratra_Peebles_1988}
Ratra B., Peebles P. J.~E., 1988, Phys. Rev., D37, 3406

\bibitem[Riess et~al.(1998)]{Riess_etal_1998}
Riess A.~G., et~al., 1998, Astron. J., 116, 1009

\bibitem[Sand et~al.(2002)Sand, Treu \& Ellis]{Sand_etal_2002}
Sand D.~J., Treu T., Ellis R.~S., 2002, Astrophys. J., 574, L129

\bibitem[Sand et~al.(2004)Sand, Treu, Smith \& Ellis]{Sand_etal_2004}
Sand D.~J., Treu T., Smith G.~P., Ellis R.~S., 2004, Astrophys. J., 604, 88

\bibitem[Sheth \& Tormen(1999)]{Sheth_Tormen_1999}
Sheth R.~K., Tormen G., 1999, Mon. Not. Roy. Astron. Soc., 308, 119

\bibitem[Simon et~al.(2003)Simon, Bolatto, Leroy \& Blitz]{Simon_etal_2003}
Simon J.~D., Bolatto A.~D., Leroy A., Blitz L., 2003, Astrophys. J., 596, 957

\bibitem[Springel(2005)]{gadget-2}
Springel V., 2005, Mon. Not. Roy. Astron. Soc., 364, 1105

\bibitem[{Springel} \& {Farrar}(2007)]{Springel2007}
{Springel} V., {Farrar} G.~R., 2007, \mnras, 380, 911

\bibitem[{Springel} et~al.(2008){Springel}, {Wang}, {Vogelsberger}
  et~al.]{Aquarius}
{Springel} V., {Wang} J., {Vogelsberger} M., et~al., 2008, \mnras, 391, 1685

\bibitem[{Springel} et~al.(2001){Springel}, {White}, {Tormen} \&
  {Kauffmann}]{Springel2001}
{Springel} V., {White} S.~D.~M., {Tormen} G., {Kauffmann} G., 2001, \mnras,
  328, 726

\bibitem[Stabenau \& Jain(2006)]{Stabenau_Jain_2006}
Stabenau H.~F., Jain B., 2006, Phys. Rev., D74, 084007

\bibitem[Sutter \& Ricker(2008)]{Sutter_Ricker_2008}
Sutter P.~M., Ricker P.~M., 2008

\bibitem[Vikhlinin et~al.(2006)]{Vikhlinin_etal_2006}
Vikhlinin A., et~al., 2006, Astrophys. J., 640, 691

\bibitem[Wang et~al.(2007)Wang, Zang, Lin, Abdalla \& Micheletti]{Wang:2006qw}
Wang B., Zang J., Lin C.-Y., Abdalla E., Micheletti S., 2007, Nucl. Phys.,
  B778, 69

\bibitem[Wetterich(1988)]{Wetterich_1988}
Wetterich C., 1988, Nucl. Phys., B302, 668

\bibitem[Wetterich(1995)]{Wetterich_1995}
Wetterich C., 1995, Astron. Astrophys., 301, 321

\bibitem[Zel'dovich(1970)]{Zeldovich_1970}
Zel'dovich Y.~B., 1970, Astron. Astrophys., 5, 84

\end{thebibliography}

\label{lastpage}

\end{document}